\documentclass[aps,twocolumn,superscriptaddress,showpacs,pra,amssymb, 
amsmath,latexsym]{revtex4}
\usepackage[dvips]{graphicx}
\newtheorem{thm}    {Theorem}%[section]
\newtheorem{lem}     {Lemma}%[section]
\newtheorem{cor}  {Corollary}%[section]
\newcommand{\defeq}{\stackrel{\rm def}{=}}

\def\Tr{\mathop{\rm Tr}\nolimits}

\def\real{\mathbb{R}}

\def\Label#1{\label{#1}\ [\ #1\ ]\ }
\def\Label{\label}

\def\cl#1{{\cal#1}}
\def\ket#1{|#1\rangle}
\def\bra#1{\langle#1|}

\begin{document}

\title{Group theoretical study of LOCC-detection of \\ maximally entangled state using hypothesis testing}
\author{Masahito Hayashi}
\email{hayashi@math.is.tohoku.ac.jp}
\address{Graduate School of Information Sciences, Tohoku University, Aoba-ku, Sendai, 980-8579, Japan}
%02.20.-a 	Group theory 
%(for algebraic methods in quantum mechanics, see 03.65.Fd; 
%for symmetries in elementary particle physics, see 11.30.-j)
%03.65.Ta Foundations of quantum mechanics; measurement theory 
%(for optical tests of quantum theory, see 42.50.Xa)
%03.65.Ud Entanglement and quantum nonlocality 
%(e.g. EPR paradox, Bell's inequalities, GHZ states, etc.) 
%(for entanglement production in quantum information, see 03.67.Mn; 
%for entanglement in Bose-Einstein condensates, see 03.75.Gg)
%03.65.Wj State reconstruction, quantum tomography
\begin{abstract}
In the asymptotic setting,
the optimal test for hypotheses testing of the maximally entangled state
is derived under several locality conditions for measurements.
The optimal test is obtained in several cases
with the asymptotic framework as well as the finite-sample framework.
In addition, the experimental scheme 
for the optimal test is presented.
\end{abstract}
\maketitle
\section{Introduction}\Label{s1}
Recently various quantum information processings are proposed, 
and many of them require maximally entangled states as 
resources\cite{bbcjpw,bennett-wiesher,ekert}.
Hence, it is often desired to generate maximally entangled states 
experimentally. 
In particular, it must be based on statistical method to decide
whether the state generated 
experimentally is really the required maximally entangled state.

Now, entanglement witness is often used as its standard 
method \cite{dariano, guhne, horodecki3, lewenstein, terhal}.
It is, however, not necessarily the optimal method 
from a viewpoint of statistics. 
On the other hand, in mathematical statistics,
the decision problem of the truth of the given hypothesis 
is called statistical hypothesis testing, and 
is systematically studied.
Hence, it is desired to treat, under the frame of 
statistical hypotheses testing, 
the problem deciding
whether the given quantum state is the required maximally 
entangled state. 
In statistical hypotheses testing,
we suppose two hypotheses (null hypothesis and alternative hypothesis) 
to be tested a priori, and assume that one of both is true.
Based on observed data, 
we decide which hypothesis is true.
Most preceding studies about 
quantum hypotheses testing
concerns only
the simple hypotheses testing, in which,
both of the null and the alternative hypotheses 
consist of a single quantum state.
For example, 
quantum Neymann Pearson lemma \cite{Ho72, helstrom} 
and quantum Stein's lemma\cite{hiai-petz, ogawa-nagaoka, Nagaoka-converse, hayashi-hypo},
quantum Chernoff bound\cite{ACMMABV, NS}, and
quantum Hoeffding bound\cite{Oga-Hay, H-expo, N-expo} 
treat simple hypotheses.

However, in a practical viewpoint,
it is unnatural to specify both hypotheses with one quantum state. 
Hence, we cannot directly apply quantum Neymann Pearson 
theorem and quantum Stein's lemma,
and we have to treat composite hypotheses,
{\it i.e.}, the case where
both hypotheses consist of plural quantum states.
It is also required to restrict our measurements for testing
among measurements based on
LOCC (local operations and classical communications)
because the tested state  
is maximally entangled state.

Recently, based on quantum statistical inference\cite{helstrom,HolP,selected},
Hayashi {\it et~al.}\cite{tsu} discussed this testing problem
under statistical hypotheses testing with a locality condition.
They treated testing problem where the null hypothesis 
consists only of the required maximally entangled state.
Their analysis has been extended to more experimental setting\cite{theory}, and its effectivity has been experimentally demonstrated \cite{exper}.
Modifying this setting,
Owari and Hayashi \cite{OH} clarified the difference in performance between 
the one-way LOCC restriction and the two-way LOCC restriction in a specific case.
Especially, Hayashi {\it et~al.}\cite{tsu}
studied the optimal test and 
the existence of the uniformly optimal test
(whose definition will be presented later)
when one or two samples of the state to be tested
are given.
Their analysis mainly concentrated the two-dimensional case.

In this paper, we treat the null hypothesis consisting of 
quantum states whose fidelity for 
the desired maximally entangled state is not
greater than $\epsilon$,
and discuss this testing problem with several given samples of 
the tested state in the following three setting
concerning the range of our measurements. 
(Note that our previous paper \cite{tsu} treats the case of $\epsilon =0$.)
In this problem, there are two kinds of locality restrictions.
{\bf L1}: One is locality concerning the two distinct parties.
{\bf L2}: The other is that concerning the samples.
{\bf M1}: All measurements are allowed.
{\bf M2}: 
There is restriction on the locality {\bf L1}, but no restriction on the locality {\bf L2}.
{\bf M3}: There is restriction on 
the locality {\bf L2} as well as {\bf L1}.
The restrction {\bf M3} for measurement is 
discussed by Virmani and Plenio \cite{virmani-plenio}, the first time.
Hayashi {\it et~al.}\cite{tsu} treated the settings {\bf M2} and 
{\bf M3}, more systematically.

This paper mainly treats the case of sufficiently many samples, 
{\it i.e.}, the first order asymptotic theory.
As a result, we find that there is no difference in
performances of both settings {\bf M1} and {\bf M2}.
Especially, the test achieving the asymptotically optimal 
performance can be realized by 
quantum measurement with quantum correlations 
between only two local samples.
That is, even if we use any higher quantum correlations among
local samples, no further improvement is available
under the first order asymptotic frame work.
In the two-dimensional case, 
the required measurement with local quantum correlations
is the four-valued Bell measurement between 
the local two samples.
In the setting {\bf M3}, we treat 
the null hypothesis consisting only of
the maximally entangled state.
Then, it is proved that
even if we use classical correlation between 
local samples for deciding local measurement,
there is no further improvement.
That is, 
the optimal protocol can be realized by repeating the optimal measurement in the one-sample case in the setting {\bf M3}.

Concerning the non-asymptotic setting,
we derive the optimal test with arbitrary finite number of samples
under a suitable group symmetry.
This result can be trivially extended to
hypothesis testing of arbitrary pure state.
Moreover, we derive 
the optimal test with two samples
under the several conditions,
and calculate its optimal performance.

Furthermore, we treat the case when
each sample system consists of two or three different quantum systems whose state
is a tensor product state of different states.
In this case, 
even if the number of samples is one,
every party consists of multiple systems.
%Then, we formulate the locality {\bf L2} as 
%the locality among different quantum systems on respective parties.
As a result, we obtain the optimal test for the one-sample case in both settings {\bf M2} and {\bf M3}.
It is proved that
repeating the optimal measurement for one sample
gives the test achieving the asymptotically optimal performance.
Moreover, 
when each sample system consists of two different system,
it is shown that
the optimal measurement for the one-sample case
can be realized by a four-valued Bell measurement
on the respective parties.
Repeating this measurement yields the optimal performance in the first order asymptotic framework.
(Indeed, it is difficult to perform the quantum measurement
with quantum correlation between two samples
because we need to prepare two samples 
from the same source at the same time.
However, in this formulation, 
it is sufficient to prepare two state from the different source.)
When each sample system consists of three different systems,
the optimal measurement can be described by the GHZ state
$\frac{1}{\sqrt{d}}\sum_i |i\rangle|i\rangle|i\rangle$,
where $d$ is the dimension of the system.
This fact seems to indicate the importance of the GHZ state 
in the three systems.

Concerning locality restriction on our measurement,
it is natural to treat two-way LOCC, but we treat one-way LOCC and 
separable measurement.
This is because the separability condition is easier to treat than 
two-way LOCC.
Hence, this paper mainly adopts separability as 
a useful mathematical condition.
It is contrast that Virmani and Plenio \cite{virmani-plenio} used 
the PPT condition and Hayashi {\it et~al.}\cite{tsu}
partially used the PPT condition.

This paper is organized as follows.
The mathematical formulation of statistical hypotheses testing is 
given in section \ref{s2} and, 
the group theoretical symmetry is explained in section \ref{s3B}.
In section \ref{s3C}, 
we explain the restrictions of our measurement for our testing,
for example, one-way LOCC, two-way LOCC, separability, {\it etc}.
In section \ref{s4}, we review the fundamental knowledge of
statistical hypotheses testing for the probability distributions
as preliminary.
In section \ref{s5}(section \ref{s6}, section \ref{s7}),
the setting {\bf M1}({\bf M2}, {\bf M3}) is discussed, respectively.
Further results in the two-dimensional case
are presented in section \ref{s6E}.
Finally, in section \ref{s8} (section \ref{s9}),
we discuss the case of two (three) different quantum states, respectively.

\section{Mathematical formulation of quantum hypothesis testing}\Label{s2}
Let ${\cl H}$ be a finite-dimensional Hilbert space
corresponding to the physical system of interest.
Then, the state is described by a density matrix on ${\cl H}$.
In the quantum hypothesis testing,
we assume that the current state $\rho$ of
the system is unknown, but 
is known to belong to a subset 
${\cl S}_0$ or ${\cl S}_1$ of the set of densities.
Hence, our task is testing
\begin{equation}
	\Label{eq:hypo}
	H_0:\rho\in{\cl S}_0\quad \mbox{versus}
\quad H_1:\rho\in{\cl S}_1
\end{equation}
based on an appropriate measurement on ${\cl H}$.
That is, we are required to decide which hypothesis is true.
We call $H_0$ {\it a null hypothesis},
and we call $H_1$ {\it an alternative hypothesis}.

A test for the hypothesis (\ref{eq:hypo})
is given by a Positive Operator Valued Measure (POVM)
$\{T_0,T_1\}$
on ${\cl H}$ composed of two elements,
where $T_0+T_1=I$.
For simplicity,
the test $\{T_0,T_1\}$ is described by the operator $T=T_0$.
Our decision should be done based on this test as follows:
We accept $H_0$ (=we reject $H_1$) if we observe $T_0$,
and
we accept $H_1$ (=we reject $H_0$) if we observe $T_1$.
In order to treat its performance, we focus on the following two kinds
of errors.:
A type 1 error is an event such that
we accept $H_1$ though $H_0$ is true.
A type 2 error is an event such that
we accept $H_0$ though $H_1$ is true.
Hence, we treat the following two kinds of error probabilities:
The type 1 error probability $\alpha(T,\rho)$
and
the type 2 error probabilities $\beta(T,\rho)$
are given by
\begin{align*}
	\alpha(T,\rho)&=\Tr(\rho T_1) = 1- \Tr (\rho T)
	\ (\rho\in{\cl S}_0),\\
	\beta (T,\rho)&=\Tr(\rho T_0) = \Tr (\rho T)
	\ (\rho\in{\cl S}_1).
\end{align*}
A quantity $1-\beta(T,\rho)$ is called {\it power}.
A test $T$ is said to be {\it level-$\alpha$}
if $\alpha(T,\rho)\le\alpha$ for any $\rho\in{\cl S}_0$.

In hypothesis testing, 
we restrict our test to tests whose first error probability
is greater than a given constant $\alpha$ for any element
$\rho \in {\cl S}_0$.
That is, 
since the type 1 error is considered to be more serious than 
the type 2 error
in hypothesis testing,
it is required to guarantee that 
the type 1 error probability 
is less than a constant which is called level of significance
or level.
Hence, a test $T$ is said to be {\it level-$\alpha$}
if $\alpha(T,\rho)\le\alpha$ for any $\rho\in{\cl S}_0$.

Then, under this condition, 
the performance of the test is given by
$1- \beta(T,\rho)$ for $\rho \in {\cl S}_1$,
which is called {\it power}.
Therefore, we often optimize 
the type 2 error probability as follows:
\begin{align*}
\beta_{\alpha}(\cl{S}_0\|\rho)& \defeq
\min_{T\in{\cl T}_{\alpha,\cl{S}_0}}
\beta(T,\rho)
 , \\
\cl{T}_{\alpha,\cl{S}_0}
& \defeq 
\{ T| 0 \le T \le I , \quad 
\alpha(T,\rho)\le \alpha \forall \rho \in {\cl S}_0\}
\end{align*}
for any $\rho\in{\cl S}_1$.
Especially, a test $T\in{\cl T}_{\alpha,\cl{S}_0}$ is called
{\it a Most Powerful (MP) test with level $\alpha$
at $\rho\in{\cl S}_1$}
if $\beta(T,\rho)\le\beta(T',\rho)$
for any level-$\alpha$ test $T'\in{\cl T}_{\alpha,\cl{S}_0}$,
that is,
\begin{align*}
\beta(T,\rho)=
\beta_{\alpha}(\cl{S}_0\|\rho).
\end{align*}
Moreover, a test $T\in{\cl T}_{\alpha,\cl{S}_0}$ is called
{\it a Uniformly Most Powerful (UMP) test}
if $T$ is MP for any level-$\alpha$ test 
$\rho\in{\cl S}_1$, that is,
\begin{align*}
\beta(T,\rho)=
\beta_{\alpha}(\cl{S}_0\|\rho), \quad \forall \rho \in {\cl S}_1.
\end{align*}
However, 
in certain instances,
it is natural to restrict our testings 
to those satisfying one or two conditions ($C_1$ or $C_1$ and $C_2$).
In such a case, 
we focus on the following quantity in stead of
$\beta(T,\rho)$:
\begin{align*}
\beta_{\alpha,C_1}^{C_2}(S_0\|\rho)& \defeq
\min_{T\in{\cl T}_{\alpha,S_0}}
\{ \beta(T,\rho)| \hbox{$T$ satisfies $C_1$ and $C_2$.}\}.
\end{align*}
If a test $T \in{\cl T}_{\alpha,S_0}$ satisfies
conditions $C_1$, $C_2$, and
\begin{align*}
\beta(T,\rho)=
\beta_{\alpha,C_1}^{C_2}(\cl{S}_0\|\rho), \quad \forall \rho \in {\cl S}_1,
\end{align*}
it is called {\it a Uniformly Most Powerful $C_1,C_2$ (UMP $C_1,C_2$) 
test}.

%there does not neceessarily exist
%a UMP test.
%Hence, we often restrect our testings as follows.
%When the hypotheses $S_0$ and $S_1$ are invariant for a
%unitary action of a group $G$,
%it is natural to restrict our test to a test invariant for this 
%action.
%In this case, we focus on the following quantities:
%\begin{align}
%\beta_{\alpha,G}(S_0\|\rho)& :=
%\min_{T\in{\cl T}_{\alpha,S_0}}
%\{\beta(T,\rho)| \hbox{} , \\
%\cl{T}_{\alpha,\cl{S}_0}& \defeq \{ T| 0 \le T \le I , \quad 
%\alpha(T,\rho)\le \alpha \forall \rho \in {\cl S}_0\}
%\end{align}
%with several types of conditions, that is,
%we focus on 
%\begin{align}
%\beta_{\alpha,C}(H_0\|\rho):=\min_{T\in{\cl T}_C}\{\beta(T,\rho)
%|\alpha(T,\rho)\le \alpha \forall \rho \in {\cl S}_0\} 
%\end{align}
%for any $\rho\in{\cl S}_1$,
%where ${\cl T}_C(\subset{\cl T}({\cl H}))$ denotes the set of 
%level-$\alpha$ tests which satisfy a condition $C$.
%However, there does not neccessarily exist 
%a UMP $C$ level-$\alpha$ test.
%That is, it is possible that
%there exist two tests $T$ and $T'$ such that
%$\beta(T,\rho)\,<\beta(T',\rho)$ for a $\rho\in{\cl S}_1$,
%but $\beta(T,\rho')\,>\beta(T',\rho')$ for another $\rho'\in{\cl S}_1$.
%In such a case, we need another concept charactering good tests.
%A test is called {\it a test $T$ dominates a test $T'$}
%if $\beta(T,\rho)\le\beta(T',\rho)$ for any $\rho\in{\cl S}_1$ and
%if the inequality is strict somewhere. A test $T\in{\cl T}_C$ is said to 
be
%{\it admissible in ${\cl T}_C$} if no other $T'\in{\cl T}_C$ dominates 
$T$.
%A UMP test is admissible. If there is no A UMP test,
%we usually seek an admissible test in hypothesis testing.

\section{Our formulations}\Label{s3}
\subsection{Hypothesis}\Label{s3A}
Our target is teasting wheather the generated state is sufficiently close to 
the maximal entangled state
\[
	\ket{\phi^0_{AB}}=
	\frac{1}{\sqrt d}\sum_{i=0}^{d-1}
	\ket{i}_{A}\otimes\ket{i}_{B}
\]
on the tensor product space ${\cal H}_{A,B}$ of 
the two $d$-dimensional systems ${\cal H}_A$ and ${\cal H}_B$
spanned by
$\ket0_A,\ket1_A,...,\ket{d-1}_A$
and
$\ket0_B,\ket1_B,...,\ket{d-1}_B$, respectively.
Note that we refer to $\{\ket i_A\}$ and $\{\ket i_B\}$
as {\it the standard basis}.
Suppose that
$n$ independent samples
are provided, that is, the state is given in the form
\begin{equation*}
	\rho=\bigotimes_{i=1}^n \sigma_i
	=
	\underbrace{\sigma_1\otimes\cdots\otimes\sigma_n}_{n}
\end{equation*}
for $n$ unknown densities $\sigma_1, \ldots \sigma_n$.
We also assume that these densities $\sigma_1, \ldots ,\sigma_n$
equal a density $\sigma$.
In this case, the state $\rho$ is called 
{\it $n$-independent and identical density ($n$-i.i.d.)}.
In the following, we consider two settings for our hypotheses:
\begin{align*}
%1)&	H_0:
%\sigma  =  \ket{\phi^0_{AB}}\bra{\phi^0_{AB}}
%	\mbox{ versus }
%	H_1: \sigma \ne \ket{\phi^0_{AB}}\bra{\phi^0_{AB}}\\
H_0:&\quad 
\sigma \in \cl{S}_{\le \epsilon}
\defeq \{ \sigma %\in \cl{S}(\cl{H})
|
1- \bra{\phi^0_{AB}} \sigma \ket{\phi^0_{AB}} \le \epsilon\} \\
&	\quad \mbox{ versus }\\
H_1: &\quad \sigma \in \cl{S}_{\le \epsilon}^c
\end{align*}
and 
\begin{align*}
H_0:&\quad 
\sigma \in \cl{S}_{\ge \epsilon}
\defeq \{ \sigma %\in \cl{S}(\cl{H})
|
1- \bra{\phi^0_{AB}} \sigma \ket{\phi^0_{AB}} \ge \epsilon\} \\
&	\quad \mbox{ versus }\\
H_1: &\quad \sigma \in \cl{S}_{\ge \epsilon}^c.
\end{align*}
When the null hypothesis is ``$\sigma \in \cl{S}_{\le \epsilon}$'',
the set of level $\alpha$-tests is given 
in the $n$-fold i.i.d. case by
\begin{align*}
\cl{T}_{\alpha,\le \epsilon}^n
\defeq
\left\{
T \left|
0 \le T \le I,\quad
\forall \sigma \in \cl{S}_{\le \epsilon}
, ~ 1- \Tr \sigma^{\otimes n}T \le \alpha
\right.
\right\}.
\end{align*}
Similarly,
when the null hypothesis is ``$\sigma \in \cl{S}_{\ge \epsilon}$'',
the set of level $\alpha$-tests is given 
in the $n$-fold i.i.d. case by
\begin{align*}
\cl{T}_{\alpha,\ge \epsilon}^n
\defeq
\left\{
T \left|
0 \le T \le I,\quad
\forall \sigma \in \cl{S}_{\ge \epsilon}
, ~ 1- \Tr \sigma^{\otimes n}T \le \alpha
\right.
\right\}.
\end{align*}
In this paper, we only treat the null hypothesis 
$\cl{S}_{\le \epsilon}$.
However, a large part of obtained results
can be trivially extended to the case of the null hypothesis 
$\cl{S}_{\ge \epsilon}$.

\subsection{Restriction I: group action}\Label{s3B}
In this paper, we treat 
these two cases
with the invariance conditions
for the following group action,
which preserve the two hypotheses $H_0$ and $H_1$.
The naturalness of this condition will be 
discussed later.

{\bf 1)$U(1)$-action:}
\begin{align*}
\phi &\mapsto 
U_\theta \phi ,\quad \phi \in {\cal H}_{A,B},
\quad \theta \in \real 
\end{align*}
where $U_\theta $ is defined by
\begin{align*}
U_\theta &\defeq e^{i \theta} 
\ket{\phi^0_{AB}}\bra{\phi^0_{AB}}
+(I- \ket{\phi^0_{AB}}\bra{\phi^0_{AB}} ).
\end{align*}
For a vector $|u\rangle$ orthogonal to $\bra{\phi^0_{AB}}$
and a positive number $0 \,< p\,<1$,
the entanglement properties of 
the two sates $\sqrt{p} \ket{\phi^0_{AB}}+ \sqrt{1-p}
\ket{u}$ 
and
$e^{i\theta}\sqrt{p} \ket{\phi^0_{AB}}+ \sqrt{1-p}\ket{u}$ 
are essentially equivalent.
Hence, this symmetry is very natural.
We can easily check that 
this action preserves our hypotheses.
The $U(1)$-action is so small that
it is not suitable to adopt this invariance as
our restriction.
However, this invariance can be, often,
treated so easily that 
it be adopted only by a technical reason.

{\bf 2)$SU(d)$-action:}
We consider the unitary action on
the tensor product space ${\cal H}_{A,B}=
{\cal H}_A \otimes {\cal H}_B$:
\begin{align*}
\phi \mapsto U(g) \phi,\quad \phi \in {\cal H}_{A,B},
\quad g \in SU(d),
\end{align*}
where
\begin{align*}
U(g)\defeq g \otimes \overline{g},
\end{align*}
and $\overline{g}$ is the complex conjugate of $g$ concerning
the standard basis
$\ket0_B,\ket1_B,...,\ket{d-1}_B$ on the system $B$.
Indeed, this action preserves
the maximally entangled state $\ket{\phi^0_{AB}}$.
Hence, this action preserves our hypotheses.
Furthermore, this action preserves the entanglement property.
Similarly to the $U(1)$-invariance, 
the $SU(1)$-action is so small that
it will be adopted only by a technical reason.

{\bf 3)$SU(d)\times U(1)$-action:}
Since the $SU(d)$ action and the $U(1)$-action
preserve the entanglement property,
the following action of the 
direct sum product group $SU(d)\times U(1)$
of $SU(d)$ and $U(1)$ also preserves this property:
\begin{align*}
\phi \mapsto U(g,\theta)\phi
\quad \phi \in {\cal H}_{A,B},
\quad (g,e^{i\theta}) \in SU(d)\times U(1),
\end{align*}
where
\begin{align*}
U(g,\theta)\defeq U(g)U_\theta= U_\theta U(g).
\end{align*}
Thus, this condition is most suitable as our restriction.

{\bf 4)$U(d^2-1)$-action:}
As a stronger invariance, 
we can consider the invariance of the $U(d^2-1)$-action, {\it i.e.},
the following 
unitary action on
the orthogonal space of $\ket{\phi^0_{AB}}\bra{\phi^0_{AB}}$,
which is a $d^2-1$-dimensional space.
\begin{align*}
\phi \mapsto 
V(g) \phi, \quad \phi \in {\cal H}_{A,B},\quad g \in U(d^2-1).
\end{align*}
where
\begin{align*}
V(g)\defeq
g (I- \ket{\phi^0_{AB}}\bra{\phi^0_{AB}} 
)+
\ket{\phi^0_{AB}}\bra{\phi^0_{AB}}.
\end{align*}
This group action contains
the $U(1)$-action and the $SU(d)$-action.
Hence, the invariance of the $U(d^2-1)$-action is stronger than 
the invariances of above three actions.
This action does not preserve the entanglement property.
Thus, based on this definition, we cannot say that this 
condition is natural for our setting
while it is natural if we are not care of entanglement.

Furthermore,
in the $n$-fold i.i.d. setting,
it is suitable to
assume the invariance of the $n$-tensor product action of 
the above actions, {\it i.e.},
$U_\theta^{\otimes n}$, $U(g)^{\otimes n}$, $U(g,\theta)^{\otimes n}$, 
$V(g)^{\otimes n}$, {\it etc}.

\subsection{Restriction II: locality}\Label{s3C}
When the system consists of two distinct parties $A$ and $B$,
it is natural to restrict our testing to LOCC measurements
between $A$ and $B$.
Hence, we can consider several restrictions concerning 
locality condition.
%However, it is difficult to optimize a LOCC test between $A$ and $B$.
Hence, in section \ref{s4}, as the first step, 
in order to discuss the hypotheses testing with the null hypothesis
$\cl{S}_{\le \epsilon}$,
we will treat 
the following optimization:
\begin{align*}
\beta_{\alpha,G}^n(\le \epsilon\| \sigma) 
\defeq 
\min_{
T \in \cl{T}_{\alpha,\le \epsilon}^{n}}
\left\{
\beta(T,\sigma^{\otimes n})
\left|
T \hbox{ is $G$-invariant.}
\right.
\right\},
\end{align*}
where $G= U(1),SU(d),SU(d)\times U(1),$ or $U(d^2-1)$.
However, since our quantum system consists of two distant system,
we cannot neccessarily use all measurements.
Hence, it is natural to restrict our test to
a class of tests.
In this paper, we focus on the following seven classes.

\begin{description}
\item[$\emptyset$:] No condition

\item[$S(A,B)$:] The test is {\it separable} between
two systems $\cl{H}_{A}^{\otimes n}$ and $\cl{H}_{B}^{\otimes n}$, 
{\it i.e.},
the test $T$ has the following form:
\begin{align*}
T= \sum_{i} a_i T_i^A \otimes T_i^B,
\end{align*}
where $a_i \ge 0$ and the matrix $T_i^A$ ($T_i^B$) is
a positive semi-definite matrix on the system
$\cl{H}_{A}^{\otimes n}$ ($\cl{H}_{B}^{\otimes n}$),
respectively.

\item[$L(A \leftrightarrows B)$:]
The test can be realized by two-way LOCC between 
two systems $\cl{H}_{A}^{\otimes n}$ and $\cl{H}_{B}^{\otimes n}$.

\item[$L(A \to B)$:]
The test can be realized by one-way LOCC from
the system $\cl{H}_{A}^{\otimes n}$ to the system $\cl{H}_{B}^{\otimes n}$.

\item[$S(A_1, \ldots, A_n,B_1, \ldots, B_n)$:]
\quad The test is separable among 
$2n$ systems $\cl{H}_{A_1}$, \ldots, $\cl{H}_{A_n}$, $\cl{H}_{B_1}$,
\ldots, $\cl{H}_{B_n}$, {\it i.e.},
the test $T$ has the following form:
\begin{align*}
T= \sum_{i} a_i T_i^{A_1}\otimes \cdots \otimes T_i^{A_n} \otimes 
T_i^{B_1}\otimes \cdots \otimes T_i^{B_n},
\end{align*}
where $a_i \ge 0$ and the matrix $T_i^{A_k}$ ($T_i^{B_k}$) is
a positive semi-definite matrix on the system
$\cl{H}_{A_k}$ ($\cl{H}_{B_k}$),
respectively.

\item[$L(A_1, \ldots, A_n,B_1, \ldots, B_n)$:]
\quad The test can be realized by two-way LOCC among 
$2n$ systems $\cl{H}_{A_1}$, \ldots, $\cl{H}_{A_n}$, $\cl{H}_{B_1}$,
\ldots, $\cl{H}_{B_n}$.

\item[$L(A_1, \ldots, A_n \to B_1, \ldots, B_n)$:]
\quad The test can be realized by LOCC among 
$2n$ systems $\cl{H}_{A_1}$, \ldots, $\cl{H}_{A_n}$, $\cl{H}_{B_1}$,
\ldots, $\cl{H}_{B_n}$.
Moreover, the classical communication among two groups
$\cl{H}_{A_1}$, \ldots, $\cl{H}_{A_n}$ and 
$\cl{H}_{B_1}$,\ldots, $\cl{H}_{B_n}$ is restricted to 
one-way from the former to the later.
\end{description}
Based on the above conditions,
we define the following quantity as the optimal second error probability:
\begin{align*}
\beta_{\alpha,n,G}^{C}(\le\epsilon\| \sigma) 
\defeq &
\min_{T\in \cl{T}_{\alpha,\le \epsilon}^{n}}
\left\{\!
\beta(T,\sigma^{\otimes n})
\left|
\begin{array}{l}
T\hbox{ is $G$-invariant,}\\
\hbox{and satisfies }C\\
\end{array}
\!\!\!
\right.
\right\} .
\end{align*}
As is easily checked,
any LOCC operation is separable.
Hence, the condition $L(A \leftrightarrows B)$ is stronger than 
the condition $S(A,B)$.
Also, the condition 
$L(A_1, \ldots, A_n \to B_1, \ldots, B_n)$ is stronger than 
the condition $S(A_1, \ldots, A_n \to B_1, \ldots, B_n)$.
The relation among these conditions can be illustrated as follows.

Next, 
we focus on the trivial relations of the optimal second error probability.
If a group $G_1$ is greater than $G_2$,
the inequality
\begin{align}
\beta_{\alpha,n,G_1}^{C}(\le\epsilon\| \sigma)
\ge \beta_{\alpha,n,G_2}^{C}(\le\epsilon\| \sigma)\Label{22-7}
\end{align}
holds.
Moreover,
if a condition $C_1$ is stronger than
another condition $C_2$,
the similar inequality
\begin{align}
\beta_{\alpha,n,G}^{C_1}(\le\epsilon\| \sigma)
\ge \beta_{\alpha,n,G}^{C_2}(\le\epsilon\| \sigma)\Label{22-9}
\end{align}
holds.

Similarly,
we define 
$\beta_{\alpha,n,G}^C(\ge \epsilon\| \sigma) $
by replacing $\le \epsilon$ by $\ge \epsilon$
in RHS.

Indeed, if 
the condition is invariant for the action of $G$,
it is very natural to restrict our test among
$G$-invariant tests, as is indicated by the following lemma.
\begin{lem}\Label{l12}
Assume that
a set of test satisfying the condition $C$
is invariant for the action of $G$,
Then
\begin{align*}
\beta_{\alpha,n,G}^C(\le \epsilon\|\sigma)
&=
\min_{T \in \cl{T}_{\alpha,\le \epsilon}^{n}}
\max_{g \in G}
\beta(T,(f(g)\sigma f(g)^{\dagger})^{\otimes n}) \\
&=
\min_{T \in \cl{T}_{\alpha,\le \epsilon}^{n}}
\int_G
\beta(T,(f(g)\sigma f(g)^{\dagger})^{\otimes n}) 
\nu_G(d g),
\end{align*}
where $\nu_G$ is the invariant measure and
$f$ denotes the action of $G$. 
\end{lem}
In the following, we sometimes abbreviate 
the invariant measure $\nu_G$ by $\nu$.
For a proof see Appendix \ref{a21}.
This lemma is a special version of quantum Hunt-Stein lemma 
\cite{HolP}.
The condition $\emptyset$ is invariant for 
the actions $U(1),SU(d),SU(d)\times U(1), U(d^2-1)$.
But, other conditions $S(A,B),L(A \leftrightarrows B)$,
$L(A \to B), S(A_1, \ldots, A_n,B_1, \ldots B_n)$,
$L(A_1, \ldots, A_n,B_1, \ldots B_n)$,
$L(A_1, \ldots, A_n \to B_1, \ldots B_n)$ are 
invariant only for $SU(d)$.
Hence, Lemma \ref{l12} cannot be applied to 
the pair of these conditions and 
the actions $U(1),SU(d),SU(d)\times U(1), U(d^2-1)$.
The following lemma is useful in such a case.
\begin{lem}\Label{l13}
Assume that the group $G_1$ includes 
another group $G_2$ which satisfies 
the condition of Lemma \ref{l12}.
If 
\begin{align*}
\beta_{\alpha,n,G_1}^C(\le \epsilon\|\sigma)
=\beta_{\alpha,n,G_2}^C(\le \epsilon\|\sigma),
\quad \forall \sigma
\end{align*}
then
\begin{align*}
&\beta_{\alpha,n,G_1}^C(\le \epsilon\|\sigma)\\
=&
\min_{T \in \cl{T}_{\alpha,\le \epsilon}^{n}}
\max_{g \in G_1}
\beta(T,(f(g)\sigma f(g)^{\dagger})^{\otimes n}) \\
=&
\min_{T \in \cl{T}_{\alpha,\le \epsilon}^{n}}
\int_{G_1}
\beta(T,(f(g)\sigma f(g)^{\dagger})^{\otimes n}) 
\nu_{G_1}(d g).
\end{align*}
\end{lem}
Its proof is given in Appendix \ref{a21}.

\section{Testing for binomial distributions}\Label{s4}
In this paper, we use several knowledges
about testing for binomial distributions
for testing for a maximally entangled state.
Hence, we review them here.

\subsection{One-sample setting:}\Label{s4A}
As a preliminary, we treat 
testing for the coin flipping probability $p$
with a single trial.
That is, we assume that 
the event $1$ happens with the probability $p$
and the event $0$ happens with the probability $1-p$,
and focus on the null hypothesis $p \in [0 ,\epsilon]$.
In this case, our test 
can be described
by a map $\tilde{T}$ from $\{0,1\}$ to $[0,1]$,
which means that 
when the data $k$ is observed,
we accept the null hypothesis
with the probability $\tilde{T}(k)$.
Then, the minimum second error probability
among level-$\alpha$ tests 
is given by
\begin{align*}
\beta^1_{\alpha}(\le \epsilon\|q)
&\defeq
\min_{\tilde{T}}
\left\{ 
q(\tilde{T})
\left| \forall p \in [0,\epsilon] , 
p(\tilde{T})
\ge 1- \alpha \right.
\right\}\\
p(\tilde{T}) &\defeq
(1-p)\tilde{T}(0)+ p \tilde{T}(1).
\end{align*}
When we define the test $\tilde{T}_{\epsilon,\alpha}^{1}$ by
\begin{align*}
\tilde{T}_{\epsilon, \alpha}^{1}(0)=
\left\{
\begin{array}{ll}
\frac{1- \alpha}{1- \epsilon} & \hbox{ if } \epsilon \le \alpha 
\\
1 & \hbox{ if } \epsilon \,> \alpha 
\end{array}
\right.  , ~
\tilde{T}_{\epsilon, \alpha}^{1}(1)=  
\left\{
\begin{array}{ll}
0 & \hbox{ if } \epsilon \le \alpha 
\\
\frac{\epsilon-\alpha}{\epsilon} & \hbox{ if } \epsilon \,> \alpha ,
\end{array}
\right.  
\end{align*}
the test $\tilde{T}_{\epsilon,\alpha}^{1}$ 
satisfies 
\begin{align}
(1-\epsilon)\tilde{T}_{\epsilon,\alpha}^{1}(0)
+\epsilon \tilde{T}_{\epsilon,\alpha}^{1}(1)= 1- \alpha \Label{22-11}.
\end{align}
Moreover, if $p \le \epsilon$,
\begin{align*}
(1-p)\tilde{T}_{\epsilon,\alpha}^{1}(0)
+p \tilde{T}_{\epsilon,\alpha}^{1}(1)\ge 1- \alpha.
\end{align*}
Hence the test $\tilde{T}_{\epsilon,\alpha}^{1}$ is 
level-$\alpha$.
Furthermore, we can easily check that
the minimum of $q(\tilde{T})$ 
with the condition (\ref{22-11}) for $\tilde{T}$
can be attained by $\tilde{T}= \tilde{T}_{\epsilon,\alpha}^{1}$
if $q \,> \epsilon$.
Hence,
\begin{align}
\beta^1_{\alpha}(\le \epsilon\|q)=
q(\tilde{T}_{\epsilon,\alpha}^{1})
=\left\{
\begin{array}{ll}
\frac{(1- \alpha)(1-q)}{1- \epsilon} & \hbox{ if } \epsilon \le \alpha 
\\
1- \frac{\alpha q}{\epsilon} & \hbox{ if } \epsilon \,> \alpha .
\end{array}
\right. \Label{22-15}
\end{align}

\subsection{$n$-sample setting:}\Label{s4B}
In the $n$-trial case, the data 
$k= 0,1,\ldots, n$ obeys the distribution
$P^n_p(k)\defeq 
\genfrac{(}{)}{0pt}{}{n}{k}
(1-p)^{n-k} p^k$ with the unknown parameter $p$.
Hence, we discuss the hypothesis testing 
with the null hypothesis
$\cl{P}_{\le \epsilon}^n
\defeq \{ P^n_p(k)| p \le \epsilon\}$
and the alternative hypothesis
$(\cl{P}_{\le \epsilon}^n)^c$.
In this case, our test $\tilde{T}$ can be
described by a function from the data set 
$\{0 ,1, \ldots, n\}$ to interval $[0,1]$.
In this case, when the data $k$ is observed,
we accept the null hypothesis
$\cl{P}_{\le \epsilon}^n$ with the probability $T(k)$.
Then, the minimum second error probability
among level-$\alpha$ tests 
is given by
\begin{align*}
\beta^n_{\alpha}(\le \epsilon\|q)
&\defeq
\min_{\tilde{T}}
\left\{ P^n_q(\tilde{T}) 
\left| \forall p \in [0,\epsilon] , 
1- P^n_p(\tilde{T}) \le \alpha \right.
\right\}\\
P^n_p(\tilde{T}) &\defeq
\sum_{k=0}^n P^n_p(k)\tilde{T}(k).
\end{align*}

We define the test $\tilde{T}^n_{\epsilon,\alpha}$ 
as follows.
\begin{align*}
\tilde{T}^n_{\epsilon,\alpha}(k)=
\left\{
\begin{array}{ll}
1 & k \,< l^n_{\epsilon,\alpha} \\
\gamma^n_{\epsilon,\alpha} & k = l^n_{\epsilon,\alpha} \\
0 & k \,> l^n_{\epsilon,\alpha} ,
\end{array}
\right.
\end{align*}
where
the integer $l_{\epsilon,\alpha}^n$
and the real number $\gamma_{\epsilon,\alpha}^n\,>0$, 
are defined by
%such that the spectral decomposition of $U_\theta^{\otimes n}$ 
%is given by $U_\theta^{\otimes n}=\sum_{k=0}^n e^{i(n-k)\theta }P_k$,
\begin{align*}
\sum_{k=0}^{l_{\epsilon,\alpha}^n-1} 
P^n_\epsilon(k)
& \,< 1- \alpha 
\le 
\sum_{k=0}^{l_{\epsilon,\alpha}^n} 
P^n_\epsilon(k)\\
\gamma_{\epsilon,\alpha}^n
P^n_\epsilon(l_{\epsilon,\alpha}^n)
& = 
1- \alpha
-\sum_{k=0}^{l_{\epsilon,\alpha}^n-1} 
P^n_\epsilon(k).
\end{align*}
\begin{thm}\Label{t-1}
The test $\tilde{T}^n_{\epsilon,\alpha}$ is
level-$\alpha$ UMP test with the null hypothesis
$\cl{P}_{\le \epsilon}^n$.
Hence, 
\begin{align*}
\beta^n_{\alpha}(\le \epsilon\|q)= 
P_q^n (\tilde{T}_{\epsilon, \alpha})=
\sum_{k=0}^{l_{\epsilon,\alpha}^n-1} 
P^n_q(k)
+
\gamma_{\epsilon,\alpha}^n
P^n_q(l_{\epsilon,\alpha}^n).
\end{align*}
\end{thm}
For a proof, see Appendix \ref{a-2}.
\subsection{Asymptotic setting}\Label{s4C}
In asymptotic theory, There are two settings at least.
One is the large deviation setting, in which
the parameter is fixed, hence we focus on the exponential 
component of the error probability.
The other is the small deviation setting, in which 
the parameter is close to a given fixed point in proportion to
the number of samples 
such that the error probability converges to a fixed number.
That is, the parameter is fixed in the former,
while the error probability is fixed in the later.

\subsubsection{Small deviation theory}\Label{s4C1}
It is useful to treat the neiborhood around 
$p=0$ as the small deviation theory of this problem
for the asymptotic discussion of testing
for an maximally entangled state.
Hence, we focus on the case that $p= \frac{t}{n}$:
Since the probability 
$P_{t/n}^n(k)=
\genfrac{(}{)}{0pt}{}{n}{k}
(1-\frac{t}{n} )^{n-k} \left(\frac{t}{n}\right)^k$ convergences to the 
Poisson distribution
$P_t(k)\defeq e^{-t}\frac{t^k}{k!}$.
Hence, 
our testing problem 
with the null hypothesis $\cl{P}_{\frac{\delta}{n}}$
and the alternative hypothesis $\frac{t'}{n}$.
is 
asymptotically equivalent with
the testing of Poisson distribution $P_t(k)$
with the null hypothesis $t\in [0,\delta]$
and the alternative hypothesis $t'$.
That is, by defining 
\begin{align*}
\beta_{\alpha}(\le \delta\|t')
&\defeq
\min_{\tilde{T}}
\left\{ P_{t'}(\tilde{T}) 
\left| \forall t \in [0,\delta] , 
1- P_t(\tilde{T}) \le \alpha \right.
\right\}\\
P_t(\tilde{T}) &\defeq
\sum_{k=0}^{\infty} P_t(k)\tilde{T}(k),
\end{align*}
the following theorem holds.
\begin{thm}\Label{t-2}
\begin{align*}
\lim \beta^n_{\alpha} \left(\le\frac{\delta}{n}\left\|\frac{t'}{n}
\right.\right)
= \beta_{\alpha}(\le \delta\|t').
\end{align*}
\end{thm}
Its proof is given in Appendix \ref{a5}.
Similarly to the test $\tilde{T}_{\epsilon,\alpha}^n$,
we define the test $\tilde{T}_{\delta,\alpha}$ as
\begin{align*}
\tilde{T}_{\delta,\alpha}(k)=
\left\{
\begin{array}{ll}
1 & k \,< l_{\delta,\alpha} \\
\gamma_{\delta,\alpha} & k = l_{\delta,\alpha} \\
0 & k \,> l_{\delta,\alpha} ,
\end{array}
\right.
\end{align*}
where
the integer $l_{\delta,\alpha}$
and the real number $\gamma_{\delta,\alpha}\,>0$, 
are defined by
%such that the spectral decomposition of $U_\theta^{\otimes n}$ 
%is given by $U_\theta^{\otimes n}=\sum_{k=0}^n e^{i(n-k)\theta }P_k$,
\begin{align*}
& \sum_{k=0}^{l_{\delta,\alpha}-1} 
P_\delta(k)
\,< 1- \alpha 
\le 
\sum_{k=0}^{l_{\delta,\alpha}} 
P_\delta(k)\\
&\gamma_{\delta,\alpha} 
P_\delta(l_{\delta,\alpha})
= 
1- \alpha
-\sum_{k=0}^{l_{\delta,\alpha}^n-1} 
P_\delta(k).
\end{align*}
Similarly to Theorem \ref{t-1}, the following theorem holds.
\begin{thm}\Label{t-3}
The test $\tilde{T}_{\delta,\alpha}$ is
level-$\alpha$ UMP test with the null hypothesis
$\cl{P}_{\le \delta}\defeq \{P_t |t \le \delta\}$.
Hence, 
\begin{align*}
\beta^n_{\alpha}(\le \delta\|t')= 
\sum_{k=0}^{l_{\delta,\alpha}-1} 
P_{t'}(k)
+
\gamma_{\delta,\alpha}
P_{t'}(l_{\delta,\alpha}).
\end{align*}
\end{thm}
%This theorem is proved in Appendix \ref{}.

\subsubsection{Large deviation theory}\Label{s4C2}
Next, we proceed to the large deviation theory.
Using the knowledge of mathematical statistics,
we can calculate the exponents of 
the 2nd error probabilities $\beta^n_\alpha(\epsilon\|p)$
and $\beta^n_\alpha(\epsilon\|p)'$
for any $\alpha \,>0$ as
\begin{align*}
\lim \frac{-1}{n} \log \beta^n_{\alpha}(\le\epsilon\|p)
&= d(\epsilon\|p), \hbox{ if }
\epsilon \,< p
\\
\lim \frac{-1}{n} \log \beta^n_{\alpha}(\ge\epsilon\|p)
&= d(\epsilon\|p), \hbox{ if }
\epsilon \,> p,
\end{align*}
where the binary relative entropy $d(\epsilon\|p)$
is defined as
\begin{align*}
d(\epsilon\|p)\defeq
\epsilon \log \frac{\epsilon}{p}
+(1-\epsilon) \log \frac{1-\epsilon}{1-p}.
\end{align*}
In the case of $\alpha=0$,
we have
\begin{align*}
\frac{-1}{n} \log \beta^n_0(\epsilon\|p)
= 
\left\{
\begin{array}{cl}
- \log (1-p) & \hbox{ if } \epsilon =0 \\
0 &  \hbox{ if } \epsilon \neq 0 .
\end{array}
\right.
\end{align*}

\section{Global tests}\Label{s5}
%\section{No locality condition}
First, we 
treat the hypotheses testing with a given group invariance condition
with no locality restriction.
\subsection{One-sample setting:}\Label{s50}
When only one sample is prepared,
the test $\ket{\phi^0_{A,B}}\bra{\phi^0_{A,B}}$
is a level-$0$ test for the null hypothesis $\cl{S}_0$.
If we perform the two-valued measurement 
$\{\ket{\phi^0_{A,B}}\bra{\phi^0_{A,B}}, 
I -\ket{\phi^0_{A,B}}\bra{\phi^0_{A,B}}\}$,
the data obeys the distribution
$\{1-p ,p\}$, where
\begin{align*}
p\defeq 1- \bra{\phi^0_{A,B}}\sigma \ket{\phi^0_{A,B}}.
\end{align*}
Hence, applying the discussion in subsection \ref{s4A},
the test $T_{\alpha}^1(\ket{\phi^0_{A,B}}\bra{\phi^0_{A,B}}, \epsilon)$
is a level-$\alpha$ test for
the null hypothesis $\cl{S}_{\le \epsilon}$,
where the operator $T_{\alpha}^1(T, \epsilon)$
is defined by
\begin{align*}
T_{\alpha}^1(T, \epsilon)\defeq
\left\{
\begin{array}{ll}
\frac{1-\alpha}{1-\epsilon}T &
\hbox{ if } \epsilon \le \alpha\\
T + \frac{\epsilon-\alpha}{\epsilon}(I-T) &
\hbox{ if } \epsilon \,> \alpha.
\end{array}
\right.
\end{align*}

\subsection{$n$-sample setting:}\Label{s5A}
In the $n$-sample setting,
we construct a test for the null hypothesis $\cl{S}_{\le \epsilon}$
as follows.
First, we perform the two-valued measurement 
$\{\ket{\phi^0_{A,B}}\bra{\phi^0_{A,B}}, 
I -\ket{\phi^0_{A,B}}\bra{\phi^0_{A,B}}\}$
for respective $n$ systems.
Then, if the number of counting $I -\ket{\phi^0_{A,B}}\bra{\phi^0_{A,B}}$
is described by $k$,
the data $k$ obeys the binomial distribution
$P^n_{p}(k)$.
%where
%\begin{align}
%p\defeq 1- \bra{\phi^0_{A,B}}\sigma \ket{\phi^0_{A,B}}.
%\end{align}
In this case, our problem can be reduced
to the hypothesis testing with the null hypothesis
$\cl{P}_{\le \epsilon}^n$,
which has been discussed in subsection \ref{s4B}.

For given $\alpha$ and $\epsilon$,
the test based on this measurement and
the classical test $\tilde{T}_{\epsilon,\alpha}^n$ is
described by the operator 
$T_{\epsilon,\alpha}^n\defeq T_{\alpha}^n(
\ket{\phi^0_{A,B}}\bra{\phi^0_{A,B}},\epsilon)$,
where  $T_{\alpha}^n(T,\epsilon)$ is defined by
\begin{align*}
T_{\alpha}^n(T,\epsilon)\defeq& 
\sum_{k=0}^{l_{\alpha}^n(\epsilon)-1}P_{k}^n(T,I-T)
 + \gamma_{\alpha}^n(\epsilon) P_{l_{\alpha}^n(\epsilon)}^n(T,I-T)\\
P_{n,k}(T,S) \defeq &
\underbrace{S\otimes \cdots \otimes
S}_{k} \otimes 
\underbrace{T\otimes \cdots \otimes T}_{n-k}\\
& %\hspace{10ex}
+\cdots \\
&+\underbrace{T\otimes \cdots \otimes T}_{n-k}
\otimes
\underbrace{S\otimes \cdots \otimes S}_{k}.  
\end{align*}
Note that the above sum contains all tensor products of 
$k$ times of $S$ and $n-k$ times of $T$.

Since the operators
$\ket{\phi^0_{A,B}}\bra{\phi^0_{A,B}}$ and
$I-\ket{\phi^0_{A,B}}\bra{\phi^0_{A,B}}$
are $U(d^2-1)$-invariant,
the test $T_{\epsilon,\alpha}^n$ is level-$\alpha$ $U(d^2-1)$-invariant 
test
with the hypothesis $\cl{S}_{\le \epsilon}$.
Hence, 
\begin{align}
\beta_{\alpha,n,U(d^2-1)}^{\emptyset}(\le \epsilon\|\sigma)
\le\beta_{\alpha}^n(\le\epsilon\|p). \Label{1-19-2}
\end{align}

On the other hand,
as is shown in Appendix \ref{a-4},
\begin{align}
\beta_{\alpha,n,U(1)}^{\emptyset}(\le \epsilon\|\sigma)
= \beta_{\alpha}^n(\le \epsilon\|p)
.\Label{1-19-1}
\end{align}
Since $U(1) \subset SU(d)\times U(1) \subset U(d^2-1)$, 
the relations (\ref{1-19-2}) and (\ref{1-19-1}) yield
the following theorem.
\begin{thm}\Label{t-4}
The equation 
\begin{align}
\beta_{\alpha,n,G}^{\emptyset}(\le \epsilon\|\sigma)
= \beta_{\alpha}^n(\le \epsilon\|p)\Label{23-4}
\end{align}
holds for $G= U(1),SU(d)\times U(1), U(d^2-1)$.
\end{thm}
Therefore,
The test $T_{\epsilon,\alpha}^n$
is the UMP $G$-invariant test, for $G= U(1), SU(d)\times U(1)$ 
or $U(d^2-1)$.
Moreover, we can derive the same results for the hypothesis
$\cl{S}_{\ge \epsilon}$.

\subsection{Asymptotic setting}\Label{s5B}
Next, we proceed to the asymptotic setting.
In the small deviation theory, 
we treat the hypothesis testing with
the null hypothesis $\cl{S}_{\le \delta/n}$.
in this setting, Theorem \ref{t-2} and Theorem \ref{t-4}
guarantee that
the limit of 
the optimal second error probability of 
the alternative hypothesis $\sigma_n$
is given by 
$\beta_\alpha(\delta\|t')$
if 
$\bra{\phi^0_{A,B}}\sigma_n \ket{\phi^0_{A,B}}=1- \frac{t'}{n}$.
That is,
\begin{align}
\lim \beta_{\alpha,G}^n \left(\left.\le 
\frac{\delta}{n}\right\|\sigma_n\right)
= \beta_\alpha(\le \delta\|t') \Label{29-1}
\end{align}
for $G=U(1),SU(d)\times U(1), U(d^2-1)$.

In the large deviation setting,
we can obtain the same results as subsection \ref{s4C}, {\it i.e.},
\begin{align}
\lim\frac{-1}{n}\log 
\beta_{\alpha,G}^n (\le \epsilon \|\sigma)
=
\left\{
\begin{array}{ll}
d(\epsilon\|p) &\hbox{ if } \alpha \,> 0\\
- \log(1- p) &\hbox{ if } \alpha =0 , \epsilon =0 \\
0 &\hbox{ if } \alpha =0 , \epsilon \,> 0 
\end{array}
\right. \Label{23-5}
\end{align}
if $\epsilon \,< p= 1- \bra{\phi^0_{A,B}}\sigma\ket{\phi^0_{A,B}}$.
Moreover, we can derive similar results with the null hypothesis
$\cl{S}_{\ge \epsilon}$.

\section{A-B locality}\Label{s6}
In this section, we treat optimization problems with 
several conditions regarding 
the locality between A and B.

\subsection{One-sample setting}\Label{s6A}
First, we focus on the simplest case,
{\it i.e.},
the case of $\epsilon =0$ and $\alpha=0$.
For this purpose, we focus on a POVM with the following
form on $\cl{H}_A$ 
\begin{align*}
M= \{ p_i \ket{u_i}\bra{u_i}\}_i,
\quad \|u_i\|=1 , \quad 0 \le p_i \le 1,
\end{align*}
where such a POVM is called {\it rank-one}.
Based on a rank-one POVM $M$,
a suitable test $T(M)$
\begin{align}
T(M)\defeq
\sum_i p_i \ket{u_i\otimes \overline{u_i}}
\bra{u_i\otimes \overline{u_i}}.\Label{2-9-3}
\end{align}
can be realized by the following one-way LOCC protocol.
From the definition, 
of course, we can easily check that $T(M)$ 
satisfies the condition of test,
{\it i.e.},
\begin{align}
0 \le T(M) \le I. \Label{24-30}
\end{align}
\par\noindent{\bf One-way LOCC protocol of $T(M)$:}
\par\noindent
{\bf 1)} Alice performs the measurement $\{ p_i \ket{u_i}\bra{u_i}\}_i$,
and sends her data $i$ to Bob.
\par\noindent
{\bf 2)}
Bob performs the two-valued measurement $\{
\ket{\overline{u_i}}\bra{\overline{u_i}},
I-\ket{\overline{u_i}}\bra{\overline{u_i}}\}$,
where
$\overline{u_i}$ is the complex conjugate of $u_i$ concerning 
the standard basis $\ket0_B,\ket1_B,...,\ket{d-1}_B$.
\par\noindent
{\bf 3)}
If Bob observes the event corresponding to 
$\ket{\overline{u_i}}\bra{\overline{u_i}}$,
the hypothesis $\ket{\phi^0_{A,B}}\bra{\phi^0_{A,B}}$
is accepted. Otherwise, it is rejected.

This test satisfies
\begin{align}
\bra{\phi^0_{A,B}} T(M)\ket{\phi^0_{A,B}} &=1, \Label{24-31}\\
\Tr T(M)& = 
\sum_i p_i \Tr \ket{u_i\otimes \overline{u_i}}
\bra{u_i\otimes \overline{u_i}}\nonumber \\
& = 
\sum_i p_i \Tr \ket{u_i}\bra{u_i}=
d. \Label{24-39}
\end{align}
Hence, it is a level-$0$ test with the null hypothesis 
$\ket{\phi^0_{A,B}}\bra{\phi^0_{A,B}}$.
In particular,
in the one-way LOCC setting,
our test can be restricted to this kind of tests
as the following sense.
\begin{lem}\Label{l-10}
Let $T$ be a one-way LOCC $(A \to B)$ level-$0$ test 
with the null hypothesis $\ket{\phi^0_{A,B}}\bra{\phi^0_{A,B}}$.
Then, there exists a POVM with the form $M= \{p_i \ket{u_i} \bra{u_i}\}_i$
such that
\begin{align}
T\ge T(M), \Label{25-1}
\end{align}
{\it i.e.}, the test $T(M)$ is better than the test $T$.
\end{lem}

Moreover, concerning the separable condition,
the following lemma holds.
Hence, Corollary \ref{c1} indicates that it 
seems natural to restrict our 
test to the test with the form (\ref{2-9-3})
even if we adopt the separable condition.
\begin{lem}\Label{l-9}
Assume that 
a separable test $T$:
satisfies
\begin{align}
\bra{\phi_{A,B}^0}T\ket{\phi_{A,B}^0}= 1.\Label{2-16-7}
\end{align}
When we describe the test $T$ as
\begin{align}
T= d\sum_i p_i \ket{u_i \otimes {u'}_i}\bra{u_i \otimes {u'}_i}
+ \sum_j q_j \ket{v_i \otimes {v'}_i}\bra{v_i \otimes {v'}_i},
\Label{2-13}
\end{align}
such that $\langle \phi_{A,B}^0 | u_i \otimes {u'}_i \rangle 
=\frac{1}{\sqrt{d}}$
and $\langle \phi_{A,B}^0 | v_i \otimes {v'}_i \rangle 
=0$,
we obtain 
\begin{align*}
\sum_i p_i u_i \otimes {u'}_i = \frac{1}{\sqrt{d}}
\phi_{A,B}^0.
\end{align*}
\end{lem}
Its proof is given in Appendix \ref{a22}.
Note that we can easily obtain 
the same statement if we replace
the summation $\sum_i$ by the integral $\int$ at (\ref{2-13}).
Since any separable test $T$ has the form (\ref{2-13}),
the following corollary holds
concerning the completely mixed state $\frac{I}{d^2}$.
\begin{cor}\Label{c1}
If a separable test $T$ satisfies the conditions
\begin{align*}
&\bra{\phi_{A,B}^0}T\ket{\phi_{A,B}^0}= 1 \\
&\Tr T \frac{I}{d^2}= d =\min_{T'\in S(A,B) }
\left\{
\left. \Tr T' \frac{I}{d^2}\right|
\bra{\phi_{A,B}^0}T'\ket{\phi_{A,B}^0}= 1
\right\},
\end{align*}
then the test $T$ has a form (\ref{2-9-3}).
\end{cor}

Next, we focus on the covariant POVM $M_{cov}^1$:
\begin{align*}
M_{cov}^1(\, d \varphi)
\defeq 
d \ket{\varphi}\bra{\varphi}\nu (\,d \varphi),
\end{align*}
where $\nu(\,d \varphi)$ is the invariant measure in the set
of pure states with the full measure is $1$.
Then, the test $T_{inv}^{1,A \to B}\defeq T(M_{cov}^1)$
has the following form
\begin{align}
T_{inv}^{1,A \to B}=&
\int d
\ket{\varphi \otimes \overline{\varphi}}
\bra{\varphi\otimes \overline{\varphi}}\nu (\,d \varphi)
\nonumber \\
=& \ket{\phi^0_{A,B}}\bra{\phi^0_{A,B}}+
\frac{1}{d+1}(I-\ket{\phi^0_{A,B}}\bra{\phi^0_{A,B}}),\Label{22-17}
\end{align}
where the last equation will be shown in Appendix \ref{a-7}.
Note that the POVM $M_{cov}^1$
can be realized as follows:
\par\noindent{\bf Realization of $M_{cov}^1$:}
\par\noindent{\bf 1)} Randomly, 
we choose $g\in SU(d)$ with the invariant measure.
\par\noindent{\bf 2)} Perform POVM 
$\{g \ket{i}_A ~_A\bra{i} g^\dagger\}_i$.
Then, the realized POVM is $M_{cov}^1$.

Since the equation (\ref{22-17}) 
guarantees the $U(d^2-1)$-invariance
of the test $T_{inv}^{1,A-B}$,
we obtain 
\begin{align*}
\Tr T_{inv}^{1,A\to B} \sigma =
1-p + \frac{p}{d+1}
= 1- \frac{d p}{d+1},
\end{align*}
which implies
\begin{align*}
\beta^{L(A\to B)}_{0,1,U(d^2-1)}( 0\| \sigma)
\le
 1- \frac{d p}{d+1}.
\end{align*}
Next, we apply the discussion in subsection \ref{s4A} 
to the probability distribution 
$\{\frac{d p}{d+1},1- \frac{d p}{d+1}\}$.
Then, 
the test $T^{1,A-B}_{\epsilon,\alpha}
\defeq T^{1}_{\alpha}(T_{inv}^{1,A-B}, \frac{d\epsilon}{d+1})$ 
is a level-$\alpha$ $U(d^2-1)$-invariant test.
Since the test $T^{1,A-B}_{\epsilon,\alpha}$ can
be performed by randomized operation with 
$T_{inv}^{1,A-B}$ and $I-T_{inv}^{1,A-B}$,
we obtain
\begin{align}
&\beta^{L(A \to B)}_{\alpha,1,U(d^2-1)}(\le \epsilon\| \sigma)
\le \Tr T^{1,A-B}_{\epsilon,\alpha} \sigma \nonumber \\
=&
\left\{
\begin{array}{ll}
\frac{(1-\alpha)\left(1- \frac{d}{d+1}p\right)}
{\left(1- \frac{d}{d+1}\epsilon\right)}
&\hbox { if } \frac{d}{d+1}\epsilon \le \alpha \\
1- \frac{\alpha p}{\epsilon}
& \hbox { if } \frac{d}{d+1}\epsilon \,> \alpha ,
\end{array}
\right.\Label{1-49}
\end{align}

On the other hand, 
concerning $SU(d)$-invariance and 
separable tests,
the equation
\begin{align}
\beta^{S(A,B)}_{\alpha,1,SU(d)}(\le \epsilon\| \sigma)=
\Tr T^{1,A-B}_{\epsilon,\alpha} \sigma \Label{22-19}
\end{align}
holds,
which is shown in Appendix \ref{a-8}.
The equation in the case of $\alpha=0,\epsilon=0$ is 
obtained by Hayashi {\it et~al.}\cite{tsu}.
A similar result with the PPT condition 
is appeared in Virmani and Plenio \cite{virmani-plenio}.

Since
$U(d^2-1)$ is a larger group action than
$SU(d)$ and
the condition $L(A\to B)$ is stricter than
the condition $S(A,B)$,
the trivial inequalities
\begin{align*}
& \beta^{S(A,B)}_{\alpha,1,SU(d)}(\le \epsilon\| \sigma)
\le \beta^{S(A,B)}_{\alpha,1,U(d^2-1)}(\le \epsilon\| \sigma)\\
\le & \beta^{L(A \to B)}_{\alpha,1,U(d^2-1)}(\le \epsilon\| \sigma)
\end{align*}
hold.
Therefore, relations (\ref{1-49}) and (\ref{22-19})
yield
\begin{align}
&\beta_{\alpha,1,G}^{C}(\le \epsilon\|\sigma)
=
\left\{
\begin{array}{ll}
\frac{(1-\alpha)\left(1- \frac{d}{d+1}p\right)}
{\left(1- \frac{d}{d+1}\epsilon\right)}
&\hbox { if } \frac{d}{d+1}\epsilon \le \alpha \\
1- \frac{\alpha p}{\epsilon}
& \hbox { if } \frac{d}{d+1}\epsilon \,> \alpha ,
\end{array}
\right.\Label{1-51},
\end{align}
for $G=SU(d),SU(d)\times U(1),U(d^2-1)$,
and $C= L(A \to B),L(A \leftrightarrows B),S(A,B)$.
That is, the test $T^{1,A-B}_{\epsilon,\alpha}$
is the UMP $G$-invariant $C$ test with level $\alpha$
for the null hypothesis $\cl{S}_{\le \epsilon}$.
Furthermore, similar results for the null
hypothesis $\cl{S}_{\ge \epsilon}$ can be also obtained.

\subsection{Two-sample case}\Label{s6B}
In this section, we construct a $SU(d)\times U(1)$-invariant test 
which is realized by LOCC between A and B, and which 
attains the asymptotically optimal bound (\ref{29-1}).
For this purpose, we focus on
the covariant POVM $M_{cov}^2$:
\begin{align*}
& M_{cov}^2(\,d g_1\,d g_2)\\
\defeq &
d^2
(g_1\otimes g_2 )\ket{ u }\bra{u}(g_1\otimes g_2 )^*
\nu (\,d g_1)\nu (\,d g_2),
\end{align*}
where the vector $u$ is maximally entangled 
and $\nu$ is the invariant measure on $SU(d)$.
Then, the operator $T_{inv}^{2,A\to B}\defeq T(M_{cov}^2)$ 
has the form:
\begin{align}
& T_{inv}^{2,A\to B}\nonumber \\
=& \ket{\phi^0_{A,B}}\bra{\phi^0_{A,B}}
\otimes
\ket{\phi^0_{A,B}}\bra{\phi^0_{A,B}}\nonumber\\
& + 
\frac{1}{d^2-1}(I-\ket{\phi^0_{A,B}}\bra{\phi^0_{A,B}})
\otimes (I-\ket{\phi^0_{A,B}}\bra{\phi^0_{A,B}}),\Label{22-20}
\end{align}
which is shown in Appendix \ref{a-9}.
This equation implies that the testing 
$T(M_{cov}^2)$ does not depend on the choice of the
maximally entangled state $u$.
It also guarantees the $U(d^2-1)$-invariance of the test
$T_{inv}^{2,A\to B}$.
We also obtain 
the equation
\begin{align}
\Tr T_{inv}^{2,A\to B} \sigma^{\otimes 2} =
(1-p)^2 + \frac{p^2}{d^2-1}
=1- 2p+ \frac{d^2 p^2}{d^2-1}.\Label{22-23}
\end{align}
Since the test $T_{inv}^{2,A\to B} $ is a level-$0$ test with 
the null hypothesis $\cl{S}_0$,
the inequality
\begin{align*}
 \beta^{L(A\to B)}_{0,2,U(d^2-1)}( 0\| \sigma) 
\le 
1- 2p+ \frac{d^2 p^2}{d^2-1}
\end{align*}
holds.
Next, we apply the discussion of subsection \ref{s4A}.
Then, the test 
$T^{2,A-B}_{\epsilon,\alpha}\defeq
T^{1}_{\alpha}(T_{inv}^{2,A\to B},2\epsilon - \frac{d^2\epsilon^2}{d^2+1})$
is a level-$\alpha$ $U(d^2-1)$-invariant test.
Since the test $T^{2,A-B}_{\epsilon,\alpha}$ can
be performed by randomized operation with 
$T_{inv}^{2,A\to B}$ and $I-T_{inv}^{2,A\to B}$,
we obtain
\begin{align*}
&\beta^{L(A \to B)}_{\alpha,2,U(d^2-1)}(\le \epsilon\| \sigma)
\le
\Tr T^{2,A-B}_{\epsilon,\alpha} \sigma^{\otimes 2} \\
%\beta(T_{\epsilon,\alpha}^{2,A-B},\sigma)\\
= &\left\{
\begin{array}{ll}
\frac{(1- \alpha)(1- 2 p + \frac{d^2 p^2}{d^2+1})}
{1- 2 \epsilon + \frac{d^2\epsilon^2}{d^2+1}}
& \hbox{if }
2 \epsilon - \frac{d^2\epsilon^2}{d^2+1} \le \alpha \\
1- 
\frac{\alpha (2p + \frac{d^2 p^2}{d^2-1} )}
{2\epsilon - \frac{d^2 \epsilon^2}{d^2-1} }
& \hbox{if }
2 \epsilon - \frac{d^2 \epsilon^2}{d^2+1} \,> \alpha .
\end{array}
\right.
\end{align*}
Furthermore, as a generalization of (\ref{22-23}),
we obtain the following lemma, which is more useful in 
the asymptotic setting from an applied viewpoint.
\begin{lem}\Label{l-1}
Let $M= \{ p_i \ket{u_i}\bra{u_i}\}(\|u_i\|=1)$ 
be a POVM on A's two-sample space
${\cal H}_A^{\otimes 2}$.
If every state $\ket{u_i}$ is a maximally entangled state on 
${\cal H}_A^{\otimes 2}$,
the test
$T(M)$
satisfies
\begin{align}
T(M)= 
 \ket{\phi^0_{A_1,B_1}\otimes \phi^0_{A_2,B_2}}
\bra{\phi^0_{A_1,B_1}\otimes \phi^0_{A_2,B_2}}
+
P T(M) P, \Label{24-38}
\end{align}
and 
\begin{align}
\bra{\phi^0_{AB}}\sigma \ket{\phi^0_{AB}}^2 &\le
\Tr \sigma^{\otimes 2} T(M) \\
&\le \bra{\phi^0_{AB}}\sigma \ket{\phi^0_{AB}}^2
+ (1- \bra{\phi^0_{AB}}\sigma \ket{\phi^0_{AB}})^2,
\label{l-1-e}
\end{align}
where
\begin{align*}
P\defeq (I- \ket{\phi^0_{A_2,B_2}}\bra{\phi^0_{A_2,B_2}})
\otimes (I- \ket{\phi^0_{A_1,B_1}}\bra{\phi^0_{A_1,B_1}}).
\end{align*}
\end{lem}
Indeed, it is difficult to realize the covariant 
POVM $M_{cov}^2$.
The Bell measurement $M^2_{Bell}\defeq
\{ \ket{\phi^{n,m}_{1,2}}\bra{\phi^{n,m}_{1,2}}\}_{(n,m)=(0,0)}
^{(d-1,d-1)}$ can be constructed more easily,
where $\phi^{n,m}_{1,2}$ is defined by 
\begin{align*}
\phi^{0,0}_{1,2}& \defeq
\frac{1}{\sqrt{d}}\sum_{j=0}^{d-1} \ket{j}_{A,1}\ket{j}_{A,2}\\
\phi^{n,m}_{1,2}& \defeq
\left( (X^n Z^m)\otimes I \right) \phi^{0,0}_{1,2}\\
X &\defeq \sum_{j=1}^{d-1} \ket{j}\bra{j-1}+ \ket{0}\bra{d-1} \\
Z &\defeq \sum_{j=0}^{d-1} e^{2\pi j i/d}\ket{j}\bra{j}.
\end{align*}
As will be mentioned in subsection \ref{s6D},
the test $T(M^2_{Bell})$ 
can be used as the alternative test of $T_{inv}^{2,A\to B}$
in an asymptotic sense.

\subsection{$n$-sample setting}\Label{s6C}
Next, we construct a $U(d^2-1)$-invariant test
when $2n$ samples of the unknown state $\sigma$ are prepared.
It follows from a discussion similar to subsection \ref{s5A} that
the test ${T'}_{\epsilon,\alpha}^{2n}
\defeq T^{2n}_\alpha(T_{inv}^{2,A\to B},
2 \epsilon - \frac{d^2 \epsilon^2}{d^2-1})$ 
is level-$\alpha$ for given $\alpha$ and $\epsilon$.
The $U(d^2-1)$-invariance of 
the test $T_{inv}^{2,A\to B}$ implies the $U(d^2-1)$-invariance 
of the test ${T'}_{\epsilon,\alpha}^{2n}$.
Since the test ${T'}_{\epsilon,\alpha}^{2n}$ can be realized by
one-way LOCC $A \to B$, 
the inequality
\begin{align}
\beta_{\alpha,2n,U(d^2-1)}^{L(A\to B)}(\le \epsilon\|\sigma)
&\le
 \Tr {T'}_{\epsilon,\alpha}^{2n} \sigma^{\otimes 2n}\nonumber \\
& =
\beta_{\alpha}^{n}\left(\le 2\epsilon- \frac{d^2 \epsilon^2}{d^2-1}
\left \|
2p- \frac{d^2 p^2}{d^2-1}\right.\right)
\Label{22-28}
\end{align}
holds.
In addition, we can derive a similar bound for the hypothesis
$\cl{S}_{\ge \epsilon}$.

Concerning the case of $\epsilon =0$, we have another bound as follows.
For this purpose, we focus on 
the test $T^{1,A\to B}_{inv}$ in the case when 
$\cl{H}_A=\cl{H}_A^{\otimes n}$ and $\cl{H}_B=\cl{H}_B^{\otimes n}$.
Denoting this test by $T^{1,A^{\otimes n}\to B^{\otimes n}}_{inv}$,
we have
\begin{align*}
T_{inv}^{1,A^{\otimes n} \to B^{\otimes n}}
=&\ket{\phi^0_{A,B}}\bra{\phi^0_{A,B}}^{\otimes n}\\
& +
\frac{1}{d^n+1}(I- \ket{\phi^0_{A,B}}\bra{\phi^0_{A,B}}^{\otimes n})
\\
\Tr T_{inv}^{1,A^{\otimes n}\to B^{\otimes n}} \sigma^{\otimes n}
=& \frac{d^n (1-p)^n+1}{d^n+1}
\end{align*}
because $\Tr \ket{\phi^0_{A,B}}\bra{\phi^0_{A,B}}^{\otimes n}
\sigma^{\otimes n}=(1-p)^n$.
Since this test is $U(d^2-1)$-invariant,
we obtain
\begin{align}
\beta_{\alpha,n,U(d^2-1)}^{L(A\to B)}(0\|\sigma)
\le
\frac{d^n (1-p)^n+1}{d^n+1}.\Label{23-10}
\end{align}

\subsection{Asymptotic setting}
\begin{figure}[htbp]
\begin{center}
\includegraphics[width=8cm]{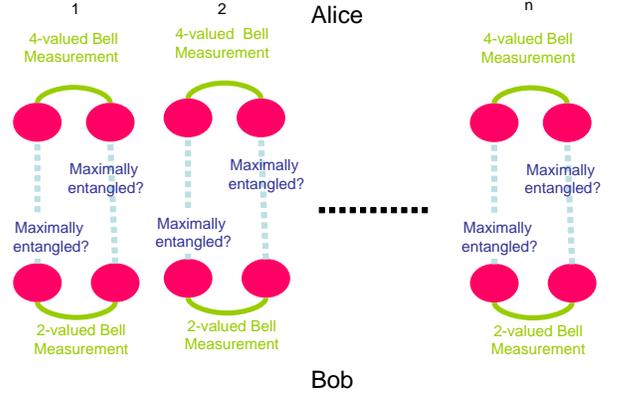}
\end{center}
\caption{
Asymptotic optimal testing scheme
when $2n$ identical copies are given}
\label{one}
\end{figure}

\Label{s6D}
We proceed to asymptotic setting.
First, we show that 
even if our test satisfies the A-B LOCC condition,
the bound (\ref{23-4}) can be attained in 
the asymptotic small deviation setting.
Indeed, since
$P^n_{2 \frac{t}{2n}- \frac{d^2}{d^2-1}\left(\frac{t}{2n}\right)^2}
(k) \to P_t(k)$,
the equation 
\begin{align*}
&\lim \beta^{n}_\alpha 
\left(\le 2\frac{\delta}{2n}- \frac{d^2}{d^2-1}
\left(\frac{\delta}{2n}\right)^2
\left \|
 2\frac{t'}{2n}- \frac{d^2}{d^2-1}\left(\frac{t'}{2n}\right)^2
\right.\right)\\
&=
\beta_{\alpha}(\le \delta\|t')
\end{align*}
can be proven similarly to Theorem \ref{t-2}.
Hence, from (\ref{22-7}) and (\ref{22-9}), we have
\begin{align*}
\lim \beta_{\alpha,2n,G}^{C}(\le \frac{\delta}{n}\|\sigma_n)
=
\beta_{\alpha}(\le \delta\|t')
\end{align*}
for $G=U(1),SU(d)\times U(1), U(d^2-1)$, $C=\emptyset,
L(A\to B), L(A \leftrightarrows B), S(A,B)$.
However, it is difficult to realized the covariant POVM
$M_{cov}^2$ on $\cl{H}_{A}^{\otimes 2}$.
Even if the test ${T'}_{\epsilon,\alpha}^{2n}$ is replaced by
${T'}_{\epsilon,\alpha,Bell}^{2n}\defeq
T^n_{\alpha}(T(M^2_{Bell}), 2\epsilon - \frac{d^2\epsilon^2}{d^2-1})$, 
the bound $\beta_\alpha(\le \delta\|t')$ can be attained 
in the following asymptotic sense.
The test ${T'}_{\frac{\delta}{2n},\alpha,Bell}^{2n}$ may be not 
level-$\alpha$ with the null hypothesis $\cl{S}_{\le \delta/2n}$,
but is asymptotically level-$\alpha$, {\it i.e.},
\begin{align}
\Tr {T'}_{\frac{\delta}{2n},\alpha,Bell}^{2n} \sigma_{2n}^{\otimes 2n}
\to 1-\delta \Label{23-11}
\end{align}
if $\bra{\phi^0_{A,B}}\sigma_n\ket{\phi^0_{A,B}}
= 1- \frac{\delta}{n}$.
Moreover, if $\bra{\phi^0_{A,B}}\sigma_n\ket{\phi^0_{A,B}}
= 1- \frac{t'}{n}$ and $t' \,> \delta$, the relation
\begin{align}
\Tr {T'}_{\frac{\delta}{2n},\alpha,Bell}^n \sigma_n^{\otimes n}
\to \beta_\alpha(\le\delta\|t') \Label{23-12} 
\end{align}
holds. 
These relations (\ref{23-11}) and (\ref{23-12}) follow from
Lemma \ref{l-1}.
Hence, there is no advantage of use of 
entanglement between $\cl{H}_A$ and $\cl{H}_B$
for this testing in the asymptotic small deviation setting.
Similar results for the null hypothesis $\cl{S}_{\ge \delta/n}$
can be obtained.
The asymptotic optimal testing scheme is illustrated as Fig. \ref{one}.

Next, we proceed to the large deviation setting.
The inequality (\ref{23-10}) yields
\begin{align}
\lim \frac{-1}{n}\log \beta_{\alpha,n,U(d^2-1)}^{L(A\to B)}(0\|\sigma)
\ge
\left\{
\begin{array}{ll}
-\log (1-p) & \hbox{ if } 1-p \ge \frac{1}{d}\\
\log d  & \hbox{ if } 1-p \,< \frac{1}{d}
\end{array}
\right. . \Label{23-15}
\end{align}
Hence, the relations (\ref{22-9}) and (\ref{23-5})
guarantee that
if $1-p \ge \frac{1}{d}$,
\begin{align*}
\lim \frac{-1}{n}\log \beta_{\alpha,n,U(d^2-1)}^{L(A\to B)}(0\|\sigma)
= -\log (1-p),
\end{align*}
for $G=U(1),SU(d)\times U(1), U(d^2-1)$, $C=\emptyset,
L(A\to B), L(A \leftrightarrows B), S(A,B)$.
Hence, we can conclude that 
if $1-p \ge \frac{1}{d}$,
there is no advantage of use of 
entanglement between $\cl{H}_A$ and $\cl{H}_B$
for this testing even in this kind of the asymptotic large deviation 
setting.

\section{A-B locality and Sample locality}\Label{s7}
In this section, we discuss 
the locality among $A_1,B_1, \ldots, A_n,B_n$.
Since the case $n=1$ of this setting is the same as 
that of the setting section \ref{s6}.
Hence, we treat the case $n=2$, at first.

\subsection{Two-sample setting}\Label{s7A}
We construct a level-$0$ $SU(d)$-invariant test for 
the null hypothesis $\cl{S}_0= \{ \ket{\phi^0_{A,B}}
\bra{\phi^0_{A,B}}\}$ as follows.
For this purpose, we define a POVM $M_{cov}^{1\to 2}$
on Alice's space $\cl{H}_{A}^{\otimes 2}$,
which can be realized by one-way LOCC $A_1 \to A_2$
from the first system $\cl{H}_{A_1}$
to the second system $\cl{H}_{A_2}$.

\par\noindent{\bf Construction of $M_{cov}^{1\to 2}$:~}

\par\noindent{\bf 1)} Alice performs the covariant POVM $M_{cov}^1$
on the first system $\cl{H}_{A_1}$, and obtain
the data corresponding to the state 
$\ket{\varphi}\bra{\varphi}$.

\par\noindent{\bf 2)}We choose the Projection-valued measure
$\{ \ket{u^i(\varphi)}\bra{u^i(\varphi)}\}_i$
satisfying that
\begin{align}
\langle u^i(\varphi)| u^j(\varphi)\rangle = 0 , \quad 
\langle u^i(\varphi)| \varphi \rangle= \frac{1}{\sqrt{d}} . \Label{22-25}
\end{align}
The existence of $\{ u^i(\varphi)\}_i$ is shown in Appendix \ref{a28}.

\par\noindent{\bf 3)} Alice randomly chooses $g \in U(d-1)$ which acts on 
the space orthogonal to $\varphi$,
and performs 
the Projection-valued measure
$\{ \ket{g u^i(\varphi)}\bra{g u^i(\varphi)}\}_i$ on the second
system $\cl{H}_{A_2}$.

Since Bob's measurement of the test $T(M_{cov}^{1\to 2})$
can be also realized by one-way LOCC on Bob's space,
this test is a $L(A_1,A_2 \to B_1, B_2)$ test.
Its POVM is given by 
\begin{align*}
M_{cov}^{1\to 2}(\,d g)
=
d^2 (g \otimes g )
\ket{u_1 \otimes u_2}\bra{u_1 \otimes u_2}
(g \otimes g )^{\dagger} \nu(\,d g),
\end{align*}
where we choose $u_1$ and $u_2$ satisfying
$|\langle u_1| u_2\rangle|^2 = \frac{1}{d}$.
Thus, the $SU(d)$-covariance of $M_{cov}^{1\to 2}$
guarantees the $SU(d)$-invariance of
the test $T_{inv}^{A_1\to A_2\to B^{\otimes 2}}
\defeq T(M_{cov}^{1\to 2})$.
Moreover, as is shown in Appendix \ref{a-11},
the test $T_{inv}^{A_1\to A_2\to B^{\otimes 2}}$ is $U(1)$-invariant.
Hence,
the inequality
\begin{align*}
\beta^{L(A_1,A_2 \to B_1,B_2)}_{0,2,SU(d)\times U(1)}
(0\| \sigma)
\le \Tr T_{inv}^{A_1\to A_2\to B^{\otimes 2}}\sigma^{\otimes 2}
\end{align*}
holds.
On the other hand, 
the equation
\begin{align}
\beta^{L(A_1,A_2 \to B_1,B_2)}_{0 ,2,SU(d)}
(0\| \sigma)
=
\Tr T_{inv}^{A_1\to A_2\to B^{\otimes 2}}
\sigma^{\otimes 2} \Label{22-27}
\end{align}
holds,
which is shown in Appendix \ref{a-14}.
Hayashi {\it et~al.}\cite{tsu}
have obtained a similar result in the two-dimensional case.
Thus,
\begin{align*}
\beta^{L(A_1,A_2 \to B_1,B_2)}_{0,2,SU(d)}
(0\| \sigma)
&=\beta^{L(A_1,A_2 \to B_1,B_2)}_{0,2,SU(d)\times U(1)}
(0\| \sigma)\\
& =
\Tr T_{inv}^{A_1\to A_2\to B^{\otimes 2}}
\sigma^{\otimes 2}.
\end{align*}
Therefore, the test $T_{inv}^{A_1\to A_2\to B^{\otimes 2}}$
is a UMP $L(A_1,A_2 \to B_1,B_2)$ $G$-invariant
test with level-$0$ for the null hypothesis $\cl{S}_0$,
where $G= SU(d),SU(d)\times U(1)$.

\subsection{$n$-sample setting}\Label{s7B}
Next, we proceed to $n$-sample setting.
Since
the test ${T''}_{\epsilon,\alpha}^n
\defeq T_\alpha^n(T_{inv}^{1,A\to B}, \frac{d \epsilon}{d+1})
$ is level-$\alpha$ $U(d^2-1)$-invariant 
test
with the hypothesis $\cl{S}_{\le \epsilon}$,
and satisfies the condition of $L(A_1,\ldots, A_n \to B_1,\ldots, B_n)$,
the inequality
\begin{align}
& \beta_{\alpha,n,U(d^2-1)}^{L(A_1,\ldots, A_n \to B_1,\ldots, B_n)}
(\le \epsilon\|\sigma) \nonumber \\
\le &
 \Tr {T''}_{\epsilon,\alpha}^n \sigma^{\otimes n}
=
\beta_{\alpha}^{n}\left(\left.\le \frac{d \epsilon}{d+1}
\right\|\frac{d p}{d+1}\right)
\Label{23-19}
\end{align}
holds.

Conversely,
as a lower bound of
$\beta_{\alpha,n,SU(d)}
^{S(A_1,\ldots, A_n ,B_1,\ldots, B_n)}(\le \epsilon\|\sigma)$,
we obtain 
\begin{align}
& \frac{1}{n}
\log \frac{\beta_{\alpha,n,SU(d)}
^{S(A_1,\ldots, A_n ,B_1,\ldots, B_n)}(0\|\sigma) }{1-\alpha}
\nonumber \\
\ge &
\min_{u,u': |\langle u | \overline{u'} \rangle|=1,\|u\|=1} \nonumber\\
& \quad \int_{SU(d)}
\log 
d \bra{gu \otimes \overline{g} u'}
\sigma 
\ket{gu \otimes \overline{g} u'}
\nu (d g) , \Label{23-20}
\end{align}
which will be shown in Appendix \ref{a-12}.

\subsection{Asymptotic setting}\Label{s7C}
Taking the limit in (\ref{23-19}), we obtain
\begin{align}
&\lim \beta_{\alpha,n,U(d^2-1)}^{L(A_1,\ldots, A_n \to B_1,\ldots, B_n)}
\left(\left. \le \frac{\delta}{n} \right\|\sigma_n\right) \nonumber \\
\le &
\beta_{\alpha}\left(\left.\le \frac{d \delta}{d+1}\right\|\frac{d 
t'}{d+1}\right)
\Label{23-24}
\end{align}
if $\bra{\phi^0_{A,B}}\sigma\ket{\phi^0_{A,B}}= 1-\frac{t'}{n}$.
Conversely,
by using the inequality (\ref{23-20}),
the compactness of the sets 
$\{u,u'| |\langle u | \overline{u'} \rangle|=1, \|u\|=1\}$
and $SU(d)$
yields
\begin{align*}
& \lim 
\log \frac{\beta_{\alpha,n,SU(d)}
^{S(A_1,\ldots, A_n ,B_1,\ldots, B_n)}(0\|\sigma_n)}{1-\alpha} 
\nonumber \\
\ge &
\min_{u,u': |\langle u | \overline{u'} \rangle|=1,\|u\|=1}
\int_{SU(d)}
\lim n \\
&\hspace{10ex} \log 
d \bra{gu \otimes \overline{g} u'}
 \sigma_n 
 \ket{gu \otimes \overline{g} u'}
\nu (d g) \nonumber \\
= &
- \min_{u,u': |\langle u | \overline{u'} \rangle|=1,\|u\|=1}
\int_{SU(d)} \\
&\hspace{10ex} \lim n 
\left(1- d \bra{gu \otimes \overline{g} u'}
 \sigma_n 
 \ket{gu \otimes \overline{g} u'}\right)
\nu (d g) \\
=&
- \min_{u,u': |\langle u | \overline{u'} \rangle|=1,\|u\|=1}
\lim n \Tr (I - T_{u,u'}) \sigma_n,
\end{align*}
where
\begin{align*}
T_{u,u'}\defeq
\int_{SU(d)}
d \ket{gu \otimes \overline{g} u'}
\bra{gu \otimes \overline{g} u'}
\nu (d g).
\end{align*}
Since $T_{u,u'}$ is $SU(d)$-invariant.
The test $T_{u,u'}$ has the form 
$t_0 \ket{\phi^0_{A,B}}\bra{\phi^0_{A,B}}+
t_1(I-\ket{\phi^0_{A,B}}\bra{\phi^0_{A,B}})$.
The condition $|\langle u | \overline{u'} \rangle|=1$
guarantees that $t_0 = 1$.
The definition of $T_{u,u'}$ guarantees that
$\Tr T_{u,u'} \ge d$, which implies
$t_1 \ge \frac{1}{d+1}$.
Hence,
\begin{align}
& \Tr (I - T_{u,u'}) \sigma_n
\le \frac{d}{d+1}\Tr (I-\ket{\phi^0_{A,B}}\bra{\phi^0_{A,B}})\sigma_n 
\nonumber \\
= &\frac{d}{d+1}
(1- \bra{\phi^0_{A,B}}\sigma \ket{\phi^0_{A,B}})
= \frac{d}{d+1}\frac{t'}{n} \Label{23-26}.
\end{align}
Thus, we have
\begin{align*}
\lim 
\log \frac{\beta_{\alpha,n,SU(d)}
^{S(A_1,\ldots, A_n ,B_1,\ldots, B_n)}(0\|\sigma_n)}{1-\alpha} 
\ge - \frac{d t'}{d+1},
\end{align*}
which implies 
\begin{align*}
\lim 
\beta_{\alpha,n,SU(d)}
^{S(A_1,\ldots, A_n ,B_1,\ldots, B_n)}(0\|\sigma_n)
\ge (1-\alpha)e^{-\frac{d t'}{d+1}}.
\end{align*}
Combining (\ref{23-24}) in the case of $\epsilon$,
we obtain
\begin{align*}
\lim 
\beta_{\alpha,n,G}
^{S(A_1,\ldots, A_n ,B_1,\ldots, B_n)}(0\|\sigma_n)
= (1-\alpha)e^{-\frac{d t'}{d+1}}
\end{align*}
for $G= SU(d), SU(d)\times U(1), U(d^2-1)$,
$C=S(A_1,\ldots, A_n ,B_1,\ldots, B_n)$,
$L(A_1,\ldots, A_n ,B_1,\ldots, B_n)$,
$L(A_1,\ldots, A_n  \to B_1,\ldots, B_n)$.
Since $(1-\alpha)e^{-\frac{d t'}{d+1}}
\,< (1-\alpha)e^{-t'}= \beta_\alpha(0\|t')$,
there is an advantage to use of quantum correlation among samples.

\section{Two-sample Two-dimensional setting}\Label{s6E}
Next, we proceed to the special case $n=2$ and $d=2$.
For the analysis of this case, we define the $3 \times 3$ 
real symmetric matrix 
$V= (v_{i,j})_{1\le i,j\le 3}$ as
\begin{align*}
v_{i,j} &\defeq \Re \bra{\phi_{A,B}^i} \sigma \ket{\phi_{A,B}^j} \\
\phi_{A,B}^1&\defeq \frac{1}{\sqrt{2}}
\left(
\ket{10}+ \ket{10}
\right) ,\quad
\phi_{A,B}^2 \defeq \frac{1}{\sqrt{2}}
\left(
-i \ket{10}+ i \ket{10}
\right) ,\\
\phi_{A,B}^3&\defeq \frac{1}{\sqrt{2}}
\left(
\ket{00}- \ket{11}
\right) .
\end{align*}
When $\sigma$ satisfies the following condition
$p \le \frac{1}{2}$,
as is shown in Appendix \ref{a19},
the equation
\begin{align}
&\beta_{0,2,SU(2)\times U(1)}^C(0\|\sigma )\nonumber \\
= &
(1-p)^2 + \frac{p^2}{3}
- \frac{3}{5}
\left(
\Tr \frac{I}{3} V^2 
- (\Tr \frac{I}{3} V )^2 
\right)\Label{2-9-2}
\end{align}
holds, where $C=L(A \to B),L(A \leftrightarrows B), S(A,B)$.
Since the quantity $
\Tr \frac{I}{3} V^2 - (\Tr \frac{I}{3} V )^2 $ 
is greater than $0$,
its $\frac{3}{5}$ times give the advantage of this optimal test against
the test introduced in subsection\ref{s6B}.
Hence, this merit vanish if and only if the real symmetric 
matrix $V$ is constant.
In addition, 
the optimal test $T$ is 
given as follows.
First, we define a covariant POVM 
\begin{align*}
M_{op}(d g)\defeq 
4 \int_{SU(2)}
g^{\otimes 2}
\ket{u_{op}}\bra{u_{op}}
(g^{\otimes 2})^{\dagger}
\nu (d g),
\end{align*}
where the vector $u_{op}$ is defined
as
\begin{align*}
u_{op} \defeq 
& \frac{1}{2}\left(\ket{01}_{A_1,A_2}- \ket{10}_{A_1,A_2}\right) \\
& \quad +
\frac{\sqrt{3}}{2}\left(\ket{00}_{A_1,A_2}+ \ket{11}_{A_1,A_2}\right).
\end{align*}
Then, as is shown in Appendix \ref{a19},
the relation 
\begin{align}
\beta_{0,2,SU(2)\times U(1)}^C(0\|\sigma )
= \Tr T(M_{op}) \sigma^{\otimes 2}\Label{2-12-10}
\end{align}
holds.
That is, the test $T(M_{op})$ is the UMP $SU(2)\times U(1)$-invariant
$C$ test with the condition $p \le \frac{1}{2}$,
where $C=L(A \to B),L(A \leftrightarrows B), S(A,B)$.

On the other hand,
as is shown in Appendix \ref{a20},
the RHS of (\ref{22-27}) is calculated as
\begin{align}
&\beta_{0,2,SU(2)}^{L(A_1,A_2 \to B_1,B_2)}
(0\|\sigma) 
=\beta_{0,2,SU(2)\times U(1)}^{L(A_1,A_2 \to B_1,B_2)}
(0\|\sigma) \nonumber \\
=& (1-\frac{2}{3}p)^2 
- \frac{1}{5}
\left(
\Tr \frac{I}{3} V^2 
- (\Tr \frac{I}{3} V )^2 
\right). \Label{2-9}
\end{align}
That is, the quantity 
$\frac{1}{5}\left(
\Tr \frac{I}{3} V^2 
- (\Tr \frac{I}{3} V )^2 
\right)$
presents the effect of use of classical communication
between $A_1$ and $A_2$.

\section{Two different systems}\Label{s8}
In section \ref{s6}, we showed that 
if we can prepare the two identical states simultaneously and
we can perform Bell measurement on this joint system,
the asymptotically optimal test can be realized.
However, it is a bit difficult to
prepare two identical states from the same source simultaneously.
However, as is discussed in this section,
if we can prepare two quantum states from the different source independently, 
%even if these are not identical,
this Bell measurement is asymptotically
optimal.

\subsection{Formulation}\Label{s8Z}
Since the state on $\cl{H}_{A,B}^{\otimes 2}$
can be described as $\sigma_1\otimes \sigma_2$,
our hypotheses are given as
\begin{align*}
H_0:&\quad \cl{S}_{\le \epsilon}^2\defeq
\left\{ \sigma_1\otimes \sigma_2
\left|
\begin{array}{ll}
(1- \bra{\phi^{0}_{A,B}}\sigma_1\ket{\phi^{0}_{A,B}}) \\
+
(1- \bra{\phi^{0}_{A,B}}\sigma_2\ket{\phi^{0}_{A,B}})
\le \epsilon 
\end{array}
\right.\right\}\\
&\hbox{ versus}\\
H_1:&\quad \cl{S}_{\le \epsilon}^{2c}
\defeq
\left\{ \sigma_1\otimes \sigma_2
\left|
\begin{array}{ll}
(1- \bra{\phi^{0}_{A,B}}\sigma_1\ket{\phi^{0}_{A,B}}) \\
+
(1- \bra{\phi^{0}_{A,B}}\sigma_2\ket{\phi^{0}_{A,B}})
> \epsilon 
\end{array}
\right.\right\}.
\end{align*}
For any group action $G$ introduced in subsection \ref{s3B},
these hypotheses are invariant for $G\times G$-action
defined as
\begin{align*}
\phi \mapsto (g_1 \otimes g_2)\phi \quad
\forall (g_1,g_2)\in G\times G.
\end{align*}

When only two particles $\cl{H}_{A_1,B_1}\otimes
\cl{H}_{A_2,B_2}$ are prepared,
similarly to subsection \ref{s3C}, we can define 
the quantities 
$\beta_{\alpha,2,G\times G}^C
(\le \epsilon\|\sigma_1\otimes \sigma_2)$
for the condition $C=\emptyset,S(A,B),L(A\leftrightarrows B),
L(A\to B),S(A_1,A_2 ,B_1,B_2),L(A_1,A_2, B_1,B_2),
L(A_1,A_2 \to B_1,B_2)$,
in which, ``$2$'' means two particles, 
{\it i.e.}, there is only one sample of
$\sigma_1 \otimes \sigma_2$.
When $n$ samples $(\sigma_1\otimes \sigma_2)^{\otimes n}$
are prepared, 
we also define the quantities
$\beta_{\alpha,2n,G\times G}^C
(\le \epsilon\|\sigma_1\otimes \sigma_2)$
for the condition $C=\emptyset,S(A,B),L(A\leftrightarrows B),
L(A\to B),S(A_1,A_2 ,B_1,B_2),L(A_1,A_2, B_1,B_2),
L(A_1,A_2 \to B_1,B_2)$.

\subsection{One-sample setting}\Label{s8A}
In this section,
we treat the case of $one-sample$ and $\epsilon=0$ case.
In the first step,
we focus on the case of $C=\emptyset$.
In this case,
%When we use the test $T_\emptyset\defeq 
%\ket{\phi_{A,B}^0\otimes\phi_{A,B}^0}
%\bra{\phi_{A,B}^0\otimes\phi_{A,B}^0}$,the second error can be calculated
%\begin{align*}
%\beta(T_\emptyset,\sigma_1\otimes \sigma_2)=(1-p_1)(1-p_2) ,
%\end{align*}
%Moreover,
the relations
\begin{align*}
\beta_{0,2,G\times G}^{\emptyset}
(0\|\sigma_1\otimes \sigma_2)
&=\bra{\phi_{A,B}^0\otimes\phi_{A,B}^0}\sigma_1\otimes \sigma_2
\ket{\phi_{A,B}^0\otimes\phi_{A,B}^0}\\
&=(1-p_1)(1-p_2) 
\end{align*}
hold for $G=\emptyset, U(1), SU(d)\times U(1), U(d^2-1)$,
where $p_i = 1- \bra{\phi_{A,B}^0}\sigma_i\ket{\phi_{A,B}^0}$.
%Hence, the test $T_\emptyset$ is UMP $G$-invariant test.

Next, we focus on the case of $C=L(A\to B),L(A\leftrightarrows B),S(A, B)$.
When we use the test $T_{inv}^{2,A\to B}$,
the second error is
\begin{align*}
\beta(T_{inv}^{2,A\to B}
,\sigma_1\otimes \sigma_2)=
(1-p_1)(1-p_2)+ \frac{p_1 p_2}{d^2-1} .
\end{align*}
Moreover, the optimal second error can also be calculated as
\begin{align}
\beta_{0,2,G\times G}^{C}
(0\|\sigma_1\otimes \sigma_2)
=(1-p_1)(1-p_2)+ \frac{p_1 p_2}{d^2-1} \Label{2-18-1}
\end{align}
for $C=L(A\to B),L(A\leftrightarrows B),S(A, B)$
when $\frac{p_1 p_2}{d^2-1}\le
(1-p_1)p_2,p_1(1-p_2)$.
Its proof is given in Appendix \ref{a24}.
Hence, the test $T_{inv}^{2,A\to B}$ is the
$C$-UMP $G$-invariant test.
Using the PPT condition,
Hayashi {\it et~al.}\cite{tsu} derived this optimal test 
in the case of $\sigma_1=\sigma_2,d=2$.

Finally, we proceed to 
the case of $C=L((A_1, A_2) \to (B_1, B_2) ),
L(A_1, A_2,B_1,B_2),S(A_1, A_2,B_1,B_2)$.
When we use the test $T_{inv}^{1,A_1\to B_1}\otimes T_{inv}^{1,A_2\to 
B_2}$,
the second error is
\begin{align*}
& \beta(T_{inv}^{1,A_1\to B_1}\otimes T_{inv}^{1,A_2\to B_2}
,\sigma_1\otimes \sigma_2)\\
=&
\left(1-\frac{d p_1}{d+1}\right)
\left(1-\frac{d p_2}{d+1}\right).
\end{align*}
In this case, as is proven in Appendix \ref{a25},
the optimal second error is calculated as
\begin{align}
\beta_{0,2,G\times G}^{C}
(0\|\sigma_1\otimes \sigma_2)
=\left(1-\frac{d p_1}{d+1}\right)
\left(1-\frac{d p_2}{d+1}\right),\Label{2-19-1}
\end{align}
for $G=SU(d), SU(d)\times U(1), U(d^2-1)$.
Thus, the test $T_{inv}^{1,A_1\to B_1}\otimes T_{inv}^{1,A_2\to B_2}$ is 
the $C$-UMP $G$-invariant test.
Hayashi {\it et~al.}\cite{tsu} derived this optimal test 
in the case of $\sigma_1=\sigma_2,d=2$.

\subsection{Asymptotic setting}\Label{s8B}
In the small deviation asymptotic setting with $n$ samples,
we focus on the case $\epsilon= \frac{\delta}{n}$ and
$\frac{t_i'}{n}= 1- \bra{\phi_{A,B}^0}
\sigma_{i,n}'\ket{\phi_{A,B}^0}$.
In this setting, as is shown in Appendix \ref{a26},
\begin{align}
\lim \beta_{\alpha,2n,G\times G}^{\emptyset}
(\le \frac{\delta}{n}\|\sigma_{1,n}'\otimes \sigma_{2,n}')
=\beta_\alpha(\le \delta\|t_1'+t_2') \Label{2-17-2}
\end{align}
for $G=U(1), SU(d)\times U(1), U(d^2-1)$.

\begin{figure}[htbp]
\begin{center}
\includegraphics[width=8cm]{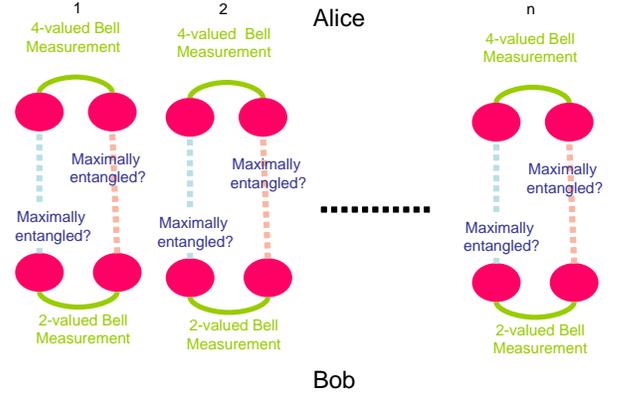}
\end{center}
\caption{
Asymptotic optimal testing scheme
when $n$ identical copies are given 
and one sample system consists of two different systems
}
\label{two}
\end{figure}

Next, we consider the case of $C=L(A\to B)$.
When we perform the test $T_{inv}^{2,A\to B}$ for all systems
$\cl{H}_{A_1}\otimes \cl{H}_{B_1}, \ldots,
\cl{H}_{A_n}\otimes \cl{H}_{B_n}$
whose state is $\sigma_{1,n}'\otimes \sigma_{2,n}'$,
the number $k$ of detecting $T_{inv}^{2,A\to B}$
almost obeys the Poisson distribution
$e^{-(t_1'+t_2')}\frac{(t_1'+t_2')^k}{k!}$.
This is because
$n\left(1- (1-\frac{t_1'}{n})(1-\frac{t_2'}{n})+ 
\frac{\frac{t_1'}{n} \frac{t_2'}{n}}{d^2-1} \right)
\to t_1'+t_2'$.
Treating the hypothesis testing of this Poisson distribution,
we can show that the $L(A\to B)$ $U(d^2-1)\times U(d^2-1)$-invariant
test 
$T^{n,2}_{\epsilon,\alpha}
\defeq T_\alpha^n( T_{inv}^{2,A\to B},
\max_{p_1+p_2 = \epsilon} p_1+p_2 - \frac{d^2 p_1 p_2}{d^2-1})$
satisfies that
\begin{align*}
\lim 
\beta(T^{n,2}_{\delta/n,\alpha}
,\sigma_{1,n}'\otimes \sigma_{2,n}')
=\beta_\alpha(\le \delta\|t_1'+t_2').
\end{align*}
Hence, combining (\ref{2-17-2}),
we obtain 
\begin{align*}
\lim \beta_{\alpha,2n,G\times G}^C
\left(\left. \le \frac{\delta}{n}\right\|
\sigma_{1,n}'\otimes \sigma_{2,n}'\right)
=\beta_\alpha(\le \delta\|t_1'+t_2').
\end{align*}
for $C=\emptyset,L(A\to B),L(A\leftrightarrows B),S(A, B)$,
$G=SU(d)\times U(1), U(d^2-1)$.
Therefore,
the test $T^{n,2}_{\epsilon,\alpha}$ is $C$-UMP $G$-invariant test
in the asymptotic small deviation setting.
The asymptotic optimal testing scheme is illustrated as Fig. \ref{two}.

Moreover, if we use the test based on the Bell measurement 
in stead of the test $T_{inv}^{2, A\to B}$,
the bound $\beta_\alpha(\le \delta\|t_1'+t_2')$ can be attained
because of a reason similar to Lemma \ref{l-1}.

\section{Three different systems}\Label{s9}
Finally,
we treat the case of three quantum states
are prepared independently.
Similarly to section \ref{s8Z},
we put two hypotheses
\begin{align*}
H_0:&~ \cl{S}_{\le \epsilon}^3\defeq
\left\{ 
%\sigma_1\otimes \sigma_2\otimes \sigma_3
\bigotimes_{i=1}^3 \sigma_i
\left|
%\begin{array}{ll}
1- \sum_{i=1}^3\bra{\phi^{0}_{A_i,B_i}}\sigma_i\ket{\phi^{0}_{A_i,B_i}}
%+(1- \bra{\phi^{0}_{A,B}}\sigma_2\ket{\phi^{0}_{A,B}})\\
%+(1- \bra{\phi^{0}_{A,B}}\sigma_3\ket{\phi^{0}_{A,B}})
\le \epsilon 
%\end{array}
\right.\right\}\\
&\hbox{ versus}\\
H_1:&~ \cl{S}_{\le \epsilon}^{3c}
\defeq
\left\{ 
%\sigma_1\otimes \sigma_2\otimes \sigma_3
\bigotimes_{i=1}^3 \sigma_i
\left|
%\begin{array}{ll}
1- \sum_{i=1}^3\bra{\phi^{0}_{A_i,B_i}}\sigma_i\ket{\phi^{0}_{A_i,B_i}}
%+(1- \bra{\phi^{0}_{A,B}}\sigma_2\ket{\phi^{0}_{A,B}})\\
%+(1- \bra{\phi^{0}_{A,B}}\sigma_3\ket{\phi^{0}_{A,B}})
> \epsilon 
%\end{array}
\right.\right\},
\end{align*}
where 
the given state is assumed to be
$\sigma_1 \otimes\sigma_2 \otimes\sigma_3$.
Similarly we define the quantities
$\beta_{\alpha,3,G\times G\times G}^C
(\le \epsilon\|\sigma_1\otimes \sigma_2\otimes\sigma_3)$
for the condition $C=\emptyset,S(A,B),L(A\leftrightarrows B),
L(A\to B),L((A_1, A_2,A_3) \to (B_1, B_2 ,B_3)),
L(A_1, A_2,A_3,B_1,B_2,B_3)$, $S(A_1, A_2,A_3,B_1,B_2,B_3)$
under the similar $G\times G\times G$-invariance.

Similarly to subsection \ref{s8A}, 
we focus on the case of $C=L(A\to B),L(A\leftrightarrows B),S(A, B)$
with one sample.
In this case, 
as is mentioned,
the GHZ state 
$\ket{GHZ}\defeq
\frac{1}{\sqrt{d}}\sum_{i=0}^{d-1} \ket{i}_{A_1}\ket{i}_{A_2}\ket{i}_{A_3}
$ plays an important role.
Since the $SU(d)\times SU(d)\times SU(d)$-action on
$\cl{H}_{A_1}\otimes\cl{H}_{A_2}\otimes\cl{H}_{A_3}$
is irreducible,
the following is a POVM:
\begin{align*}
&M_{cov}^3(d g_1,d g_2,d g_3)\\
\defeq&
d^3
g_1\otimes g_2\otimes g_3
\ket{GHZ}\bra{GHZ}
(g_1\otimes g_2\otimes g_3)^{\dagger}\\
& \quad \nu(d g_1)
\nu(d g_2)
\nu(d g_3).
\end{align*}
As is proved in Appendix \ref{a27},
the test $T_{inv}^{3,A\to B}
\defeq T(M_{cov}^3)$
has the form
\begin{align}
&T_{inv}^{3,A \to B}\nonumber \\
= &P_1 \otimes P_2 \otimes P_3 
+\frac{(d+2)P_1^c \otimes P_2^c \otimes P_3^c}{(d+1)^3(d-1)}\nonumber \\
&\quad +
\frac{
P_1 \otimes P_2^c \otimes P_3^c 
+P_1^c \otimes P_2 \otimes P_3^c
+P_1^c \otimes P_2^c \otimes P_3 
}{(d+1)^2(d-1)},\Label{2-19-2}
\end{align}
where $P_i= \ket{\phi_{A_i,B_i}^0}\bra{\phi_{A_i,B_i}^0},
P_i^c= I-P_i$.
Thus, this test is 
$U(d^2-1)\times U(d^2-1)\times U(d^2-1)$-invariant.
Hence, when we use the test $T_{inv}^{3,A\to B}$,
the second error is
\begin{align*}
&\beta(T_{inv}^{3,A\to B}
,\sigma_1\otimes \sigma_2\otimes \sigma_3)\\
=&
(1-p_1)(1-p_2)(1-p_3)
+ \frac{(d+2)p_1 p_2 p_3}{(d+1)^2(d-1)}\\
& \quad 
+ \frac{p_1 p_2(1-p_3)+p_1 (1-p_2)p_3 + (1-p_1) p_2 p_3}{(d+1)^2(d-1)} .
\end{align*}
Moreover, the optimal second error can be also calculated as
\begin{align}
&\beta_{0,3,G\times G\times G}^{C}
(0\|\sigma_1\otimes \sigma_2\otimes \sigma_3)\nonumber\\
=&(1-p_1)(1-p_2)(1-p_3)
+ \frac{(d+2)p_1 p_2 p_3}{(d+1)^2(d-1)}\nonumber\\
& \quad 
+ \frac{p_1 p_2(1-p_3)+p_1 (1-p_2)p_3 + (1-p_1) p_2 p_3}{(d+1)^2(d-1)} 
\Label{2-19-3}
\end{align}
for $C=L(A\to B),L(A\leftrightarrows B),S(A, B)$
when $p_i \le \frac{d-1}{d}$.
Its proof is given in Appendix \ref{a27}.
Hence, the test $T_{inv}^{3,A\to B}$ is the $C$-UMP $G$-invariant test.

On the other hand, the case of $C=L(A_1, A_2,A_3 \to B_1, B_2 ,B_3)$,
$L(A_1, A_2,A_3,B_1,B_2,B_3)$,
$S(A_1, A_2,A_3,B_1,B_2,B_3)$.
Similarly to (\ref{2-19-1}), we can show
the optimality of the test $T_{inv}^{1,A\to B}\otimes T_{inv}^{1,A\to B}
\otimes T_{inv}^{1,A\to B}$.
Moreover,
we can derive the same result in the small deviation asymptotic setting 
with $n$ samples.

\section{Conclusion}
%We derived the optimal performance in several settings.
In this paper, we treated the hypotheses testing problem
when the null hypothesis consists only of 
the required entangled state or is its neighboor hood.
%本研究では帰無仮説が required entangled state もしくは，
%その近傍である場合での仮説検定を扱った．
In order to treat the structure of entanglement,
we consider three settings concerning the range of 
accessible measurements as follows:
{\bf M1}: All measurements are allowed.
{\bf M2}: A measurement is forbidden if it requires
the quantum correlation between two distinct parties.
{\bf M3}: A measurement is forbidden if it requires
the quantum correlation between two distinct parties,
or that among local samples.
%そして，エンタングルメントの構造に注意し，
%検定のために用いる測定の範囲について，3種の設定を考えた．
%M1 すなわち，全く制限を課さない場合，
%M2 two distinct parties の間の量子相関を許さない場合，
%M3 local samples 間の量子相関も許さない場合である．
As a result, we found that there is difference between
the accuracies of {\bf M1} and {\bf M2}
in the first order asymptotics.
%その結果，first order asymptotics の枠組では，
%M1 と M2 との間に差が生じないことが分かった．
The protocol achieving the asymtotic bound has been
proposed in the setting {\bf M2}.
%そして，M2 の枠組で限界を達成するプロトコルを提案した．
In this setting, it is required to prepare two identical samples
at the same time.
%この設定では，2つの系に対する相関を用いるために，
%同時に同一のサンプルを２つ準備する必要がある．
However, it is difficult to prepare the two states from the same source.
In order to avoid this difficulty,
we proved that even if 
the two states is prepared from the different source,
this proposed protocol works effectively.
%さらに，同時に準備された2つのサンプルが同一ではない
%場合であっても，ここで提案したプロトコルが有効に働くことを示した．
In particular, this protocol can be realized in the two-dimensional system
if the four-valued Bell measurement can be realized.
%なお，two-dimensional system では
%four-valued Bell measurement が構成できれば
%このプロトコルは実現できる．
Moreover, concerning the finite samples case,
we derived optimal testing in several examples.
%さらに，サンプルが有限の場合には，
%幾つかの具体例について，最適な検定を構成した．
Thus, as was demonstrated by Hayashi {\it et al.} \cite{exper},
it is a future target to demonstrate the proposed testing experimentally.

In this paper, the optimal test is constructed 
based on continuous valued POVM.
However, any realizable POVM is 
finite valued.
%本論文では最適なプロトコルを
%連続値の POVM を用いて構成した．
%しかし，実際に実現可能な測定は
%離散値の POVM である．
Hence, it is desired to construct the optimal test 
based on the finite valued POVM.
This problem is partially discussed by Hayashi {\it et~al.},
and will be more deeply discussed by another paper \cite{ha1}.
%最適な測定を離散値の POVM で構成する問題は
%forcoming paper \cite{} で論じられる予定である．

The obtained protocol is 
essentially equivalent with the following 
procedure based on the quantum teleportation.
First, we perform quantum teleportation 
from the system $A$ to the system $B$, which succeed when the true state is 
the required maximally entangled state.
%ここで与えた最適な検定プロトコルは
%基本的に真の状態が
%required maximally entangled state である場合には，
%量子テレポーテーションを行ない，
Next, we check whether the state on the system $B$
is the initial state on the system $A$.
%終状態が所望の状態であるか否か検定する
%実験とみなすことができる．
Hence, an interesting relation between 
the obtained results and the quantum teleportation
is expected, and it will be treated in 
a forthcoming paper \cite{ha2}.
%したがって，本研究成果と
%量子テレポーテーションとの間には interesting 関係が
%あると期待される．これについても，
%another forcoming paper \cite{} で論じられる予定である．

As a related research, 
the following testing problem has been discussed \cite{LC,BS}.
Assume that $N$ qubits state are given,
and we can measure only $M$ qubits.
The required problem is testing whether 
the remaining $N-M$ qubits are the desired maximally entangled
state.
%なお，関連研究として
%$N$ qubit の maximally entangled state と思われる量子状態が
%得られた時，
%そのうち，$M$ qubit だけ測定し，
%それから残りの $N-M$ qubit の状態を検定するという問題が
%扱われている．
Indeed, this problem is important not only for
gurarantee of the quality of the prepared 
maximally entangled state,
but also for the security for the quantum key distribution.
%この問題は大変重要で，量子暗号の安全性から，
%実際に生成された量子状態の品質を保証するために重要である．
The problem discussed in this paper is 
different from the preceding probelem in testing
the given state by measuring the whole system.
%本論文は，$n$ qubit 与えられた設定の元で，
%それがmaximally entangled state であるか否か
%判定する問題を扱っており，若干異なる．
In order to apply our result to the preceding problem,
we have to randomly choose $M$ qubits among the given 
$N$ qubits, and test the $N$ qubits.
%この問題に本論文の成果を適用するには
%$N$ qubits の中から，
%ランダムに $M$ qubit 取り出し，
%それに対して，本論文の成果を適用することになる．
When the given $N$ qubits do not satisfy the
independent and identical condition,
their method \cite{LC,BS} is better than our method.
%最初の $N$ qubit の量子状態が独立同一性を満たさない
%例では，このように本論文を手法を用いる方法と，
%論文\cite{}での提案手法を比較すると，後者の方が優れている．
Since their method \cite{LC,BS} requires the 
the quantum correlation among whole $M$ qubits,
it is difficult to realize their method for testing 
the prepared maximally entangled state,
but it is possible to apply their method to testing 
the security of quantum key distribution \cite{LC}.
This is because the maximally entangled state is only virtually 
discussed in the latter case.
%しかし，これらの論文で述べられている測定では
%$M$ qubit だけ測定するにも関わらず，
%元の $N$ qubit の系に
%跨る測定を用いており，
%状況によってはその実現が困難である．
%論文では
%量子暗号への応用を考えており，
%そのような系では
%仮想的に maximally entangled state やその検定を考えているに過ぎず，
%実際に扱っている問題はそれとは別の問題である．
%そのため，彼らが考えている検定プロトコルに対応する
%操作を行なうことは実現容易である．
Hence, for testing the prepared maximally entangled state,
it is natural from the practical viewpoint 
to restrict our test among random sampling method.
%だが，実際にmaximally entangled state そのものを準備する設定では，
%元の $N$ qubit の系に跨る測定を行なうことは
%極めて困難であり，実用的には，
%$N$ qubit の中から，
%ランダムに $M$ qubit 取り出す手法に限ることが有効である．
Since our results can be applied this setting,
they can be expected to be applied to the check 
of the quality of maximally entangled state.
%このような制限の下では本論文の成果はその最適化を行なっていることになり，
%極めて有効である．
%一方，これらの成果は maximally entangled state の品質管理に
%用いられると期待できる．

As another problem, Ac\'{i}n {\it et~al.} \cite{acin-tarrach-vidal}
discussed the problem testing whether the given $n$-i.i.d. state
of the unknown pure state
is the $n$-tensor product of a pure maximally entangled state
(not the specific maximally entangled state) in the two-dimensional
system.
This problem is closely related to 
universal entanglement concentration\cite{MH}.
Its $d$-dimensional case 
is a future problem.

%他の問題設定として，与えられた状態が
%特定の (required) maximally entangled stateではなく，
%単にmaximally entangled state であるか否か判定する問題がある．
%この問題は始めに，Two-dimensional case が Ac\'{i}n {\it et~al.} によって
%議論された．本質的にはこの問題は，一般の次元へ拡張可能であり，
%universal enatnglement concentration と同じ問題である．

\section*{Acknowledgment}
This research
was partially supported by a Grant-in-Aid for Scientific Research on Priority Area `Deepening and Expansion of Statistical Mechanical Informatics (DEX-SMI)', No. 18079014 and
a MEXT Grant-in-Aid for Young Scientists (A) No. 20686026.

\appendix
\section{Proof of Lemma \ref{l12} and Lemma \ref{l13}}\Label{a21}
Assume that
a set of test satisfying the condition $C$
is invariant for the action of $G_2$.
Let $T \in \cl{T}_{\alpha,\le \epsilon}^{n}$ be a test 
satisfying the condition $C$,
then the test 
$\overline{T}\defeq
(f(g)^{\dagger})^{\otimes n}T f(g)^{\otimes n}$
also satisfies the condition $C$ and belongs to
the set $\cl{T}_{\alpha,\le \epsilon}^{n}$.
Since
\begin{align*}
\beta(T,(f(g)\sigma f(g)^{\dagger})^{\otimes n}) 
=
\beta(f(g)^{\dagger}T f(g),\sigma ^{\otimes n}) ,
\end{align*}
we obtain 
\begin{align*}
&\max_{g \in G}
\beta(T,(f(g)\sigma f(g)^{\dagger})^{\otimes n})\\
\ge &
\int_{G_2}
\beta(T,(f(g)\sigma f(g)^{\dagger})^{\otimes n})
\nu_G(d g) \\
= &
\beta(\overline{T},(f(g)\sigma f(g)^{\dagger})^{\otimes n})
\ge
\beta_{\alpha,n,G_2}^C(\le \epsilon\|\sigma).
\end{align*}
Hence,
\begin{align}
&\min_{T \in \cl{T}_{\alpha,\le \epsilon}^{n}}
\max_{g \in G}
\beta(T,(f(g)\sigma f(g)^{\dagger})^{\otimes n})\nonumber  \\
&\ge
\min_{T \in \cl{T}_{\alpha,\le \epsilon}^{n}}
\int_G
\beta(T,(f(g)\sigma f(g)^{\dagger})^{\otimes n}) 
\nu_G(d g) \nonumber \\
&\ge
\beta_{\alpha,n,G}^C(\le \epsilon\|\sigma).\Label{2-16-1}
\end{align}
On the other hand, 
if the $G_2$-invariant test $T\in \cl{T}_{\alpha,\le \epsilon}^{n}$ 
satisfies the condition $C$ and 
\begin{align*}
\beta(T,\sigma ^{\otimes n}) 
=
\beta_{\alpha,n,G_2}^C(\le \epsilon\|\sigma),
\end{align*}
then 
\begin{align*}
&
\beta(T,(f(g)\sigma f(g)^{\dagger})^{\otimes n})\\
= &
\beta(T,\sigma ^{\otimes n}) 
= \beta_{\alpha,n,G_2}^C(\le \epsilon\|\sigma) \quad \forall g \in G_2,
\end{align*}
which implies
\begin{align*}
\max_{g \in G}
\beta(T,(f(g)\sigma f(g)^{\dagger})^{\otimes n})
= \beta_{\alpha,n,G_2}^C(\le \epsilon\|\sigma).
\end{align*}
Thus, we obtain the inequality opposite to (\ref{2-16-1}).
Therefore, the proof of Lemma \ref{l12} is completed.

Next, we proceed to prove Lemma \ref{l13}.
Since
the equation
\begin{align*}
&\int_{G_1}
\beta(T,(f(g)\sigma f(g)^{\dagger})^{\otimes n}) 
\nu_{G_1}(d g)\\
=&
\int_{G_1}
\beta((f(g')^\dagger)^{\otimes n}T f(g')^{\otimes n}
,(f(g)\sigma f(g)^{\dagger})^{\otimes n}) 
\nu_{G_1}(d g)
\end{align*}
holds for $\forall g'\in G_2$,
we obtain
\begin{align}
&\min_{T \in \cl{T}_{\alpha,\le \epsilon}^{n}}
\max_{g \in G_1}
\beta(T,(f(g)\sigma f(g)^{\dagger})^{\otimes n})\nonumber  \\
&\ge
\min_{T \in \cl{T}_{\alpha,\le \epsilon}^{n}}
\int_{G_1}
\beta(T,(f(g)\sigma f(g)^{\dagger})^{\otimes n}) 
\nu_{G_1}(d g)\nonumber \\
&=
\min_{T \in \cl{T}_{\alpha,\le \epsilon}^{n}}
\int_{G_2}
\int_{G_1}
\beta(
(f(g')^\dagger)^{\otimes n}T f(g')^{\otimes n}
,(f(g)\sigma f(g)^{\dagger})^{\otimes n}) \nonumber \\
& \hspace{5ex}\nu_{G_1}(d g)
\nu_{G_2}(d g')\nonumber \\
& =
\min_{T \in \cl{T}_{\alpha,\le \epsilon}^{n}}
\int_{G_1}
\beta(
(\overline{T}
,(f(g)\sigma f(g)^{\dagger})^{\otimes n}) 
\nu_{G_1}(d g)\nonumber \\
&\ge 
\int_{G_1}
\beta_{\alpha,n,G_2}^C(\le \epsilon\|f(g)\sigma f(g)^{\dagger})
\nu_{G_1}(d g)\nonumber \\
&=
\int_{G_1}
\beta_{\alpha,n,G_1}^C(\le \epsilon\|f(g)\sigma f(g)^{\dagger})
\nu_{G_1}(d g).\Label{2-16-5}
\end{align}
Since 
$\beta(T',\sigma)= \beta(T',f(g)\sigma f(g)^{\dagger})$
for any $G_1$-invariant test $T'$,
we have
\begin{align*}
\beta_{\alpha,n,G_1}^C(\le \epsilon\|\sigma)
=\beta_{\alpha,n,G_1}^C(\le \epsilon\|f(g)\sigma f(g)^{\dagger}),
\end{align*}
which implies
\begin{align*}
\int_{G_1}
\beta_{\alpha,n,G_1}^C(\le \epsilon\|f(g)\sigma f(g)^{\dagger})
\nu_{G_1}(d g)=
\beta_{\alpha,n,G_1}^C(\le \epsilon\|\sigma).
\end{align*}
We choose a $G_1$-invariant test $T \in \cl{T}_{\alpha,\le \epsilon}^{n}$
satisfying the condition $C$ and
\begin{align*}
\beta_{\alpha,n,G_1}^C(\le \epsilon\|\sigma)
=\beta(T,\sigma).
\end{align*}
Then,
\begin{align*}
\max_{g \in G_1}
\beta(T,(f(g)\sigma f(g)^{\dagger})^{\otimes n}) 
=
\beta(T,\sigma )
=\beta_{\alpha,n,G_1}^C(\le \epsilon\|\sigma).
\end{align*}
Thus, we obtain the inequality opposite to (\ref{2-16-5}),
which yields Lemma \ref{l13}.

\section{Basic properties of classical tests}
In the classical hypotheses testing,
Neymann Pearson Lemma 
plays a central role.
\begin{lem}\Label{l-20}
Assume that 
the null hypothesis is one probability distribution 
$P_0$ and the alternative one is another probability distribution 
$P_1$.
For any $\,> \alpha \,>0$,
we choose 
$r$ and $\gamma$ such that
\begin{align*}
P_0 \left\{ x\left| \frac{P_0(x)}{P_1(x)}\,> r\right.\right\} &\le 1- \alpha \\
P_0 \left\{ x\left| \frac{P_0(x)}{P_1(x)}\,< r\right.\right\} &\le \alpha \\
\gamma P_0 \left\{ x \left| \frac{P_0(x)}{P_1(x)}= r \right.\right\}
&= 1- \alpha - P_0 \left\{ x\left| \frac{P_0(x)}{P_1(x)}\,> r  \right.\right\},
\end{align*}
and define the test $\tilde{T}_{P_0,P_1,\alpha}$ as
\begin{align}
\tilde{T}_{P_0,P_1,\alpha}(x)=
\left\{
\begin{array}{ll}
1 & \hbox{ if } \frac{P_0(x)}{P_1(x)}\,> r \\
\gamma & \hbox{ if } \frac{P_0(x)}{P_1(x)}= r \\
0 & \hbox{ if } \frac{P_0(x)}{P_1(x)}\,< r .
\end{array}
\right. \Label{24-1}
\end{align}
Then, the test $T_{P_0,P_1,\alpha}$
is the MP level-$\alpha$ test.
\end{lem}
In classical statistics, the function $\frac{P_0(x)}{P_1(x)}$
is called the {\it likelihood ratio},
which plays an important role.

\begin{proof}
Assume that $\tilde{T}^*$ is a level-$\alpha$ test.
We focus on the weighted sum of two kinds of error probabilities
$\sum_x P_0(x) (1-\tilde{T}^*(x))
+ r \sum_x P_1(x) \tilde{T}^*(x)$.
Since 
$\sum_x (P_0 (x) - r P_1(x))\tilde{T}^*(x)
\le \sum_x (P_0 (x) - r P_1(x))\tilde{T}_r(x)$, we can see
\begin{align*}
& \sum_x P_0(x) (1-\tilde{T}^*(x))
+ r \sum_x P_1(x) \tilde{T}^*(x)\\
= & 1 - \sum_x (P_0 (x) - r P_1(x))\tilde{T}^*(x) \\
\ge & 1 - \sum_x (P_0 (x) - r P_1(x))\tilde{T}_{P_0,P_1,\alpha}(x)\\
= &\sum_x P_0(x) (1-\tilde{T}_{P_0,P_1,\alpha}(x))
+ r \sum_x P_1(x) \tilde{T}_{P_0,P_1,\alpha}(x).
\end{align*}
Hence, the relation
$\sum_x P_0(x) (1-\tilde{T}^*(x))=\sum_x P_0(x) 
(1-\tilde{T}_{P_0,P_1,\alpha}(x))=\alpha$
yields that
\begin{align*}
\sum_x P_1(x) \tilde{T}^*(x)
\ge \sum_x P_1(x) \tilde{T}_{P_0,P_1,\alpha}(x).
\end{align*}
\end{proof}
We have the following corollary.
\begin{cor}\Label{c-1}
If the test $\tilde{T}$ has the form (\ref{24-1}),
{\it i.e.}, a likelihood ratio test,
then the inequality
\begin{align*}
\sum_x P_0(x) (1-\tilde{T}(x)) 
\le \sum_x P_1(x) (1-\tilde{T}(x))
\end{align*}
holds.
\end{cor}

\begin{proof}
We focus on the test $\tilde{T}'(x)\defeq 1- \alpha$.
Since the test $\tilde{T}'$ is trivially level-$\alpha$,
Lemma \ref{l-20} guarantees that
$\sum_x P_1(x) \tilde{T}(x)
\le \sum_x P_1(x) \tilde{T}'(x)= 1-\alpha$,
which implies that
$\sum_x P_0(x) (1-\tilde{T}(x)) = \alpha
\le \sum_x P_1(x) (1-\tilde{T}(x))$.
\end{proof}

\section{Proof of Theorem \ref{t-1}}\Label{a-2}
Since the likelihood ratio
$\frac{P_{\epsilon}^n(k)}{P_{q}^n(k)}$
is the monotone decreasing function of $k$,
the test $\tilde{T}_{\epsilon, \alpha}$ 
equals the test $\tilde{T}_{P_{\epsilon}^n,P_{q}^n,\alpha}$.
Lemma \ref{l-20} guarantees that
the test $\tilde{T}_{\epsilon, \alpha}$ is the MP level-$\alpha$
test with the null hypothesis $P_{\epsilon}^n$.
Since a level-$\alpha$ test with the null hypothesis 
$\cl{P}^n_{\le \epsilon}$
is a level-$\alpha$ test with the null hypothesis $P_{\epsilon}^n$,
\begin{align}
\beta_\alpha^n (\le \epsilon\|q)
\ge P_q^n (\tilde{T}_{\epsilon, \alpha}).\Label{24-3}
\end{align}

Since the likelihood ratio
$\frac{P_{p_0}^n(k)}{P_{p_1}^n(k)}$
is the monotone decreasing function of $k$ for $p_0 \,< p_1$,
the test $\tilde{T}_{\epsilon, \alpha}$
is a likelihood ratio test of $P_{p_0}^n$ and $P_{p_1}^n$.
Hence, Corollary \ref{c-1} guarantees that
$P_{p_0}^n (\tilde{T}_{\epsilon, \alpha})
\ge P_{p_1}^n (\tilde{T}_{\epsilon, \alpha})$.
That is,
the probability
$P_{p}^n (\tilde{T}_{\epsilon, \alpha})$ is a monotone decreasing function of
$p$.
Since the definition of the test $\tilde{T}_{\epsilon, \alpha}$
implies that $P_{\epsilon}^n (\tilde{T}_{\epsilon, \alpha}) = 1- \alpha$,
$P_{p}^n (\tilde{T}_{\epsilon, \alpha}) \le
1- \alpha$ if $p \le \epsilon$.
In other words, the test $\tilde{T}_{\epsilon, \alpha}$ is level-$\alpha$
with the null hypothesis $\cl{P}^n_{\le \epsilon}$.
Hence, it follows from the inequality (\ref{24-3}) that
the test $\tilde{T}_{\epsilon, \alpha}$ is 
level-$\alpha$ UMP test with the null hypothesis $\cl{P}^n_{\le \epsilon}$.

\section{Proof of Theorem \ref{t-2}}\Label{a5}
Since $\tilde{T}_{\delta/n,\alpha}^n$
is a level-$\alpha$ test 
the null hypothesis $\cl{P}_{\frac{\delta}{n}}$,
$\forall t \in [0,\delta] \quad 
\limsup 1- 
\sum_k P_t(k) \tilde{T}_{\delta/n,\alpha}^n(k)
= 
\limsup 1- \sum_k P_{t/n}^n(k) \tilde{T}_{\delta/n,\alpha}^n(k)
\le \alpha$.
Hence, for $\forall \epsilon \,>0$
there exists $N$ such that
$\forall n \ge N, \forall t \in [0,\delta] \quad
1- P_t(\tilde{T}_{\delta/n,\alpha}^n) \le \alpha + \epsilon$.
Hence,
\begin{align*}
P_{t'}(\tilde{T}_{\delta/n,\alpha}^n) 
\ge
 \beta_{\alpha+ \epsilon}(\le \delta\|t')
\end{align*}
Since
$\liminf P_{t'}(\tilde{T}_{\delta/n,\alpha}^n) 
=\liminf P_{t'/n}^n(\tilde{T}_{\delta/n,\alpha}^n)
=\liminf \beta^n_{\alpha} \left(\le\frac{\delta}{n}\left\|\frac{t'}{n}
\right.\right)$,
\begin{align*}
\liminf \beta^n_{\alpha} \left(\le\frac{\delta}{n}\left\|\frac{t'}{n}
\right.\right)\ge
 \beta_{\alpha+ \epsilon}(\le \delta\|t').
\end{align*}
Since the continuity of $\alpha \mapsto 
\beta_{\alpha}(\le \delta\|t')$ follows from
Theorem \ref{t-3},
\begin{align*}
\liminf \beta^n_{\alpha} \left(\le\frac{\delta}{n}\left\|\frac{t'}{n}
\right.\right)\ge
 \beta_{\alpha}(\le \delta\|t').
\end{align*}

Since $\tilde{T}_{\delta,\alpha- \epsilon}$
is level-$\alpha$ test 
the null hypothesis $t \in [0,\delta]$ for $\forall \epsilon \,>0$,
we have $\forall t \in [0,\delta] \quad 
\limsup 1- \sum_k P_{t/n}^n(k) \tilde{T}_{\delta,\alpha- \epsilon}(k)
=
\limsup 1- \sum_k P_t(k) \tilde{T}_{\delta,\alpha- \epsilon}(k)
\le \alpha- \epsilon$.
Hence, 
there exists $N$ such that
$\forall n \ge N, \forall t \in [0,\delta] \quad
1- P_{t/n}^n(\tilde{T}_{\delta,\alpha- \epsilon}) \le \alpha$.
Hence,
\begin{align*}
P_{t'/n}^n(\tilde{T}_{\delta,\alpha- \epsilon}) 
\ge
 \beta_{\alpha}^n(\le \delta/n\|t'/n)
\end{align*}
Thus,
\begin{align*}
\beta_{\alpha}^n(\le \delta/n\|t'/n)
\le
 \beta_{\alpha- \epsilon}(\le \delta\|t'),
\end{align*}
which implies 
\begin{align*}
\limsup \beta_{\alpha}^n(\le \delta/n\|t'/n)
\le
 \beta_{\alpha- \epsilon}(\le \delta\|t').
\end{align*}
The continuity of $\alpha \mapsto 
\beta_{\alpha}(\le \delta\|t')$ guarantees that
\begin{align*}
\limsup \beta_{\alpha}^n(\le \delta/n\|t'/n)
\le
 \beta_{\alpha}(\le \delta\|t').
\end{align*}

\section{Proof of (\ref{1-19-1})}\Label{a-4}
For a fixed density matrix $\sigma$ on $\cl{H}_{A,B}$,
we define a density matrix $\sigma_q$ as
\begin{align*}
\sigma_q & \defeq 
\left(
\sqrt{\frac{1-q}{1- p}}
\ket{\phi^0_{A,B}}\bra{\phi^0_{A,B}}
+ \sqrt{\frac{q}{p}}
(I-\ket{\phi^0_{A,B}}\bra{\phi^0_{A,B}})
\right)\cdot\\
&\quad \sigma 
\left(
\sqrt{\frac{1-q}{1- p}}
\ket{\phi^0_{A,B}}\bra{\phi^0_{A,B}}
+ \sqrt{\frac{q}{p}}
(I-\ket{\phi^0_{A,B}}\bra{\phi^0_{A,B}})
\right),
\end{align*}
where $p \defeq 1- \bra{\phi^0_{A,B}}\sigma \ket{\phi^0_{A,B}}$.
We also define the matrix $\sigma'$ by
\begin{align*}
\frac{1}{p}
(I-\ket{\phi^0_{A,B}}\bra{\phi^0_{A,B}})
\sigma (I-\ket{\phi^0_{A,B}}\bra{\phi^0_{A,B}}).
\end{align*}

Let $T$ be a $U(1)$-invariant test with level-$\alpha$.
The $U(1)$-invariance yields that
\begin{align*}
\Tr T (U_\theta^{\otimes n})^\dagger \rho_q^{\otimes n} U_\theta ^{\otimes n}
=\Tr U_\theta^{\otimes n} T (U_\theta^{\otimes n})^\dagger 
\rho_q ^{\otimes n}
= \Tr T \rho_q^{\otimes n} .
\end{align*}
Hence,
\begin{align}
\Tr T \rho_q^{\otimes n} = \frac{1}{2\pi}\int_0^{2\pi} \Tr T 
(U_\theta^{\otimes n})^\dagger \rho_q^{\otimes n} 
U_\theta^{\otimes n} \,d \theta
= \Tr T \overline{\rho}_{q}^n, \Label{24-10}
\end{align}
where we define $\overline{\rho}_{\theta}$ as
\begin{align*}
\overline{\rho}_{q}^n &\defeq
\frac{1}{2\pi}\int_0^{2\pi} 
(U_\theta^{\otimes n})^\dagger \rho_q^{\otimes n} 
U_\theta^{\otimes n} \,d\theta \\
& =
\sum_{k=0}^n 
q^k (1-q)^{n-k}
P_k^n(\ket{\phi^0_{A,B}}\bra{\phi^0_{A,B}},\sigma').
\end{align*}
Thus, 
the test $T$ is a level-$\alpha$ test with 
the null hypothesis $\overline{\rho}_{\epsilon}^n$.

In the following, we focus on the hypotheses testing
with the null hypothesis $\overline{\rho}_{\epsilon}^n$
and the alternative hypothesis $\overline{\rho}_{p}^n$.
Since these two states are commutative with each other,
there exists a basis
$\{e_{k,l}\}$ diagonalizing them.
As they are written as
$\overline{\rho}_{\epsilon}= 
\sum_k \sum_{l} P^{k,l}_0 \ket{e_{k,l}}\bra{e_{k,l}}$
and $\overline{\rho}_{p}= 
\sum_k \sum_{l} P^{k,l}_1 \ket{e_{k,l}}\bra{e_{k,l}}$,
our problem is essentially 
equivalent with the classical hypothesis testing with 
the null hypothesis $P_0\defeq (P_0^{k,l})$ 
and the alternative hypothesis $P_1\defeq (P_1^{k,l})$.
Since the likelihood ratio 
is given by the ratio 
$\frac{\epsilon^k (1-\epsilon)^{n-k}}{p^k (1-p)^{n-k}}$,
we have 
\begin{align*}
T^n_{\epsilon,\alpha}
= \sum_{k} \sum_{l} 
\tilde{T}_{P_0,P_1,\alpha} \ket{e_{k,l}}\bra{e_{k,l}}.
\end{align*}
Hence, Lemma \ref{l-20} guarantees that
\begin{align*}
\Tr T \overline{\rho}_{p}^n
\ge T^n_{\epsilon,\alpha} \overline{\rho}_{p}^n
\end{align*}
because the test $T$ is a level-$\alpha$ test with 
the null hypothesis $\overline{\rho}_{\epsilon}^n$.
Since $T^n_{\epsilon,\alpha}$ is $U(1)$-invariant,
the equation (\ref{24-10}) guarantees that
\begin{align*}
\Tr T \sigma^{\otimes n}
\ge T^n_{\epsilon,\alpha} \sigma^{\otimes n}.
\end{align*}
The equation $T^n_{\epsilon,\alpha} \sigma^{\otimes n}
= \beta_{\alpha}^n(\le \epsilon\|p)$ yields
(\ref{1-19-1}).

\section{Proof of Lemma \ref{l-10}}\Label{a-13}
Let $T$ be a one-way LOCC $(A \to B)$ level-$0$ test 
with the null hypothesis $\ket{\phi^0_{A,B}}\bra{\phi^0_{A,B}}$.
We denote Alice's first measurement by $M=\{M_i\}$.
In this case, Bob's measurement can be described by two-valued measurement
$\{T_0^i, I- T_0^i\}$,
where $T_0^i$ corresponds to the decision accepting the state 
$\ket{\phi^0_{A,B}}\bra{\phi^0_{A,B}}$.
Hence the test $T$ can be described as
\begin{align*}
T= \sum_i M_i \otimes T_0^i.
\end{align*}
When Alice observes the data $i$,
the Bob's state is $\frac{1}{\Tr \overline{M_i}}\overline{M_i}$.
Since this test is level-$0$,
\begin{align*}
T_0^i \ge P_i,
\end{align*}
where $P_i$ is the projection to the range of 
the matrix $\overline{M_i}$.

Here, we diagonalize $M_i$ as
$M_i = \sum_j p_{i,j}\ket{u_{i,j}}\bra{u_{i,j}}$.
Since $P_i \ge \ket{\overline{u_{i,j}}}\bra{\overline{u_{i,j}}}$,
the POVM $M'= 
\{p_{i,j}\ket{u_{i,j}}\bra{u_{i,j}}\}_{i,j}$,
satisfies
\begin{align*}
T(M') 
&=\sum_{i,j} p_{i,j}
\ket{u_{i,j}\otimes \overline{u_{i,j}}}
\bra{u_{i,j}\otimes \overline{u_{i,j}}}\\
& =
\sum_{i,j} p_{i,j}
\ket{u_{i,j}}\bra{u_{i,j}}
\otimes \ket{\overline{u_{i,j}}}\bra{\overline{u_{i,j}}}\\
& \le 
\sum_{i,j} p_{i,j}
\ket{u_{i,j}}\bra{u_{i,j}}
\otimes P_i
=\sum_i M_i \otimes P_i \\
& \le \sum_i M_i \otimes T_0^i = T.
\end{align*}

\section{Proof of Lemma \ref{l-9}}\Label{a22}
It follows from the condition (\ref{2-16-7})
that $\sum_i p_i=1$.
We choose the vector $\varphi\defeq
\sqrt{d} \sum_i p_i u_i \otimes {u'}_i$.
Since the function $x \to |x|^2$ is
convex, we obtain
\begin{align*}
&\bra{\varphi}T\ket{\varphi}
\ge 
d \sum_i p_i \bra{\varphi}u_i \otimes {u'}_i \rangle
\langle u_i \otimes {u'}_i \ket{\varphi}\\
=&
d \sum_i p_i |\bra{\varphi}u_i \otimes {u'}_i \rangle|^2
\ge
d 
|\sum_i p_i 
\bra{\varphi}u_i \otimes {u'}_i \rangle|^2\\
=&
d\left| \frac{1}{\sqrt{d}}\bra{\varphi}\varphi\rangle\right|^2
=\| \varphi\|^4.
\end{align*}
Hence,
\begin{align*}
1 \ge \Tr T \frac{\ket{\varphi}\bra{\varphi}}{\| \varphi\|^2}
= \| \varphi\|^2.
\end{align*}
On the other hand,
\begin{align*}
\langle \phi^0_{A,B}\ket{\varphi}
=\sqrt{d}  \sum_i p_i \langle \phi^0_{A,B}\ket{u_i \otimes {u'}_i}
= 1.
\end{align*}
Since $\|\phi^0_{A,B}\|=1$,
we obtain 
\begin{align*}
\varphi=\phi^0_{A,B}.
\end{align*}

\section{Proof of (\ref{22-17})}\Label{a-7}
The representation space $\cl{H}_A \otimes \cl{H}_B$
of $SU(d)$-action can be irreducibly decomposed to
two subspaces:
One is the one-dimensional space $<\phi^0_{A,B}>$
spanned by $\phi^0_{A,B}$.
The other is its orthogonal space
$<\phi^0_{A,B}>^\perp$.
Since the $T_{inv}^{1,A\to B}$ is $SU(d)$-invariant,
it has the form
\begin{align*}
T_{inv}^{1,A\to B}= t_0 \ket{\phi^0_{A,B}}\bra{\phi^0_{A,B}}+
t_1(I-\ket{\phi^0_{A,B}}\bra{\phi^0_{A,B}}).
\end{align*}
The equation $\bra{\phi^0_{A,B}} T_{inv}^{1,A\to B}\ket{\phi^0_{A,B}}=1$
implies 
$t_0 =1$.
Its trace can be calculated as
\begin{align*}
\Tr T_{inv}^{1,A\to B}= 
\int d
\Tr \ket{\varphi \otimes \overline{\varphi}}
\bra{\varphi\otimes \overline{\varphi}}\nu (\,d \varphi)
=d .
\end{align*}
Hence, $t_1 = \frac{d-1}{d^2-1}= \frac{1}{d+1}$.

\section{Proof of (\ref{22-19})}\Label{a-8}
\begin{lem}\Label{l-3}
If the test $T$ is a separable on the space $\cl{H}_A
\otimes \cl{H}_B$,
then
\begin{align}
\Tr T \ge d \bra{\phi^0_{A,B}} T \ket{\phi^0_{A,B}} \Label{24-20},
\end{align}
where $d$ is the dimension of $\cl{H}_A$.
\end{lem}
\begin{proof}
Since $T$ is separable, 
$T$ has the form 
$T=\sum_{l}\ket{u_i\otimes u'_i}\bra{u_i\otimes u'_i} $.
For any two vectors $u,u'$,
Schwarz inequality yields that
\begin{align*}
\langle u\otimes u'| u \otimes u'\rangle
&= \langle u | u \rangle
\langle u'| u'\rangle
=\langle u | u \rangle
\langle \overline{u'}| \overline{u'}\rangle
\ge |\langle u| \overline{u'}\rangle|^2 \\
& = d |\langle \phi_{A,B}^0| u \otimes u'\rangle|^2.
\end{align*}
Hence, we have
\begin{align*}
\Tr \ket{u_i\otimes u'_i}\bra{u_i\otimes u'_i} 
\ge d \langle\phi^0_{A,B} \ket{u_i\otimes u'_i}\bra{u_i\otimes u'_i}
\phi^0_{A,B}\rangle.
\end{align*}
Taking the sum, we obtain (\ref{24-20}).
\end{proof}

Assume that $T$ is a $SU(d)$-invariant separable test.
From the discussion in section \ref{a-7},
the test $T$ has the form
\begin{align*}
T= t_0 \ket{\phi^0_{A,B}}\bra{\phi^0_{A,B}}+
t_1(I-\ket{\phi^0_{A,B}}\bra{\phi^0_{A,B}}).
\end{align*}
Lemma \ref{l-3} implies that
$t_1 (d^2-1)+ t_0 \ge d t_0$, {\it i.e.},
$t_1 \ge \frac{1}{d+1} t_0$.
Hence, the test $T$ has another form
\begin{align*}
T=&  t_0 '
\left(\ket{\phi^0_{A,B}}\bra{\phi^0_{A,B}}+
\frac{1}{d+1}(I-\ket{\phi^0_{A,B}}\bra{\phi^0_{A,B}})
\right)\\
& +t_1'\frac{d}{d+1}(I-\ket{\phi^0_{A,B}}\bra{\phi^0_{A,B}})\\
=& t_0 'T_{inv}^{1,A\to B} +t_1 '(I-T_{inv}^{1,A\to B}).
\end{align*}
Since
\begin{align*}
\Tr \sigma T_{inv}^{1,A\to B}
= 1-\frac{dp }{d+1},\quad
\Tr \sigma
(I-T_{inv}^{1,A\to B})
 = \frac{dp }{d+1},
\end{align*}
our problem is equivalent with 
the hypotheses testing 
concerning the probability
$(1-\frac{dp }{d+1},\frac{dp }{d+1})$.
Thus, we obtain (\ref{22-19}).

\section{Proof of Lemma \ref{l-1} and (\ref{22-20})}\Label{a-9}
\begin{lem}\Label{l-5}
A state $u \in \cl{H}_{A_1}\otimes \cl{H}_{A_2}$
is maximally entangled 
if and only if 
\begin{align}
\ket{\phi^0_{A_1,B_1}}\bra{\phi^0_{A_1,B_1}}
\otimes (I- \ket{\phi^0_{A_2,B_2}}\bra{\phi^0_{A_2,B_2}})
u \otimes \overline{u} &= 0 \Label{24-20-1}\\
(I- \ket{\phi^0_{A_1,B_1}}\bra{\phi^0_{A_1,B_1}})
\otimes \ket{\phi^0_{A_2,B_2}}\bra{\phi^0_{A_2,B_2}}
u \otimes \overline{u} &= 0 .\Label{24-21}
\end{align}
\end{lem}
\begin{proof}
The condition (\ref{24-20-1})
equivalent to the condition that 
$\langle \phi^0_{A_1,B_1}|
u \otimes \overline{u}\rangle$
equals the constant times of
$\ket{\phi^0_{A_2,B_2}}$.
When we choose a matrix $U$ as
$u=\sum_{i,j} \frac{U_{i,j}}{\sqrt{d}} \ket{i}_{A_1}\ket{j}_{A_2}$,
this condition equals to the condition
that 
$U I U^\dagger$ is a constant matrix.
Thus, if and only if
$u$ is maximally entangled, $U$ is unitary,
which is equivalent with the condition (\ref{24-20-1}).
Similarly, we can show that
the maximally entangledness of $u$ equivalent with the condition
(\ref{24-21}).
Hence, the desired argument is proved.
\end{proof}

The relations (\ref{24-30}) and (\ref{24-31}) 
guarantee that
$\phi^0_{A_1,B_1}\otimes \phi^0_{A_2,B_2}$ is 
an eigenvector of $T(M)$ with the eigenvalue $1$.
Hence, Lemma \ref{l-5} implies (\ref{24-38}).
On the other hand,
\begin{align*}
& \Tr P \sigma^{\otimes 2} P =
\left(
\Tr \sigma (I- \ket{\phi^0_{A_1,B_1}}\bra{\phi^0_{A_1,B_1}})
\right)^2 \\
= & \left( 1- \bra{\phi^0_{A_1,B_1}}\sigma \ket{\phi^0_{A_1,B_1}}
\right)^2.
\end{align*}
Since $0 \le P T(M) P\le I$,
we obtain (\ref{l-1-e}).

Next, we consider the case of $M=M^2_{cov}$.
The test $T(M^2_{cov})$ is invariant the following action,
{\it i.e.},
\begin{align*}
U(g_1) \otimes U(g_2)
T(M^2_{cov}) (U(g_1) \otimes U(g_2))^{\dagger}
=
T(M^2_{cov})
\end{align*}
Since the subspace $<\phi^0_{A_1,B_1}>^{\perp}\otimes
<\phi^0_{A_2,B_2}>^{\perp}$ is irreducible subspace,
the equation (\ref{24-38}) implies 
\begin{align*}
T(M^2_{cov})=
\ket{\phi^0_{A_1,B_1}\otimes \phi^0_{A_2,B_2}}
\bra{\phi^0_{A_1,B_1}\otimes \phi^0_{A_2,B_2}}
+ t P ,
\end{align*}
where $t$ is a constant.
Since the equation (\ref{24-39}) implies that
$\Tr T(M^2_{cov})= d^2$,
we obtain $t= \frac{d^2-1}{(d^2-1)^2}= \frac{1}{d^2-1}$.

\section{Proof of (\ref{22-25})}\Label{a28}
We focus on the vertex of the simplex of 
the $d-1$-dimensional subspace orthogonal to
$\varphi$.
That is, there exist $d$ vectors $u_1(\varphi), \ldots , u_{d}(\varphi)$
such that
\begin{align*}
\langle u_i(\varphi)| u_j(\varphi)\rangle=
\left\{
\begin{array}{ll}
-\frac{1}{d} & i\neq j\\
\frac{d-1}{d} & i=j.
\end{array}
\right.
\end{align*}
Hence, the $d$ vectors
$u^i(\varphi)\defeq u_i(\varphi)+ \frac{1}{\sqrt{d}}\varphi$
satisfy the condition (\ref{22-25}).

\begin{widetext}
\section{Proof of the $U(1)$-invariance of $T_{inv}^{1\to 2}$}\Label{a-11}
As is proved later, the test $T_{inv}^{1\to 2}$
has the form
\begin{align}
&T_{inv}^{A_1\to A_2\to B^{\otimes 2}}\nonumber \\
=& 
d^2 \int_G 
(g \otimes \overline{g} )
\otimes (g \otimes \overline{g} )
\ket{
 u_1\otimes \overline{u_1}
\otimes
 u_2\otimes \overline{u_2}
}
\bra{
u_1\otimes \overline{u_1}
\otimes
u_2\otimes \overline{u_2}
}
(g \otimes \overline{g} )^\dagger
\otimes (g \otimes \overline{g} )^\dagger
\nu(\,d g)\nonumber \\
=&
d^2 \int_G 
(g \otimes \overline{g} )
\otimes (g \otimes \overline{g} )
\Biggl(
\left\ket{\frac{1}{d^2}
\phi_{A_1,B_1}^0 \otimes \phi_{A_2,B_2}^0 \right}
\left\bra{\frac{1}{d^2}
\phi_{A_1,B_1}^0 \otimes \phi_{A_2,B_2}^0 \right} \nonumber \\
& +
\left\ket{\frac{1}{d}
\phi_{A_1,B_1}^0 \otimes
\left( u_2 \otimes \overline{u_2}- 
\frac{1}{d} \phi_{A_2,B_2}^0\right)\right}
\left\bra{\frac{1}{d}
\phi_{A_1,B_1}^0 \otimes
\left( u_2 \otimes \overline{u_2}- 
\frac{1}{d} \phi_{A_2,B_2}^0\right)\right}\nonumber \\
&+
\left\ket{\left( u_1 \otimes \overline{u_1}- 
\frac{1}{d} \phi_{A_1,B_1}^0\right)
\otimes
\frac{1}{d}\phi_{A_2,B_2}^0 \right}
\left\bra{\left( u_1 \otimes \overline{u_1}- 
\frac{1}{d} \phi_{A_1,B_1}^0\right)
\otimes
\frac{1}{d}
\phi_{A_2,B_2}^0\right}\nonumber \\
&+
\left\ket{\left( u_1 \otimes \overline{u_1}- 
\frac{1}{d} \phi_{A_1,B_1}^0\right)
\otimes
\left( u_2 \otimes \overline{u_2}- 
\frac{1}{d} \phi_{A_2,B_2}^0\right)\right}
\left\bra{\left( u_1 \otimes \overline{u_1}- 
\frac{1}{d} \phi_{A_1,B_1}^0\right)
\otimes
\left( u_2 \otimes \overline{u_2}- 
\frac{1}{d} \phi_{A_2,B_2}^0\right)\right}\nonumber \\
& \hspace{10ex} \Biggr)
(g \otimes \overline{g} )^\dagger
\otimes (g \otimes \overline{g} )^\dagger
\nu(\,d g)\Label{2-3-7}.
\end{align}
Thus, we can easily check that
the matrix 
$T_{inv}^{1\to 2}$ is commutative with
the matrix
\begin{align*}
U_\theta^{\otimes 2}=&
e^{2\theta i }
\ket{\phi_{A_1,B_1}^0 }\bra{\phi_{A_1,B_1}^0 }\otimes 
\ket{\phi_{A_2,B_2}^0 }\bra{\phi_{A_2,B_2}^0 } \\
&+
e^{\theta i }
\left(
\ket{\phi_{A_1,B_1}^0 }\bra{\phi_{A_1,B_1}^0 }\otimes 
\left(
I_{A_2,B_2} -\ket{\phi_{A_2,B_2}^0 }\bra{\phi_{A_2,B_2}^0 }
\right)
+
\left(I_{A_1,B_1} -\ket{\phi_{A_1,B_1}^0 }\bra{\phi_{A_1,B_1}^0 }
\right)
\otimes 
\ket{\phi_{A_2,B_2}^0 }\bra{\phi_{A_2,B_2}^0 }
\right)\\
&+
\left(I_{A_1,B_1} -\ket{\phi_{A_1,B_1}^0 }\bra{\phi_{A_1,B_1}^0 }
\right)
\otimes 
\left(I_{A_2,B_2} -\ket{\phi_{A_2,B_2}^0 }\bra{\phi_{A_2,B_2}^0 }
\right).
\end{align*}
, we obtain $U(1)$-invariance.

Next, we prove (\ref{2-3-7}).
Since 
\begin{align*}
& u_1 \otimes \overline{u_1}
\otimes 
u_2 \otimes \overline{u_2} \\
= &
\frac{1}{d^2}
\phi_{A_1,B_1}^0 \otimes \phi_{A_2,B_2}^0 
+
\frac{1}{d}
\phi_{A_1,B_1}^0 \otimes
\left( u_2 \otimes \overline{u_2}- 
\frac{1}{d} \phi_{A_2,B_2}^0\right) \\
&+
\left( u_1 \otimes \overline{u_1}- 
\frac{1}{d} \phi_{A_1,B_1}^0\right)
\otimes
\frac{1}{d}
\phi_{A_2,B_2}^0 
+
\left( u_1 \otimes \overline{u_1}- 
\frac{1}{d} \phi_{A_1,B_1}^0\right)
\otimes
\left( u_2 \otimes \overline{u_2}- 
\frac{1}{d} \phi_{A_2,B_2}^0\right),
\end{align*}
it is sufficient to prove that 
the integrals of the cross terms equal to $0$.
We denote the invariant subgroup of $u$ by $G_u$
and its invariant measure by $\nu_u$.
Then, 
we can calculate
\begin{align*}
\int_{G_{u_1}}
(g' \otimes \overline{g'})
\left( u_2 \otimes \overline{u_2}- 
\frac{1}{d} \phi_{A_2,B_2}^0\right)
\nu_u(d g')
= 
\frac{1}{d} \phi_{A_2,B_2}^0-
\frac{1}{d} \phi_{A_2,B_2}^0=0.
\end{align*}
Hence, the integral of one cross term 
can be calculated as
\begin{align*}
&d^2\int_G 
(g \otimes \overline{g})\otimes (g \otimes \overline{g})
\left\ket{\left( u_1 \otimes \overline{u_1}- 
\frac{1}{d} \phi_{A_1,B_1}^0\right)\right}
\left\bra{\left( u_1 \otimes \overline{u_1}- 
\frac{1}{d} \phi_{A_1,B_1}^0\right)\right}
\otimes \\
& \hspace{10ex} \left\ket{\frac{1}{d}\phi_{A_2,B_2}^0 \right}
\left\bra{\left( u_2 \otimes \overline{u_2}- 
\frac{1}{d} \phi_{A_2,B_2}^0\right)\right}
(g \otimes \overline{g})^\dagger\otimes (g \otimes \overline{g})^\dagger
\nu(d g) \\
=&d^2\int_G 
(g \otimes \overline{g})
\left\ket{\left( u_1 \otimes \overline{u_1}- 
\frac{1}{d} \phi_{A_1,B_1}^0\right)\right}
\left\bra{\left( u_1 \otimes \overline{u_1}- 
\frac{1}{d} \phi_{A_1,B_1}^0\right)\right}
(g \otimes \overline{g})^\dagger
\otimes \\
&\hspace{10ex} \int_{G_{u_1}}
(g' \otimes \overline{g'})
\left\ket{\frac{1}{d}\phi_{A_2,B_2}^0 \right}
\left\bra{\left( u_2 \otimes \overline{u_2}- 
\frac{1}{d} \phi_{A_2,B_2}^0\right)\right}
(g' \otimes \overline{g'})^\dagger \nu_{u_1}(d g')
\nu(d g)\\
=&0.
\end{align*}
Similarly, we can check that the integrals of other cross term is $0$.
\end{widetext}

\section{Proof of (\ref{22-27})}\Label{a-14}
Let $T$ be an $SU(d)$-invariant $L(A_1,A_2 \to B_1,B_2)$
test with level-$0$.
Using the discussion of Proof of Lemma \ref{l-10},
we can find a POVM $M'=\{d^2 \ket{u_x^1 \otimes u_x^2}
\bra{u_x^1 \otimes u_x^2}\mu(d x)\}$ satisfying 
the condition (\ref{25-1}),
where $\mu$ is a probability measure.
We define the covariant POVM $M$ as
\begin{align*}
M(d g d x)= 
d^2 \ket{ g^{\otimes 2} u_x^1 \otimes u_x^2}
\bra{ g^{\otimes 2} u_x^1 \otimes u_x^2} \nu(d g) \mu(d x).
\end{align*}
The $SU(d)$-invariance of $T$ guarantees that
\begin{align*}
T \ge \int_{SU(d)} U(g)^{\otimes 2}T(M')
(U(g)^{\otimes 2})^{\dagger} \nu(d g)
= T(M).
\end{align*}
Note test $T(M)$ can be expressed as
\begin{align}
T(M)=&
\int \int_{SU(d)}
d^2 \ket{ U(g) u_x^1 \otimes \overline{u_x^1}}
\bra{ U(g) u_x^1 \otimes \overline{u_x^1}}\nonumber \\
& \otimes 
\ket{ U(g) u_x^2 \otimes \overline{u_x^2}}
\bra{ U(g) u_x^2 \otimes \overline{u_x^2}}
\nu(d g) \mu(d x).
\Label{25-2}
\end{align}

Thus, we can restrict our tests to the tests $T(M)$
with the form (\ref{25-2}).
First, we calculate the following value:
\begin{align*}
\Tr 
 \int_{SU(d)} &
\ket{ U(g) u_x^1 \otimes \overline{u_x^1}}
\bra{ U(g) u_x^1 \otimes \overline{u_x^1}}\\
& \otimes 
\ket{ U(g) u_x^2 \otimes \overline{u_x^2}}
\bra{ U(g) u_x^2 \otimes \overline{u_x^2}}
\nu(d g) 
\sigma^{\otimes 2}.
\end{align*}
Indeed, from the $SU(d)$-invariance, 
this value depends only on the 
inner product $r\defeq |\langle u_x^1 , u_x^2 \rangle|^2$.
Hence, we can denote it by
$f(r)$.
Without loss of generality,
we can assume that
$u_x^1= \ket{0}$,$u_x^2= \sqrt{p}\ket{0}
+ \sqrt{1-p}\ket{1}$.
The group $SU(d)$ has the subgroup:
\begin{widetext}
\begin{align*}
G'\defeq
\left\{\left.g_\theta \defeq e^{i\theta/2}\ket{0}\bra{0}
+e^{-i\theta/2}\ket{1}\bra{1}
+ \sum_{i=2}^{d-1} \ket{i}\bra{i}
\right| \theta \in [0,2\pi]
\right\}.
\end{align*}
Hence,
\begin{align*}
& \int_{0}^{2\pi}
\ket{ U(g_\theta) u_x^1 \otimes \overline{u_x^1}}
\bra{ U(g_\theta) u_x^1 \otimes \overline{u_x^1}}
\otimes 
\ket{ U(g_\theta) u_x^2 \otimes \overline{u_x^2}}
\bra{ U(g_\theta) u_x^2 \otimes \overline{u_x^2}}
d \theta \\
=&\ket{00}\bra{00}\otimes \Bigl(
p(1-p) \ket{01}\bra{01}
+p(1-p) \ket{10}\bra{10}\\
&+ (p \ket{00}+ (1-p)\ket{11} \bra{00})
(p \bra{00}+ (1-p)\bra{11} \bra{00})
\Bigr)\\
=&\ket{00}\bra{00}\otimes \Bigl(
 p(1-p) \ket{10}\bra{10}
+ p(1-p) \ket{01}\bra{01}\\
&+ p^2 \ket{00}\bra{00}
+p(1-p) (\ket{00}\bra{11}+\ket{11}\bra{00})
+(1-p)^2 \ket{11}\bra{11}
\Bigr)\\
=&
\ket{00}\bra{00}
\otimes \Bigl( 
\ket{11}\bra{11}
+ p^2 \left(
- \ket{01}\bra{01}-  \ket{10}\bra{10}
+ \ket{00}\bra{00}
+ \ket{11}\bra{11}
-\ket{00}\bra{11}-\ket{11}\bra{00}\right)\\
& +p\left( \ket{01}\bra{01}+  \ket{10}\bra{10}
-2 \ket{11}\bra{11}+ 
\ket{00}\bra{11}+\ket{11}\bra{00}\right)
\Bigr).
\end{align*}
We put 
\begin{align*}
C_1(\sigma) & \defeq 
\int_{SU(d)} 
\bra{00}U(g) \sigma U(g)^{\dagger}\ket{00}
\Bigl(
- \bra{01}U(g) \sigma U(g)^{\dagger}\ket{01}
- \bra{10}U(g) \sigma U(g)^{\dagger}\ket{10}
+ \bra{00}U(g) \sigma U(g)^{\dagger}\ket{00}\\
&\hspace{20ex} + \bra{11}U(g) \sigma U(g)^{\dagger}\ket{11}
- \bra{11}U(g) \sigma U(g)^{\dagger}\ket{00}
- \bra{00}U(g) \sigma U(g)^{\dagger}\ket{11}
\Bigr)
\nu(d g) \\
C_2(\sigma) & \defeq 
\int_{SU(d)} 
\bra{00}U(g) \sigma U(g)^{\dagger}\ket{00}
 \Bigl(
  \bra{01}U(g) \sigma U(g)^{\dagger}\ket{01}
+ \bra{10}U(g) \sigma U(g)^{\dagger}\ket{10}\\
& \hspace{20ex} -2\bra{11}U(g) \sigma U(g)^{\dagger}\ket{11}
+ \bra{11}U(g) \sigma U(g)^{\dagger}\ket{00}
+ \bra{00}U(g) \sigma U(g)^{\dagger}\ket{11}
\Bigr)
\nu(d g) \\
C_3(\sigma) & \defeq 
\int_{SU(d)} 
\bra{00}U(g) \sigma U(g)^{\dagger}\ket{00}
\bra{11}U(g) \sigma U(g)^{\dagger}\ket{11}
\nu(d g) .
\end{align*}
\end{widetext}
Hence, putting
$p(x)\defeq |\langle u_x^1 | u_x^2\rangle|^2$,
we have
\begin{align*}
& \Tr T(M)\sigma^{\otimes 2} \\
= &\int d^2 
\left(
C_1(\sigma)p(x)^2 + C_2(\sigma)p(x) + C_3(\sigma)
\right) \mu(d x).
\end{align*}

Denoting the projection to the symmetric subspace of
$\cl{H}_A^{\otimes 2}$ by $P_S$,
we obtain
\begin{align*}
& \frac{d(d+1)}{2}
=
\Tr P_S I P_S \\
= &
\int \int_{SU(d)}
d^2 \Tr P_S \ket{ g u_x^1 \otimes g u_x^2}
\bra{ g u_x^1 \otimes g u_x^2}P_S
\nu(d g) \mu(d x)\\
= &
\int \int_{SU(d)}
d^2 
\frac{1+ |\langle g u_x^1 | g u_x^2\rangle|^2}{2}
\nu(d g) \mu(d x)\\
= &
\int d^2 
\frac{1+ |\langle u_x^1 | u_x^2\rangle|^2}{2}
\mu(d x),
\end{align*}
which implies
\begin{align*}
\int  p(x)
\mu(d x)=
\frac{1}{d}
\end{align*}
because $|\langle u_x^1 | u_x^2\rangle|^2=p(x)$.
As is shown later, $C_1(\sigma)$ is positive.
Since
$\int p(x)^2 \mu(d x)\ge \frac{1}{d^2}$,
\begin{align*}
\Tr T(M)\sigma^{\otimes 2}
\ge d^2 
\left(
\frac{C_1(\sigma)}{d^2}
 + \frac{C_2(\sigma)}{d} + C_3(\sigma)
\right).
\end{align*}
The equality holds if
$p(x)= \frac{1}{d}$ for all $x$.
That is, if $T= T_{inv}^{1\to 2}$,
the equality holds.
Therefore, we obtain (\ref{22-27}).

Letting 
\begin{align*}
g'= 
\left(
\begin{array}{ccc}
0 & -1 & \\
1 & 0  & \\
  &    & I
\end{array}
\right) g,
\end{align*} 
we obtain
\begin{widetext}
\begin{align*}
&\int_{SU(d)}
\bra{00}U(g) \sigma U(g)^{\dagger}\ket{00}
\Bigl(
- \bra{01}U(g) \sigma U(g)^{\dagger}\ket{01}
- \bra{10}U(g) \sigma U(g)^{\dagger}\ket{10}
+ \bra{00}U(g) \sigma U(g)^{\dagger}\ket{00}\\
&\hspace{20ex} + \bra{11}U(g) \sigma U(g)^{\dagger}\ket{11}
- \bra{11}U(g) \sigma U(g)^{\dagger}\ket{00}
- \bra{00}U(g) \sigma U(g)^{\dagger}\ket{11}
\Bigr)
\nu(dg ) \\
=
&\int_{SU(d)}
\bra{11}U(g') \sigma U(g')^{\dagger}\ket{11}
\Bigl(
- \bra{01}U(g') \sigma U(g')^{\dagger}\ket{01}
- \bra{10}U(g') \sigma U(g')^{\dagger}\ket{10}
+ \bra{00}U(g') \sigma U(g')^{\dagger}\ket{00}\\
&\hspace{20ex} + \bra{11}U(g') \sigma U(g')^{\dagger}\ket{11}
- \bra{11}U(g') \sigma U(g')^{\dagger}\ket{00}
- \bra{00}U(g') \sigma U(g')^{\dagger}\ket{11}
\Bigr)
\nu(dg' ) .
\end{align*}
Hence,
\begin{align}
C_1(\sigma)=&
\int_{SU(d)}
\frac{1}{2}
\left(\bra{00}U(g) \sigma U(g)^{\dagger}\ket{00}
+
\bra{11}U(g) \sigma U(g)^{\dagger}\ket{11}\right) \nonumber\\
& \hspace{15ex} \Bigl(
- \bra{01}U(g) \sigma U(g)^{\dagger}\ket{01}
- \bra{10}U(g) \sigma U(g)^{\dagger}\ket{10}
+ \bra{00}U(g) \sigma U(g)^{\dagger}\ket{00}\nonumber \\
&\hspace{20ex} + \bra{11}U(g) \sigma U(g)^{\dagger}\ket{11}
- \bra{11}U(g) \sigma U(g)^{\dagger}\ket{00}
- \bra{00}U(g) \sigma U(g)^{\dagger}\ket{11}
\Bigr)
\nu(dg ) . \Label{2-3-8}
\end{align}
By using the notations
\begin{align*}
\varphi_{A,B}^0\defeq \frac{1}{\sqrt{2}}
\left(
\ket{00}+ \ket{11}
\right) ,\quad
\varphi_{A,B}^1\defeq \frac{1}{\sqrt{2}}
\left(
\ket{10}+ \ket{10}
\right) ,\quad
\varphi_{A,B}^2\defeq \frac{1}{\sqrt{2}}
\left(
-i \ket{10}+ i \ket{10}
\right) ,\quad
\varphi_{A,B}^3\defeq \frac{1}{\sqrt{2}}
\left(
\ket{00}- \ket{11}
\right) ,
\end{align*}
$C_1(\sigma)$ can be calculated as
\begin{align*}
C_1(\sigma)=&
\int_{SU(d)}
\frac{1}{2}
\left(\bra{\phi_{A,B}^0}U(g) \sigma U(g)^{\dagger}\ket{\phi_{A,B}^0}
+
\bra{\phi_{A,B}^3}U(g) \sigma U(g)^{\dagger}\ket{\phi_{A,B}^3}
\right)\\
& \hspace{10ex}\Bigl(
2 \bra{\phi_{A,B}^3}U(g) \sigma U(g)^{\dagger}\ket{\phi_{A,B}^3}
- \bra{\phi_{A,B}^1}U(g) \sigma U(g)^{\dagger}\ket{\phi_{A,B}^1}
- \bra{\phi_{A,B}^2}U(g) \sigma U(g)^{\dagger}\ket{\phi_{A,B}^2}
\Bigr)
\nu(dg ) .
\end{align*}
Similarly to (\ref{2-3-8}),
focusing the elements $g_{1,2},g_{2,3},g_{3,1}$ of $SU(d)$
such that
\begin{align*}
g_{1,2}:& (\phi_{A,B}^1,\phi_{A,B}^2,\phi_{A,B}^3)
\to (\phi_{A,B}^2,-\phi_{A,B}^1,\phi_{A,B}^3) \\
g_{2,3}:& (\phi_{A,B}^1,\phi_{A,B}^2,\phi_{A,B}^3)
\to (\phi_{A,B}^1,\phi_{A,B}^3,-\phi_{A,B}^2) \\
g_{3,1}:& (\phi_{A,B}^1,\phi_{A,B}^2,\phi_{A,B}^3)
\to (-\phi_{A,B}^3,\phi_{A,B}^2,\phi_{A,B}^1) ,
\end{align*}
we can prove
\begin{align*}
C_1(\sigma)=&
\int_{SU(d)}
\frac{1}{3}
\left(\sum_{i=1}^{3}
\bra{\phi_{A,B}^i}U(g) \sigma U(g)^{\dagger}\ket{\phi_{A,B}^i}^2
-
\sum_{i>j}
\bra{\phi_{A,B}^i}U(g) \sigma U(g)^{\dagger}\ket{\phi_{A,B}^i}
\bra{\phi_{A,B}^j}U(g) \sigma U(g)^{\dagger}\ket{\phi_{A,B}^j}
\right)
\nu(dg ) \\
=&\int_{SU(d)}
\frac{1}{6}
\sum_{i>j}
\left(\bra{\phi_{A,B}^i}U(g) \sigma U(g)^{\dagger}\ket{\phi_{A,B}^i}
-
\bra{\phi_{A,B}^j}U(g) \sigma U(g)^{\dagger}\ket{\phi_{A,B}^j}\right)^2
\nu(dg ) \ge 0.
\end{align*}
\end{widetext}

\section{Proof of (\ref{23-20})}\Label{a-12}
Let $T$ be a $SU(d)$-invariant separable level-$\alpha$ test among 
$A_1, \ldots, A_n,B_1, \ldots, B_n$
with the null hypothesis $\cl{S}_0$.
Then, $T$ has the following form:
\begin{align*}
T=
\sum_k a_k
\ket{u_1^k\otimes {u'}_1^k}\bra{u_1^k\otimes {u'}_1^k}
\otimes \cdots \otimes 
\ket{u_n^k\otimes {u'}_n^k}\bra{u_n^k\otimes {u'}_n^k},
\end{align*}
where $\|u_i^k\|=1, \langle u_i^k| {u'}_i^k\rangle=1$.
Since $T$ is level-$\alpha$
and $\bra{ \phi_{A,B}^0}\ket{u_1^k\otimes {u'}_1^k}
\bra{u_1^k\otimes {u'}_1^k}\ket{ \phi_{A,B}^0}= \frac{1}{d}$,
we have
\begin{align*}
1- \alpha = \bra{ {\phi_{A,B}^0}^{\otimes n}}
T \ket{ {\phi_{A,B}^0}^{\otimes n}}
=\sum_k a_k
\frac{1}{d^n}.
\end{align*}
It follows from the $SU(d)$-invariance of $T$ that
\begin{align*}
T=
&\int_{SU(d)}
\sum_k a_k
\ket{g u_1^k\otimes \overline{g} {u'}_1^k}
\bra{g u_1^k\otimes \overline{g}{u'}_1^k} \\
& \quad \otimes \cdots \otimes 
\ket{g u_n^k\otimes \overline{g}{u'}_n^k}
\bra{g u_n^k\otimes \overline{g}{u'}_n^k}
\nu(d g).
\end{align*}
The concavity of the function $x \mapsto \log x$
implies that
\begin{align*}
& \log \Tr 
\Biggl(\int_{SU(d)}
\ket{g u_1^k\otimes \overline{g} {u'}_1^k}
\bra{g u_1^k\otimes \overline{g}{u'}_1^k}
\otimes \cdots \\
&\quad \otimes 
\ket{g u_n^k\otimes \overline{g}{u'}_n^k}
\bra{g u_n^k\otimes \overline{g}{u'}_n^k}
\nu(d g)\Biggr)
\sigma^{\otimes n}\\
=& \log \int_{SU(d)}
\Tr \bra{g u_1^k\otimes \overline{g}{u'}_1^k}
\sigma \ket{g u_1^k\otimes \overline{g} {u'}_1^k} \\
& \quad \cdots 
\Tr \bra{g u_n^k\otimes \overline{g}{u'}_n^k}
\sigma \ket{g u_n^k\otimes \overline{g} {u'}_n^k}
\nu(d g)\\
\ge &
\int_{SU(d)}
\sum_{i=1}^n
\log \Tr \bra{g u_i^k\otimes \overline{g}{u'}_i^k}
\sigma \ket{g u_i^k\otimes \overline{g} {u'}_i^k}
 \nu(d g) \\
\ge  &
n 
\min_{u,u': |\langle u | \overline{u'} \rangle|=1,\|u\|=1} \\
&\quad \int_{SU(d)}
\log \Tr \bra{g u \otimes \overline{g}{u'}}
\sigma \ket{g u\otimes \overline{g} {u'}}
 \nu(d g) .
\end{align*}
Denoting the RHS by $C$,
we obtain 
\begin{align*}
\Tr T \sigma^{\otimes n}
\ge \sum_{k} a_k e^{C}
= e^{C} \sum_{k} a_k
= e^{C}d^n (1-\alpha).
\end{align*}
Hence,
\begin{align*}
\log \frac{\Tr T \sigma^{\otimes n}}{1-\alpha}
\ge n \log d + C,
\end{align*}
which implies (\ref{23-20}).

\begin{widetext}
\section{Proof of (\ref{2-9-2}) and (\ref{2-12-10})}\Label{a19}
Let $T$
be an $SU(2) \times U(1)$-invariant $A$-$B$ separable test.
Then, the $SU(2)$-invariance guarantees that
$T= U(g)^{\otimes 2} T
(U(g)^{\otimes 2})^{\dagger}$
for $\forall g \in SU(2)$.
Hence,
$T= \int_{SU(2)}
U(g)^{\otimes 2} T
(U(g)^{\otimes 2})^{\dagger}
 \nu(dg )$.
Thus,
the test $T$ has the form
\begin{align*}
T= \int
4
\int_{SU(2)}
U(g,\theta)^{\otimes 2}
\ket{u_x \otimes {u'}_x} \bra{u_x \otimes {u'}_x}
(U(g,\theta)^{\otimes 2})^{\dagger}
\nu(dg )
\mu(d x),
\end{align*}
where $u_x \in \cl{H}_{A_1}\otimes \cl{H}_{A_1}$,
${u'}_x \in \cl{H}_{B_1}\otimes \cl{H}_{B_1}$,
$\langle \phi_{A_1,B_1}^0 \otimes  \phi_{A_2,B_2}^0
\ket{u_x \otimes {u'}_x}= \frac{1}{2}$,
and
$\mu$ is arbitrary probability measure.
Since our purpose is calculating the minimum value of
the second error probability
$\Tr T \sigma^{\otimes 2}$,
we can assume that the second term of (\ref{2-13}) is $0$
without loss of generality.
Therefore, Lemma \ref{l-9} implies that
\begin{align}
\int 
4
\int_{SU(2)}
(g^{\otimes 2} u_x) \otimes (\overline{g}g^{\otimes 2} {u'}_x )
\nu(d g)\mu(d x) = 
2\phi_{A_1,B_1}^0 \otimes  \phi_{A_2,B_2}^0
\Label{2-12}
\end{align}

Moreover, the $SU(2) \times U(1)$-invariance guarantees that
$T= U(g,\theta)^{\otimes 2} T
(U(g,\theta)^{\otimes 2})^{\dagger}$
for $\forall g \in SU(2)$ and $\forall \theta \in \real$.
Hence,
\begin{align*}
\Tr T \sigma^{\otimes 2}
= \Tr T (U(g,\theta)^{\otimes 2})^{\dagger}\sigma^{\otimes 2}
U(g,\theta)^{\otimes 2}.
\end{align*}
Taking the integral,
we obtain 
\begin{align*}
\Tr T \sigma^{\otimes 2}
= \Tr T 
\frac{1}{2\pi}
\int_0^{2\pi}
\int_{SU(2)}
(U(g,\theta)^{\otimes 2})^{\dagger}\sigma^{\otimes 2}
U(g,\theta)^{\otimes 2}
\nu(dg )d \theta. 
\end{align*}
Therefore,
the RHS can be written by use of projections
of the irreducible spaces regarding the action of the group
$SU(2) \times U(1)$.
Indeed, the tensor product space
$\cl{H}_{A_1} \otimes \cl{H}_{A_2} \otimes 
\cl{H}_{B_1} \otimes \cl{H}_{B_2}$
is decomposed to the direct sum product the following 
irreducible spaces regarding the action of the group
$SU(2) \times U(1)$:
\begin{align}
\Sigma^0_5 \defeq < &
\frac{1}{\sqrt{2}}\left(
  |1,2\rangle_{1,2}
+ |2,1\rangle_{1,2}
\right),\quad
\frac{1}{\sqrt{2}}\left(
  |2,3\rangle_{1,2}
+ |3,2\rangle_{1,2}
\right),\quad
 \frac{1}{\sqrt{2}}\left(
  |3,1\rangle_{1,2}
+ |1,3\rangle_{1,2}
\right),\nonumber \\
&\frac{1}{\sqrt{3}}\left(
           |1,1\rangle_{1,2}
+ \omega   |2,2\rangle_{1,2}
+ \omega^2 |3,3\rangle_{1,2}
\right),\quad
\frac{1}{\sqrt{3}}\left(
           |1,1\rangle_{1,2}
+ \omega^2 |2,2\rangle_{1,2}
+ \omega   |3,3\rangle_{1,2}
\right)> \nonumber \\
\Sigma_3^1 \defeq
<&
\frac{1}{\sqrt{2}}\left(
  |0,1\rangle_{1,2}
+ |1,0\rangle_{1,2}
\right),\quad
\frac{1}{\sqrt{2}}\left(
  |0,2\rangle_{1,2}
+ |2,0\rangle_{1,2}
\right),\quad
 \frac{1}{\sqrt{2}}\left(
  |0,3\rangle_{1,2}
+ |3,0\rangle_{1,2}
\right) >\nonumber \\
\Sigma_1^2 \defeq
<&  |0,0\rangle_{1,2}>\Label{2-12-2}\\
\Sigma_1^0 \defeq
<&
\frac{1}{\sqrt{3}}\left(
   |1,1\rangle_{1,2}
+  |2,2\rangle_{1,2}
+  |3,3\rangle_{1,2}
\right)>\nonumber \\
\Lambda_3^0 \defeq
<&
\frac{1}{\sqrt{2}}\left(
  |1,2\rangle_{1,2}
- |2,1\rangle_{1,2}
\right),\quad
\frac{1}{\sqrt{2}}\left(
  |2,3\rangle_{1,2}
- |3,2\rangle_{1,2}
\right),\quad
 \frac{1}{\sqrt{2}}\left(
  |3,1\rangle_{1,2}
- |1,3\rangle_{1,2}
\right)>\nonumber \\
\Lambda_3^1 \defeq
<&
\frac{1}{\sqrt{2}}\left(
  |0,1\rangle_{1,2}
- |1,0\rangle_{1,2}
\right),\quad
\frac{1}{\sqrt{2}}\left(
  |0,2\rangle_{1,2}
- |2,0\rangle_{1,2}
\right),\quad
 \frac{1}{\sqrt{2}}\left(
  |0,3\rangle_{1,2}
- |3,0\rangle_{1,2}
\right) >\nonumber 
\end{align}
where 
$|i,j\rangle_{1,2}$ denotes the vector 
$\phi^i_{A_1,B_1}\otimes \phi^j_{A_2,B_2}$, and
$\omega = \frac{-1+\sqrt{3}i}{2}$.
The meaning of this notation is given as follows.
The superscript $k=0,1,2$ denotes the 
$U(1)$-action, {\it i.e.},
the element $e^{i\theta}$ acts on this space as
$e^{k \theta i}$.
The subscript $l=1,3,5$ denotes the dimension of the space.
In the spaces labeled as $\Sigma$,
the action $|i,j\rangle_{1,2} \to |j,i\rangle_{1,2}$
is described as the action of the constant $1$.
But, in the spaces labeled as $\Lambda$,
it is described as the action of the constant $-1$.
In the following, for simplicity,
we abbreviate the projection to the subspace $\Sigma^k_l$ and
$\Lambda^k_l$ as $\Sigma^k_l$ and $\Lambda^k_l$, respectively.
Hence, we obtain
\begin{align*}
& \frac{1}{2\pi}
\int_0^{2\pi}
\int_{SU(2)}
(U(g,\theta)^{\otimes 2})^{\dagger}\sigma^{\otimes 2}
U(g,\theta)^{\otimes 2}
\nu(dg )d \theta \\
=&
 \frac{\Tr \sigma^{\otimes 2}\Sigma^0_5}{5}\Sigma^0_5   
+\frac{\Tr \sigma^{\otimes 2}\Sigma^1_3}{3}\Sigma^1_3  
+(\Tr \sigma^{\otimes 2}\Sigma^2_1)\Sigma^2_1  
+(\Tr \sigma^{\otimes 2}\Sigma^0_1)\Sigma^0_1  
+\frac{\Lambda^0_3 \Tr \sigma^{\otimes 2}\Lambda^0_3}{3}
+\frac{\Lambda^1_3 \Tr \sigma^{\otimes 2}\Lambda^1_3 }{3}
.
\end{align*}
In order to calculate the quantities $\Tr \sigma^{\otimes 2}\Sigma^k_l$
and $\Tr \sigma^{\otimes 2}\Lambda^k_l$,
we describe the matrix elements of $\sigma$ 
with the basis $<\phi_{A,B}^0,\ldots,\phi_{A,B}^3>$ by
$x_{i,j} \defeq \bra{\phi_{A,B}^i} \sigma \ket{\phi_{A,B}^i}$.
For our convenience, we treat this matrix by use of the notation 
\begin{align*}
(x_{i,j})= \left(
\begin{array}{cc}
a & b^{\dagger}  \\
b& C
\end{array}
\right),
\end{align*}
where $a$ is a real number, $b$ is 
a $3$-dimensional complex-valued vector,
$C$ is a $3\times 3$ Hermitian matrix.
Thus, the quantities $\Tr \sigma^{\otimes 2}\Sigma^k_l$
and $\Tr \sigma^{\otimes 2}\Lambda^k_l$ are calculated as
\begin{align*}
\Tr \sigma^{\otimes 2}\Sigma^0_5 &=
\frac{2}{3} \sum_{i=1}^3 x_{i,i}+
\sum_{1\le i<j\le 3} x_{i,i}x_{j,j} + |x_{i,j}|^2
- \frac{1}{3} (x_{i,j}^2 + x_{j,i}^2) 
= \frac{1}{2}\left( (\Tr C^2)+ (\Tr C)^2 \right)
-\frac{1}{3} \Tr C \overline{C} \\
\Tr \sigma^{\otimes 2}\Sigma^1_3 &=
\sum_{i=1}^3 (x_{0,0}x_{i,i} + |x_{0,i}|^2 )
= a \Tr C + | b|^2 \\
\Tr \sigma^{\otimes 2}\Sigma^2_1 &=
x_{0,0}^2= a^2\\
\Tr \sigma^{\otimes 2}\Sigma^0_1 &=
\frac{1}{3} \sum_{1\le i , j \le 3}
x_{i,j}^2 
= \frac{1}{3} \Tr C \overline{C}
\\
\Tr \sigma^{\otimes 2}\Lambda^0_3 &=
\sum_{i< j} x_{i,i}x_{j,j} -|x_{i,j}|^2 
=\frac{1}{2}\left((\Tr C)^2 - (\Tr C^2) \right)
\\
\Tr \sigma^{\otimes 2}\Lambda^1_3 &=
\sum_{i=1}^3
(x_{0,0}x_{i,i} -| x_{0,i}|^2)
= a \Tr C - | b|^2 ,
\end{align*}
where $\overline{C}$ is the complex conjugate of $C$.
As is proven later,
the inequalities
\begin{align}
\Tr \sigma^{\otimes 2}\Lambda^1_3 - \Tr \sigma^{\otimes 2}\Lambda^0_3 
&\ge 0 \Label{2-12-5}\\
5 \Tr \sigma^{\otimes 2}\Sigma^1_3 - 
3 \Tr \sigma^{\otimes 2}\Sigma^0_5
&\ge 0 \Label{2-12-6}\\
10 \Tr \sigma^{\otimes 2}\Sigma^0_1 
+\Tr \sigma^{\otimes 2}\Sigma^0_5 
-5 \Tr \sigma^{\otimes 2}\Lambda^0_3 
&\ge 0 \Label{2-12-9}
\end{align}
hold, when $p = \Tr C \le \frac{1}{2}$.

On the other hand, 
we focus on the 
following basis of the space $\cl{H}_{A_1}\otimes \cl{H}_{A_2}$:
\begin{align*}
\varphi_{A_1,A_2}^0&\defeq \frac{1}{\sqrt{2}}
\left(
\ket{01}_{A_1,A_2}- \ket{10}_{A_1,A_2}
\right) ,\quad
\varphi_{A_1,A_2}^1 \defeq \frac{1}{\sqrt{2}}
\left(
\ket{00}_{A_1,A_2}+ \ket{11}_{A_1,A_2}
\right) ,\\
\varphi_{A_1,A_2}^2 &\defeq \frac{i}{\sqrt{2}}
\left(
\ket{00}_{A_1,A_2}- \ket{11}_{A_1,A_2}
\right) ,\quad
\varphi_{A_1,A_2}^3\defeq \frac{i}{\sqrt{2}}
\left(
\ket{01}_{A_1,A_2}+ \ket{10}_{A_1,A_2}
\right) .
\end{align*}
The other space $\cl{H}_{B_1}\otimes \cl{H}_{B_2}$
is spanned by the complex conjugate basis:
\begin{align*}
\varphi_{B_1,B_2}^0&\defeq \frac{1}{\sqrt{2}}
\left(
\ket{01}_{B_1,B_2}- \ket{10}_{B_1,B_2}
\right) ,\quad
\varphi_{B_1,B_2}^1 \defeq \frac{1}{\sqrt{2}}
\left(
\ket{00}_{B_1,B_2}+ \ket{11}_{B_1,B_2}
\right) ,\\
\varphi_{B_1,B_2}^2 &\defeq \frac{-i}{\sqrt{2}}
\left(
\ket{00}_{B_1,B_2}- \ket{11}_{B_1,B_2}
\right) ,\quad
\varphi_{B_1,B_2}^3\defeq \frac{-i}{\sqrt{2}}
\left(
\ket{01}_{B_1,B_2}+ \ket{10}_{B_1,B_2}
\right) .
\end{align*}
By using this basis, the irreducible subspaces of
$\cl{H}_{A_1}\otimes \cl{H}_{A_2}\otimes 
\cl{H}_{B_1}\otimes \cl{H}_{B_2}$
are written as
\begin{align*}
\Sigma_1^2 =
<&  \frac{1}{2}\left(
%\sum_{k=0}^3
|0,0\rangle_{A,B}+|1,1\rangle_{A,B}+|2,2\rangle_{A,B}+|3,3\rangle_{A,B}
\right)>\\
\Sigma^0_5 = < &
\frac{1}{\sqrt{2}}\left(
  |1,2\rangle_{A,B}
+ |2,1\rangle_{A,B}
\right),\quad
\frac{1}{\sqrt{2}}\left(
  |2,3\rangle_{A,B}
+ |3,2\rangle_{A,B}
\right),\quad
 \frac{1}{\sqrt{2}}\left(
  |3,1\rangle_{A,B}
+ |1,3\rangle_{A,B}
\right),\\
&\frac{1}{\sqrt{3}}\left(
           |1,1\rangle_{A,B}
+ \omega   |2,2\rangle_{A,B}
+ \omega^2 |3,3\rangle_{A,B}
\right),\quad
\frac{1}{\sqrt{3}}\left(
           |1,1\rangle_{A,B}
+ \omega^2 |2,2\rangle_{A,B}
+ \omega   |3,3\rangle_{A,B}
\right)> \\
\Sigma_3^1 =
<&
\frac{1}{\sqrt{2}}\left(
  |1,2\rangle_{A,B}
- |2,1\rangle_{A,B}
\right),\quad
\frac{1}{\sqrt{2}}\left(
  |2,3\rangle_{A,B}
- |3,2\rangle_{A,B}
\right),\quad
 \frac{1}{\sqrt{2}}\left(
  |3,1\rangle_{A,B}
- |1,3\rangle_{A,B}
\right)>\\
\Sigma_1^0 =
<&
\frac{3}{\sqrt{12}}  |0,0\rangle_{A,B}
- \frac{1}{\sqrt{12}}\left(
   |1,1\rangle_{A,B}
+  |2,2\rangle_{A,B}
+  |3,3\rangle_{A,B}
\right)>\\
\Lambda_3^0 =
<&
\frac{1}{\sqrt{2}}\left(
  |0,1\rangle_{A,B}
+ |1,0\rangle_{A,B}
\right),\quad
\frac{1}{\sqrt{2}}\left(
  |0,2\rangle_{A,B}
+ |2,0\rangle_{A,B}
\right),\quad
 \frac{1}{\sqrt{2}}\left(
  |0,3\rangle_{A,B}
+ |3,0\rangle_{A,B}
\right) >\\
\Lambda_3^1 =
<&
\frac{1}{\sqrt{2}}\left(
  |0,1\rangle_{A,B}
- |1,0\rangle_{A,B}
\right),\quad
\frac{1}{\sqrt{2}}\left(
  |0,2\rangle_{A,B}
- |2,0\rangle_{A,B}
\right),\quad
 \frac{1}{\sqrt{2}}\left(
  |0,3\rangle_{A,B}
- |3,0\rangle_{A,B}
\right) >,
\end{align*}
where
$|i,j\rangle_{A,B}$ denotes the vector 
$\varphi^i_{A_1,A_2}\otimes \varphi^j_{B_1,B_2}$.

In the following,
we denote the vectors $u_x \in \cl{H}_{A_1,A_2}$ 
and $u_x'\in \cl{H}_{B_1,B_2}$ by
use of scalars $a_x$,${a'}_x$ and
three-dimensional vectors $w_x$, ${w'}_x$
as 
\begin{align*}
u_x= (a_x,w_x)\defeq a_x \varphi_{A_1,A_2}^0
+ \sum_{i=1}^3 w_{x,i}\varphi_{A_1,A_2}^i, \quad
{u'}_x=({a'}_x,{w'}_x)
\defeq {a'}_x \varphi_{B_1,B_2}^0
+ \sum_{i=1}^3 {w'}_{x,i}\varphi_{B_1,B_2}^i.
\end{align*}
The condition (\ref{2-12}) implies that
\begin{align*}
\int a_x {a'}_x \mu(d x)= \frac{1}{4} , \quad 
\int (w_x |{w'}_x) \mu(d x)= \frac{3}{4}, 
\end{align*}
where the inner product $(w_x |{w'}_x)$ is defined by
$(w_x |{w'}_x)\defeq \sum_{i=1}^3 w_{x,i} {w'}_{x,i}$.
the 
condition $\langle \phi_{A_1,B_1}^0 \otimes \phi_{A_2,B_2}^0
| u_x \otimes {u'}_x\rangle = \frac{1}{2}$ 
yields
\begin{align*}
a_x {a'}_x + (w_x |{w'}_x)=1
\end{align*}
because of (\ref{2-12-2}).
Using this notation, we obtain
\begin{align*}
\bra{u_x\otimes {u'}_x}
\Sigma^0_5 \ket{u_x\otimes {u'}_x}
&=
\left\|
\frac{1}{2}\left(
w_x\otimes {w'}_x +{w'}_x\otimes {w}_x 
\right)
-\frac{(w_x|{w'}_x)}{3}I_{3\times 3}
\right\|^2
\\
\bra{u_x\otimes {u'}_x}
\Sigma^1_3 \ket{u_x\otimes {u'}_x}
&=
\left\|
\frac{1}{2}\left(
w_x\otimes {w'}_x -{w'}_x\otimes {w}_x 
\right)
\right\|^2
\\
\bra{u_x\otimes {u'}_x}
\Sigma^2_1 \ket{u_x\otimes {u'}_x}
&=
\left|
\frac{a_x{a'}_x + (w_x |{w'}_x)}{2}
\right|^2= \frac{1}{4}
\\
\bra{u_x\otimes {u'}_x}
\Sigma^0_1 \ket{u_x\otimes {u'}_x}
&=
\left|
\frac{3}{\sqrt{12}}a_x{a'}_x 
- \frac{1}{\sqrt{12}}
(w_x |{w'}_x)
\right|^2
\ge \left(
\frac{3}{\sqrt{12}}\Re a_x{a'}_x 
- \frac{1}{\sqrt{12}}
\Re (w_x |{w'}_x)
\right)^2
\\
\bra{u_x\otimes {u'}_x}
\Lambda^0_3 \ket{u_x\otimes {u'}_x}
&=
\left\|
\frac{1}{2}\left(a_x{w'}_x +{a'}_x w_x \right)
\right\|^2
\\
\bra{u_x\otimes {u'}_x}
\Lambda^1_3 \ket{u_x\otimes {u'}_x}
&=
\left\|
\frac{1}{2}\left(a_x{w'}_x -{a'}_x w_x \right)
\right\|^2,
\end{align*}
where $\Re x$ denotes the real part of $x$.
Since we can evaluate 
\begin{align*}
&\left\|
\frac{1}{2}\left(a_x{w'}_x +{a'}_x w_x \right)
\right\|^2
+
\left\|
\frac{1}{2}\left(a_x{w'}_x -{a'}_x w_x \right)
\right\|^2
=
\| a_x{w'}_x\|^2 + \| {a'}_x w_x\|^2
\ge 2 \Re a_x {a'}_x \Re (w_x|{w'}_x) \\
& \left\|
\frac{1}{2}\left(
w_x\otimes {w'}_x +{w'}_x\otimes {w}_x 
\right)
-\frac{(w_x|{w'}_x)}{3}I_{3\times 3}
\right\|^2
+
\left\|
\frac{1}{2}\left(
w_x\otimes {w'}_x -{w'}_x\otimes {w}_x 
\right)
\right\|^2 \\
=&
\left\|
w_x\otimes {w'}_x 
-\frac{(w_x|{w'}_x)}{3}I_{3\times 3}
\right\|^2
\ge
\frac{2}{3}
\left|
(w_x|{w'}_x)
\right|^2\ge
\frac{2}{3}
\left(\Re
(w_x|{w'}_x)
\right)^2,
\end{align*}
the inequalities (\ref{2-12-5}) and (\ref{2-12-6}) yield
\begin{align*}
\frac{\Tr \sigma^{\otimes 2}\Lambda^0_3}{3}
\bra{u_x\otimes {u'}_x}
\Lambda^0_3 \ket{u_x\otimes {u'}_x}
+
\frac{\Tr \sigma^{\otimes 2}\Lambda^1_3}{3}
\bra{u_x\otimes {u'}_x}
\Lambda^1_3 \ket{u_x\otimes {u'}_x}
&\ge
\Tr \sigma^{\otimes 2}\Lambda^0_3
\cdot 2 \Re a_x {a'}_x \cdot \Re (w_x|{w'}_x)  \\
\frac{\Tr \sigma^{\otimes 2}\Sigma^0_5}{5}
\bra{u_x\otimes {u'}_x}
\Sigma^0_5 \ket{u_x\otimes {u'}_x}
+
\frac{\Tr \sigma^{\otimes 2}\Sigma^1_3}{3}
\bra{u_x\otimes {u'}_x}
\Sigma^1_3 \ket{u_x\otimes {u'}_x}
&\ge
\frac{\Tr \sigma^{\otimes 2}\Sigma^0_5}{5}
\cdot \frac{2}{3}
\left(\Re
(w_x|{w'}_x)
\right)^2.
\end{align*}
Letting $r(x)=\Re a_x {a'}_x $,
we have
\begin{align}
& \frac{1}{4}\Tr T \sigma^{\otimes 2}\nonumber \\
\ge &
\int
\frac{\Tr \sigma^{\otimes 2}\Sigma^2_1}{4}
+
\Tr \sigma^{\otimes 2}\Sigma^0_1
\left(
\frac{3}{\sqrt{12}}\Re a_x{a'}_x 
- \frac{1}{\sqrt{12}}
\Re (w_x |{w'}_x)
\right)^2 \nonumber\\
& \quad +\frac{2 \Tr \sigma^{\otimes 2}\Lambda^0_3}{3}
\Re a_x {a'}_x 
\Re (w_x|{w'}_x)
+\frac{2\Tr \sigma^{\otimes 2}\Sigma^0_5}{15}
\left(\Re
(w_x|{w'}_x)\right)^2
\mu(d x) \nonumber\\
=&
\frac{\Tr \sigma^{\otimes 2}\Sigma^2_1}{4}
+\frac{\Tr \sigma^{\otimes 2}\Sigma^0_1}{12}
+\frac{2\Tr \sigma^{\otimes 2}\Sigma^0_5}{15}\nonumber\\
&+
\int
\left(
-\frac{2\Tr \sigma^{\otimes 2}\Sigma^0_1}{3}
-\frac{4\Tr \sigma^{\otimes 2}\Sigma^0_5}{15}
+\frac{2\Tr \sigma^{\otimes 2}\Lambda^0_3}{3}
\right)r(x)
+
\left(
\frac{4\Tr \sigma^{\otimes 2}\Sigma^0_1}{3}
+\frac{2\Tr \sigma^{\otimes 2}\Sigma^0_5}{15}
-\frac{2\Tr \sigma^{\otimes 2}\Lambda^0_3}{3}
\right)r(x)^2
\mu(d x)\nonumber\\
=&
\frac{\Tr \sigma^{\otimes 2}\Sigma^2_1}{4}
+\frac{\Tr \sigma^{\otimes 2}\Sigma^0_1}{12}
+\frac{2\Tr \sigma^{\otimes 2}\Sigma^0_5}{15}\nonumber\\
&+
\frac{1}{4}
\left(
-\frac{2\Tr \sigma^{\otimes 2}\Sigma^0_1}{3}
-\frac{4\Tr \sigma^{\otimes 2}\Sigma^0_5}{15}
+\frac{2\Tr \sigma^{\otimes 2}\Lambda^0_3}{3}
\right)
+
\left(
\frac{4\Tr \sigma^{\otimes 2}\Sigma^0_1}{3}
+\frac{2\Tr \sigma^{\otimes 2}\Sigma^0_5}{15}
-\frac{2\Tr \sigma^{\otimes 2}\Lambda^0_3}{3}
\right)
\int
r(x)^2
\mu(d x)\nonumber\\
\stackrel{\rm (*)}{\ge}&
\frac{\Tr \sigma^{\otimes 2}\Sigma^2_1}{4}
+\frac{\Tr \sigma^{\otimes 2}\Sigma^0_1}{12}
+\frac{2\Tr \sigma^{\otimes 2}\Sigma^0_5}{15}\nonumber\\
&+
\frac{1}{4}
\left(
-\frac{2\Tr \sigma^{\otimes 2}\Sigma^0_1}{3}
-\frac{4\Tr \sigma^{\otimes 2}\Sigma^0_5}{15}
+\frac{2\Tr \sigma^{\otimes 2}\Lambda^0_3}{3}
\right)
+
\frac{1}{16}
\left(
\frac{4\Tr \sigma^{\otimes 2}\Sigma^0_1}{3}
+\frac{2\Tr \sigma^{\otimes 2}\Sigma^0_5}{15}
-\frac{2\Tr \sigma^{\otimes 2}\Lambda^0_3}{3}
\right)\nonumber\\
\stackrel{\rm (**)}{=}&
\frac{1}{4}- 
\frac{1}{2} \Tr C + \frac{7}{20}(\Tr C)^2 - \frac{1}{20}\Tr (\Re C)^2.
\Label{2-12-8}
\end{align}
Note that the inequality $(*)$ follows from
the inequality (\ref{2-12-9}) and the inequality
$\int
r(x)^2
\mu(d x)
\ge \left(\int r(x)
\mu(d x)\right)^2 = \frac{1}{16}$,
and the equation $(**)$
follows from the equation $a= 1- \Tr C$.
Since RHS of (\ref{2-12-8}) equals $\frac{\mbox{\rm RHS of }(\ref{2-9-2})}{4}$,
we obtain the part $\ge$ of (\ref{2-9-2}).

Conversely, the vector $u_{op}\otimes \overline{u_{op}}$
satisfies that
\begin{align*}
\bra{u_{op}\otimes \overline{u_{op}}}
\Sigma^0_5 \ket{u_{op}\otimes \overline{u_{op}}}
&=\bra{u_{op}\otimes \overline{u_{op}}}
\Lambda^0_3 \ket{u_{op}\otimes \overline{u_{op}}}
=\frac{3}{8}, \quad 
\bra{u_{op}\otimes \overline{u_{op}}}
\Sigma^2_1 \ket{u_{op}\otimes \overline{u_{op}}}
= \frac{1}{4}\\
\bra{u_{op}\otimes \overline{u_{op}}}
\Sigma^1_3 \ket{u_{op}\otimes \overline{u_{op}}}
&=\bra{u_{op}\otimes \overline{u_{op}}}
\Sigma^0_1 \ket{u_{op}\otimes \overline{u_{op}}}
=\bra{u_{op}\otimes \overline{u_{op}}}
\Lambda^1_3 \ket{u_{op}\otimes \overline{u_{op}}}
=0.
\end{align*}
Hence,
\begin{align*}
\frac{1}{4}\Tr T(M_{op}) \sigma^{\otimes 2}
=  
\frac{1}{4} \cdot \Tr \sigma^{\otimes 2} \Sigma^2_1
+ \frac{3}{8}\cdot\frac{\Tr \sigma^{\otimes 2} \Sigma^0_5 }{5}
+ \frac{3}{8}\cdot\frac{\Tr \sigma^{\otimes 2} \Lambda^0_3 }{3} 
= 
\frac{1}{4}- 
\frac{1}{2} \Tr C + \frac{7}{20}(\Tr C)^2 - \frac{1}{20}\Tr (\Re C)^2.
\end{align*}
Therefore, we obtain 
(\ref{2-12-10}),
which implies the part $\le$ of (\ref{2-9-2}).
 
Finally, we proceed to prove the inequalities (\ref{2-12-5}),
(\ref{2-12-6}), and (\ref{2-12-9}).
The inequality (\ref{2-12-9}) is shown as
\begin{align*}
10 \Tr \sigma^{\otimes 2}\Sigma^0_1 
+\Tr \sigma^{\otimes 2}\Sigma^0_5 
-5 \Tr \sigma^{\otimes 2}\Lambda^0_3 
= 3 \Tr C^2+C\overline{C})- 2 (\Tr C)^2 
= 2 \left( 3 \Tr (\Re C)^2 -  (\Tr C)^2 \right)
\ge 0 .
\end{align*}

In order to prove (\ref{2-12-5}),
we denote the eigenvalues of $C$ by $\lambda_1, \lambda_2, \lambda_3$
with the decreasing order, {\it i.e.},
$\lambda_1 > \lambda_2 > \lambda_3$.
First, we prove that $a \lambda_1 \ge |b|^2$ as follows.
Let $s$ be a arbitrary real number. Then,
\begin{align*}
0 \le \bra{(s,b)}\sigma\ket{(s,b)}
= a s^2 + 2 s \|b\|^2 + \bra{b}C \ket{b}
\end{align*}
Since the discriminant is positive,
we have $\|b\|^4 \le a \bra{b}C \ket{b}$, {\it i.e.},
$\|b\|^2 \le a \frac{\bra{b}C \ket{b}}{\|b\|^2}\le
a \lambda_1$.
Hence, 
using the relation $a= 1- \Tr C$, we have
\begin{align*}
& \Tr \sigma^{\otimes 2}\Lambda^1_3 - \Tr \sigma^{\otimes 2}\Lambda^0_3 
\ge
a \left( \Tr C - \lambda_1\right)- 
\frac{1}{2}\left((\Tr C)^2 - \Tr C^2 \right)
= 
(1- 2 \lambda_1 - (\lambda_2 +\lambda_3 ))(\lambda_2 +\lambda_3 )
- \lambda_2 \lambda_3  \\
\ge &
(1- 2 \lambda_1 - (\lambda_2 +\lambda_3 ))(\lambda_2 +\lambda_3 )
- \left(\frac{\lambda_2 +\lambda_3}{2}\right)^2
= 
\frac{\lambda_2 +\lambda_3}{2}
\left(2- 4\lambda_1 - 5 \frac{\lambda_2 +\lambda_3}{2}\right) \\
\ge &
\frac{\lambda_2 +\lambda_3}{2}
\left(
2- 4\lambda_1 - 8 \frac{\lambda_2 +\lambda_3}{2}\right)
= 4 \frac{\lambda_2 +\lambda_3}{2}
\left(\frac{1}{2} - \Tr C \right) \ge 0,
\end{align*}
which implies (\ref{2-12-5}).
Next, we proceed to (\ref{2-12-6}).
For this proof, we focus on the relations
\begin{align*}
\Tr C^2 \le (\Tr C)^2 , \quad
\Tr (\Im C)^2 \le (\Tr C)^2,
\end{align*}
which follow from $C\ge 0$,
where $\Im x$ denotes the imaginary part of $x$.
Hence,
\begin{align*}
& 5 \Tr \sigma^{\otimes 2}\Sigma^1_3 - 
3 \Tr \sigma^{\otimes 2}\Sigma^0_5
=
5 (1- \Tr C) \Tr C + 5 \|b\|^2 
- 3 \left( 
\frac{1}{2}\left( \Tr C^2 + (\Tr C)^2\right)
- \frac{1}{3}\Tr C \overline{C}\right)\\
\ge &
5 (1- \Tr C) \Tr C 
- 3 \left( 
\frac{1}{2}\left( \Tr C^2 + (\Tr C)^2\right)
- \frac{1}{3}\Tr C \overline{C}\right)
=
5\Tr C - \frac{13}{2}(\Tr C)^2 
- \frac{1}{2}\Tr C^2 - 2 \Tr (\Im C)^2\\
\ge &5 \Tr C - 8 (\Tr C)^2 
= \Tr C (5 - 8 \Tr C)\ge 0,
\end{align*}
which implies (\ref{2-12-6}).
\end{widetext}

\section{Proof of (\ref{2-9})}\Label{a20}
In this section, we use the same notation as section \ref{a19}.
by using the vector $u_1= |0\rangle_{A_1}, u_2= \frac{1}{\sqrt{2}}
(|0\rangle_{A_1}+|0\rangle_{A_2})$,
the POVM $M_{cov}^{1\to 2}$ is written
as
\begin{align*}
M_{cov}^{1\to 2}(\,d g)
=
d^2 (g \otimes g )
\ket{u_1 \otimes u_2}\bra{u_1 \otimes u_2}
(g \otimes g )^{\dagger} \nu(\,d g),
\end{align*}
Since $\ket{u_1 \otimes u_2}
= \frac{1}{2}
(\varphi_{A_1,A_2}^0+\varphi_{A_1,A_2}^1-i
\varphi_{A_1,A_2}^2-i\varphi_{A_1,A_2}^3)$,
we have
\begin{align*}
\bra{u_1 \otimes u_2\otimes \overline{u_1 \otimes u_2}}
\Sigma^0_5 
\ket{u_1 \otimes u_2\otimes \overline{u_1 \otimes u_2}}=& \frac{1}{8}\\
\bra{u_1 \otimes u_2\otimes \overline{u_1 \otimes u_2}}
\Lambda^0_3 
\ket{u_1 \otimes u_2\otimes \overline{u_1 \otimes u_2}}
=& \frac{1}{8}\\
\bra{u_1 \otimes u_2\otimes \overline{u_1 \otimes u_2}}
\Sigma^0_1 
\ket{u_1 \otimes u_2\otimes \overline{u_1 \otimes u_2}}
=&0 \\
\bra{u_1 \otimes u_2\otimes \overline{u_1 \otimes u_2}}
\Sigma^2_1 
\ket{u_1 \otimes u_2\otimes \overline{u_1 \otimes u_2}}
=&\frac{1}{4}\\
\bra{u_1 \otimes u_2\otimes \overline{u_1 \otimes u_2}}
\Sigma^1_3
\ket{u_1 \otimes u_2\otimes \overline{u_1 \otimes u_2}}
=&\frac{1}{4}\\
\bra{u_1 \otimes u_2\otimes \overline{u_1 \otimes u_2}}
\Lambda^1_3
\ket{u_1 \otimes u_2\otimes \overline{u_1 \otimes u_2}}
=&\frac{1}{4}.
\end{align*}
Hence,
\begin{align*}
&\Tr T(M_{cov}^{1\to 2})\sigma^{\otimes 2}\\
=& 
(1- \Tr C)^2
+ \frac{2}{3}\Tr C
- \frac{8}{15}(\Tr C)^2
- \frac{1}{15}\Tr (\Re C)^2,
\end{align*}
which implies (\ref{2-9}).

\section{Proof of (\ref{2-18-1})}\Label{a24}
Let $T$
be an $SU(d)\times SU(d)$-invariant $A$-$B$ separable test
with level $0$.
Then, similarly to proof of \ref{2-2-2},
the $SU(d)\times SU(d)$-invariance implies that
the test $T$ has the following form
\begin{align*}
&T\\
=& 
a_{0,0}\ket{\phi_{A_1,B_1}^0}\bra{\phi_{A_1,B_1}^0}
\otimes \ket{\phi_{A_2,B_2}^0}\bra{\phi_{A_2,B_2}^0}\\
& \quad +
a_{1,0}
(I- \ket{\phi_{A_1,B_1}^0}\bra{\phi_{A_1,B_1}^0})
\otimes \ket{\phi_{A_2,B_2}^0}\bra{\phi_{A_2,B_2}^0}\\
& \quad +
a_{0,1}
\ket{\phi_{A_1,B_1}^0}\bra{\phi_{A_1,B_1}^0}
\otimes (I- \ket{\phi_{A_2,B_2}^0}\bra{\phi_{A_2,B_2}^0})\\
& \quad +
a_{1,1}
(I- \ket{\phi_{A_1,B_1}^0}\bra{\phi_{A_1,B_1}^0})
\otimes (I- \ket{\phi_{A_2,B_2}^0}\bra{\phi_{A_2,B_2}^0}).
\end{align*}
Since the test $T$ is level $0$,
$a_{0,0}=1$.
Lemma \ref{l-3} yields that $\Tr T= d^2$.
Hence,
\begin{align*}
d^2-1 = (a_{1,0}+a_{0,1})(d^2-1)+ a_{1,1}(d^2-1)^2.
\end{align*}
In this case, the second error 
$\Tr \sigma_1\otimes \sigma_2 T$ can be calculated
as
\begin{align*}
&\Tr \sigma_1\otimes \sigma_2 T\\
=&
(1-p_1)(1-p_2)
+ (1-p_1)p_2 a_{0,1}\\
&\quad + p_1(1-p_2)a_{1,0}
+ p_1 p_2 a_{1,1}\\
=& (1-p_1)(1-p_2)
+ \frac{(1-p_1)p_2}{d^2-1}t_{0,1}\\
&\quad + \frac{p_1(1-p_2)}{d^2-1}t_{1,0}
+ \frac{p_1 p_2}{(d^2-1)^2}t_{1,1},
\end{align*}
where $t_{0,1}\defeq a_{0,1}(d^2-1),
t_{1,0}\defeq a_{1,0}(d^2-1),
t_{1,1}\defeq a_{1,1}(d^2-1)^2$.
Since $\frac{p_1 p_2}{(d^2-1)^2}\le
\frac{(1-p_1)p_2}{d^2-1},\frac{p_1(1-p_2)}{d^2-1}$,
\begin{align*}
\Tr \sigma_1\otimes \sigma_2 T\le
(1-p_1)(1-p_2)+\frac{p_1 p_2}{(d^2-1)^2}d^2-1.
\end{align*}

\begin{widetext}
\section{Proof of (\ref{2-19-1})}\Label{a25}
Let $T$
be an $SU(d)\times SU(d)$-invariant $A_1,A_2,B_1,B_2$ separable test
with level $0$.
The $SU(d)\times SU(d)$-invariance implies that
the test $T$ has the form
\begin{align*}
T
=& 
\sum_i p_i
\int_{SU(d)}\int_{SU(d)}
g_1\otimes \overline{g_1}\otimes
g_2\otimes \overline{g_2}
\ket{u_{i,A_1}\otimes u_{i,B_1} \otimes
u_{i,A_2}\otimes u_{i,B_2}}
\bra{u_{i,A_1}\otimes u_{i,B_1} \otimes
u_{i,A_2}\otimes u_{i,B_2}}\\
&\quad (g_1\otimes \overline{g_1}\otimes
g_2\otimes \overline{g_2})^\dagger
\nu(d g_1)\nu(d g_2),
\end{align*}
where
$\langle \phi_{A_1,B_1}^0\ket{u_{i,A_1}\otimes u_{i,B_1}}
=\langle \phi_{A_2,B_2}^0\ket{u_{i,A_2}\otimes u_{i,B_2}}
=\frac{1}{\sqrt{d}}$.
In this case, $\sum_i p_i= d^2$.
Thus,
\begin{align*}
T
=&
\sum_i p_i
\int_{SU(d)}
g_1\otimes \overline{g_1}
\ket{u_{i,A_1}\otimes u_{i,B_1}}
\bra{u_{i,A_1}\otimes u_{i,B_1}}
(g_1\otimes \overline{g_1})^\dagger
\nu(d g_1) \\
&\quad \otimes
\int_{SU(d)}
g_2\otimes \overline{g_2}
\ket{u_{i,A_2}\otimes u_{i,B_2}}
\bra{u_{i,A_2}\otimes u_{i,B_2}}
(g_2\otimes \overline{g_2})^\dagger
\nu(d g_2)\\
=&
\sum_i p_i
\left(
\frac{1}{d}
\ket{\phi_{A_1,B_1}^0}\bra{\phi_{A_1,B_1}^0}
+ a_i(I- \ket{\phi_{A_1,B_1}^0}\bra{\phi_{A_1,B_1}^0})\right)
\otimes
\left(\frac{1}{d}
\ket{\phi_{A_2,B_2}^0}\bra{\phi_{A_2,B_2}^0}
+ b_i(I- \ket{\phi_{A_2,B_2}^0}\bra{\phi_{A_2,B_2}^0})\right).
\end{align*}
Lemma \ref{l-3} implies $a_i,b_i \ge \frac{1}{d(d+1)}$.
Thus,
\begin{align*}
T\le&
\sum_i p_i
\left(
\frac{1}{d}
\ket{\phi_{A_1,B_1}^0}\bra{\phi_{A_1,B_1}^0}
+ \frac{1}{d(d+1)}(I- \ket{\phi_{A_1,B_1}^0}\bra{\phi_{A_1,B_1}^0})\right)\\
&\quad \otimes
\left(\frac{1}{d}
\ket{\phi_{A_2,B_2}^0}\bra{\phi_{A_2,B_2}^0}
+ \frac{1}{d(d+1)}(I- \ket{\phi_{A_2,B_2}^0}\bra{\phi_{A_2,B_2}^0})\right)\\
= &
\left(\ket{\phi_{A_1,B_1}^0}\bra{\phi_{A_1,B_1}^0}
+ \frac{1}{d+1}
(I- \ket{\phi_{A_1,B_1}^0}\bra{\phi_{A_1,B_1}^0})\right)
\otimes
\left(
\ket{\phi_{A_2,B_2}^0}\bra{\phi_{A_2,B_2}^0}
+ \frac{1}{d+1}(I- \ket{\phi_{A_2,B_2}^0}\bra{\phi_{A_2,B_2}^0})\right)\\
= & T_{inv}^{1,A_1\to B_1}\otimes T_{inv}^{1,A_2\to B_2}.
\end{align*}
Hence
\begin{align*}
\Tr T \sigma_1\otimes \sigma_2
\le
\Tr T_{inv}^{1,A_1\to B_1}\otimes T_{inv}^{1,A_2\to B_2}
 \sigma_1\otimes \sigma_2
= (1-\frac{d p_1}{d+1})
(1-\frac{d p_2}{d+1}).
\end{align*}

\end{widetext}
\section{Proof of (\ref{2-17-2})}\Label{a26}
Similarly to proof of Theorem \ref{t-4}, 
the $U(1)\times U(1)$-invariance implies that
this testing problem can be resulted in 
the testing problem of the probability distribution
$P^n_{p_1,p_2}(k_1,k_2)\defeq P^n_{p_1}(k_1)P^n_{p_2}(k_2)$
with the null hypothesis
$p_1 + p_2 \le \frac{\delta}{n}$.
When $n$ is large enough,
the probability distribution
$P^n_{t_1/n,t_2/n}(k_1,k_2)$
can be approximated by
the Poisson distribution
$P_{t_1,t_2}(k_1,k_2)
= e^{-t_1}e^{-t_2}\frac{t_1^{k_1}t_2^{k_2}}{k_1!k_2!}
=e^{-(t_1+t_2)}\frac{(t_1+t_2)^{k_1+k_2}}{(k_1+k_2)!}
\genfrac{(}{)}{0pt}{}{k_1+k_2}{k_1}
\left(\frac{t_1}{t_1+t_2}\right)^{k_1}
\left(\frac{t_2}{t_1+t_2}\right)^{k_2}$.

In order to calculate the lower bound of 
the optimal second error probability of the probability distribution
$P_{t_1',t_2'}(k_1,k_2)$,
we treat the hypothesis testing with null hypothesis
$t_1+t_2 \le \delta$ only on 
the one-parameter probability distribution family
$\{P_{st_1',s t_2'}(k_1,k_2)| 0 \le s \,< \infty\}$.
In this case, the  probability distribution
$P_{st_1',s t_2'}(k_1,k_2)$
has the form
\begin{align*}
&P_{st_1',s t_2'}(k_1,k_2)\\
= &
e^{-s(t_1'+t_2')}\frac{(s(t_1'+t_2'))^{k_1+k_2}}{(k_1+k_2)!}
\genfrac{(}{)}{0pt}{}{k_1+k_2}{k_1}\\
& \quad \left(\frac{t_1'}{t_1'+t_2'}\right)^{k_1}
\left(\frac{t_2'}{t_1'+t_2'}\right)^{k_2}.
\end{align*}
Hence, the likelihood ratio 
$\frac{P_{st_1',s t_2'}(k_1,k_2)}{P_{s't_1',s t_2'}(k_1,k_2)}$
depends only on the sum $k_1+k_2$.
Since
\begin{align*}
\sum_{k_1=0}^{k}
P_{st_1',s t_2'}(k_1,k-k_1)= 
e^{-s(t_1'+t_2')}
\frac{(s(t_1'+t_2'))^{k}}{k!},
\end{align*}
this hypothesis testing can be resulted in 
the hypothesis testing of Poisson distribution 
$e^{-t}
\frac{t^{k}}{k!}$ with the null hypothesis $t\le \delta$.
In this case, when the true distribution
is $e^{-(t_1'+t_2')}
\frac{(t_1'+t_2')^{k}}{k!}$,
the second error is greater than
$\beta_\alpha(\le \delta\|t_1'+t_2')$.
Therefore, we can conclude that
\begin{align*}
\lim \beta_{\alpha,2n,G\times G}^{\emptyset}
(\le \frac{\delta}{n}\|\sigma_{1,n}'\otimes \sigma_{2,n}')
\ge 
\beta_\alpha(\le \delta\|t_1'+t_2').
\end{align*}
Conversely,
we only focus on the random variable
$k=k_1+k_2$,
we obtain probability distribution
$e^{-(t_1+t_2)}
\frac{(t_1+t_2)^{k}}{k!}$.
Using the optimal hypothesis testing 
of the Poisson distribution,
we can construct test achieving the lower bound 
$\beta_\alpha(\le \delta\|t_1'+t_2')$.

\begin{widetext}
\section{Proof of (\ref{2-19-2}) and (\ref{2-19-3})}\Label{a27}
Let $T$
be an $SU(d)\times SU(d)\times SU(d)$-
invariant $A-B$ separable test
with level $0$.
The $SU(d)\times SU(d)\times SU(d)$-invariance implies that
the test $T$ has the form
\begin{align*}
T
=& 
\sum_i q_i
d^3 
\int_{SU(d)}\int_{SU(d)}\int_{SU(d)}
g_1\otimes g_3\otimes g_3\otimes 
\overline{g_1}\otimes\overline{g_2}\otimes\overline{g_3}
\ket{u_{i,A}\otimes u_{i,B}}
\bra{u_{i,A}\otimes u_{i,B}}\\
&\quad 
(
g_1\otimes g_3\otimes g_3\otimes 
\overline{g_1}\otimes\overline{g_2}\otimes\overline{g_3}
)^\dagger
\nu(d g_1)\nu(d g_2)\nu(d g_3),
\end{align*}
where
$\langle \phi_{A_1,B_1}^0\otimes
\phi_{A_2,B_2}^0\otimes \phi_{A_3,B_3}^0
\ket{u_{i,A}\otimes u_{i,B}}
=\frac{1}{\sqrt{d^3}}$.
In this case, $\sum_i q_i= 1$.
First, we focus on 
\begin{align*}
T_i \defeq &d^3 
\int_{SU(d)}\int_{SU(d)}\int_{SU(d)}
g_1\otimes g_3\otimes g_3\otimes 
\overline{g_1}\otimes\overline{g_2}\otimes\overline{g_3}
\ket{u_{i,A}\otimes u_{i,B}}
\bra{u_{i,A}\otimes u_{i,B}}\\
&\quad 
(
g_1\otimes g_3\otimes g_3\otimes 
\overline{g_1}\otimes\overline{g_2}\otimes\overline{g_3}
)^\dagger
\nu(d g_1)\nu(d g_2)\nu(d g_3)\\
=& P_1 \otimes P_2 \otimes P_3 
+a_{0,0,1}^i P_1 \otimes P_2 \otimes P_3^c 
+a_{0,1,0}^i P_1^c \otimes P_2 \otimes P_3
+a_{1,0,0}^i P_1^c \otimes P_2 \otimes P_3 \\
&\quad 
+a_{0,1,1}^i P_1 \otimes P_2^c \otimes P_3^c 
+a_{1,0,1}^i P_1^c \otimes P_2 \otimes P_3^c
+a_{1,1,0}^i P_1^c \otimes P_2^c \otimes P_3 
+a_{1,1,1}^i P_1^c \otimes P_2^c \otimes P_3^c.
\end{align*}
In order to calculate 
the coefficients 
$a^i_{j,k,l}$,
we treat 
the quantities
$\|\langle \phi_{A_1,B_1}^0\ket{u_{i,A}\otimes u_{i,B}}\|^2$,
$\|\langle \phi_{A_2,B_2}^0\otimes \phi_{A_3,B_3}^0
\ket{u_{i,A}\otimes u_{i,B}}\|^2$, etc.
In the following, we omit the subscript $i$.
Let $X=(x_{k,l})_{1\le k \le d,
1\le l \le d^2}$ ($Y$) be a $d\times d^2$ matrix corresponding
the vector $u_{A}$ ($u_B$) on the
entangled state between two systems $\cl{H}_{A_1}$ and 
$\cl{H}_{A_2}\otimes \cl{H}_{A_3}$
($\cl{H}_{B_1}$ and 
$\cl{H}_{B_2}\otimes \cl{H}_{B_3}$), respectively.
Then,
\begin{align*}
\langle \phi_{A_1,B_1}^0\ket{u_{A}\otimes u_{B}}
&= \frac{1}{\sqrt{d}}
\sum_{l_A=1}^{d^2}\sum_{l_B=1}^{d^2}
(Y^tX)_{l_B,l_A}\ket{l_A,l_B},\\
\langle  \phi_{A_2,B_2}^0\otimes \phi_{A_3,B_3}^0
\ket{u_{A}\otimes u_{B}}
&= \frac{1}{\sqrt{d^2}}
\sum_{k_A=1}^{d}\sum_{k_B=1}^{d}
(X^t Y)_{k_A,k_B}\ket{k_A,k_B}.
\end{align*}
Hence,
\begin{align*}
\|\langle \phi_{A_1,B_1}^0\ket{u_{A}\otimes u_{B}}\|^2
&= \frac{1}{d}
\Tr (Y^tX)(Y^tX)^{\dagger},\\
\|\langle  \phi_{A_2,B_2}^0\otimes \phi_{A_3,B_3}^0
\ket{u_{A}\otimes u_{B}}\|^2
&= \frac{1}{d^2}
\Tr (X^t Y)(X^t Y)^\dagger=
\frac{1}{d^2}\Tr (Y^tX)(Y^tX)^{\dagger}.
\end{align*}
That is, when we put $
\beta_1\defeq|\langle \phi_{A_1,B_1}^0\ket{u_{A}\otimes u_{B}}\|^2$,
$\frac{\beta_1}{d}= 
\|\langle  \phi_{A_2,B_2}^0\otimes \phi_{A_3,B_3}^0
\ket{u_{A}\otimes u_{B}}\|^2$.
Since $\frac{1}{\sqrt{d^3}}= 
\langle  \phi_{A_1,B_1}^0\otimes \phi_{A_2,B_2}^0\otimes 
\phi_{A_3,B_3}^0\ket{u_{A}\otimes u_{B}}=
\frac{1}{\sqrt{d^3}} \Tr (Y^tX)$,
\begin{align*}
d \beta_1 = \Tr (Y^tX)(Y^tX)^{\dagger} \ge \frac{1}{d}.
\end{align*}
The equality holds if and only if
$(Y^tX)$ is the completely mixed state.
Hence, the equality holds when
$u_A=u_B =\ket{GHZ}$.
Similarly, we define the quantities $\beta_2$ and $\beta_3$.
We also define $\gamma \defeq \|u_A\|^2\|u_B\|^2$, which satisfies 
the inequality
\begin{align*}
\gamma \ge 1.
\end{align*}
Indeed, when $u_A=u_B =\ket{GHZ}$,
$\gamma=1$.
Thus, 
by calculating the trace of 
the products of corresponding projections and
$\ket{u_{i,A}\otimes u_{i,B}}
\bra{u_{i,A}\otimes u_{i,B}}$,
the coefficients
can be calculated as
\begin{align*}
a_{0,0,1} &=\frac{d^3}{d^2-1}\left(\frac{\beta_3}{d}-\frac{1}{d^3}\right),
\quad
a_{0,1,1} =\frac{d^3}{(d^2-1)^2}
\left(\beta_1- \frac{\beta_2+\beta_3}{d}+\frac{1}{d^3}\right)\\
a_{0,1,0} &=\frac{d^3}{d^2-1}\left(\frac{\beta_2}{d}-\frac{1}{d^3}\right),
\quad
a_{1,0,1} =\frac{d^3}{(d^2-1)^2}
\left(\beta_2- \frac{\beta_1+\beta_3}{d}+\frac{1}{d^3}\right)\\
a_{1,0,0} &=\frac{d^3}{d^2-1}\left(\frac{\beta_1}{d}-\frac{1}{d^3}\right),
\quad
a_{1,1,0} =\frac{d^3}{(d^2-1)^2}
\left(\beta_3- \frac{\beta_2+\beta_1}{d}+\frac{1}{d^3}\right)\\
a_{1,1,1} &=\frac{d^3}{(d^2-1)^3}
\left(\gamma - \frac{d-1}{d}(\beta_1+\beta_2+\beta_3)
- \frac{1}{d^3}
\right).
\end{align*}
Therefore,
substituting $\beta_i=\frac{1}{d^2},\gamma=1$,
we obtain (\ref{2-19-2}).

Moreover,
\begin{align*}
\Tr T_i \sigma_1 \otimes \sigma_2 \otimes \sigma_3
= C_0+ C_1 \beta_1+C_2 \beta_2+C_3 \beta_3
+\frac{p_1 p_2 p_3 d^3}{(d^2-1)^3)} \gamma,
\end{align*}
where
\begin{align*}
C_0 \defeq &
(1-p_1)(1-p_2)(1-p_3)
+ \frac{3 p_1 p_2 p_3 -2(p_1 p_2+ p_2 p_3 + p_3 p_1)
p_1+p_2+p_3}{d^2-1}\\
&\quad + \frac{-3 p_1 p_2 p_3 +(p_1 p_2+ p_2 p_3 + p_3 p_1)
}{(d^2-1)^2}
+
\frac{p_1 p_2 p_3}{(d^2-1)^3}\\
C_1 \defeq &
\frac{d^2}{(d^2-1)^2(d+1)}
\left(
d(d+1)(1-\frac{d}{d+1}p_1)p_2 p_3
+ p_1(d+1)^2(d-1)
(1- \frac{d}{d-1}p_2)(1- \frac{d}{d+1}p_3)
\right) \\
C_2 \defeq &
\frac{d^2}{(d^2-1)^2(d+1)}
\left(
d(d+1)(1-\frac{d}{d+1}p_2)p_3 p_1
+ p_2(d+1)^2(d-1)
(1- \frac{d}{d-1}p_3)(1- \frac{d}{d+1}p_1)
\right) \\
C_3 \defeq &
\frac{d^2}{(d^2-1)^2(d+1)}
\left(
d(d+1)(1-\frac{d}{d+1}p_3)p_1 p_2
+ p_3(d+1)^2(d-1)
(1- \frac{d}{d-1}p_1)(1- \frac{d}{d+1}p_2)
\right) .
\end{align*}
It follows from the condition
$p_i \le \frac{d-1}{d}$ that
these coefficients $C_1,C_2$ and $C_3$ are positive.
Hence,
\begin{align*}
&\Tr T_i \sigma_1 \otimes \sigma_2 \otimes \sigma_3
\ge 
 C_0+ (C_1+C_2 +C_3 )\frac{1}{d^2}
+\frac{p_1 p_2 p_3 d^3}{(d^2-1)^3} \\
= & 
(1-p_1)(1-p_2)(1-p_3)
+ \frac{(d+2)p_1 p_2 p_3}{(d+1)^2(d-1)}
+ \frac{p_1 p_2(1-p_3)+p_1 (1-p_2)p_3 + (1-p_1) p_2 p_3}{(d+1)^2(d-1)} .
\end{align*}
Therefore,
\begin{align*}
&\Tr T \sigma_1 \otimes \sigma_2 \otimes \sigma_3
= \sum_i q_i 
\Tr T_i \sigma_1 \otimes \sigma_2 \otimes \sigma_3\\
\ge &
(1-p_1)(1-p_2)(1-p_3)
+ \frac{(d+2)p_1 p_2 p_3}{(d+1)^2(d-1)}
+ \frac{p_1 p_2(1-p_3)+p_1 (1-p_2)p_3 + (1-p_1) p_2 p_3}{(d+1)^2(d-1)}. 
\end{align*}
Thus, we obtain (\ref{2-19-3})
for $G=SU(d)$.
Moreover, since this bound can be
attained by 
a $U(d^2-1)\times U(d^2-1)\times U(d^2-1)$-invariant test,
the equation (\ref{2-19-3}) holds for $G=U(d^2-1)$.
\end{widetext}

\end{document}